\documentclass[prd,twocolumn,groupedaddress]{revtex4-1}

\usepackage{graphicx}
\usepackage{xcolor}
\usepackage{amsmath}

\usepackage{hyperref}
\newcommand{\be}{\begin{eqnarray}}
\newcommand{\ee}{\end{eqnarray}}

\begin{document}

%Title of paper
%\title{The viability of halo models for the 21-cm power spectrum at cosmic dawn}
\title{A halo model approach for the 21-cm power spectrum at cosmic dawn}

\author{Aurel Schneider}
\email{aurel.schneider@uzh.ch}
 \affiliation{Center for Cosmology and Theoretical Astrophysics, Institute for Computational Science, University of Zurich, Switzerland.}
\author{Sambit K. Giri}
 \email{sambitkumar.giri@uzh.ch}
\affiliation{Center for Cosmology and Theoretical Astrophysics, Institute for Computational Science, University of Zurich, Switzerland.}
\author{Jordan Mirocha}
\email{jordan.mirocha@mcgill.ca}
\affiliation{McGill University, Department of Physics \& McGill Space Institute, Montr\'eal, Canada.}

\date{\today}

\begin{abstract}
Prior to the epoch of reionisation, the 21-cm signal of the cosmic dawn is dominated by the Lyman-$\alpha$ coupling and gas temperature fluctuations caused by the first sources of radiation. While early efforts to model this epoch relied on analytical techniques, the community quickly transitioned to more expensive semi-numerical models. Here, we re-assess the viability of simpler approaches that allow for rapid explorations of the vast astrophysical parameter space. We propose a new analytical method to calculate the 21-cm power spectrum based on the framework of the halo model. Both the Lyman-$\alpha$ coupling and temperature fluctuations are described by overlapping radiation flux profiles that include spectral red-shifting and source attenuation due to look-back (light-cone) effects. The 21-cm halo model is compared to the semi-numerical code {\tt 21cmFAST} exhibiting generally good agreement, i.e., the power spectra differ by less than a factor of three over a large range of $k$-modes and redshifts. We show that the remaining differences between the two methods are comparable to the expected variations from modelling uncertainties associated with the abundance, bias, and accretion rates of haloes. While these current uncertainties must be reduced in the future, our work suggests that inference at acceptable accuracy will become feasible with very efficient halo models of the cosmic dawn.
\end{abstract}

\maketitle

\section{\label{sec:intro}Introduction}
Current and upcoming radio interferometer telescopes such as LOFAR \citep[][]{Haarlem:2006aaa, Mertens:2020llj}, MWA \citep[][]{Tingay:2013aaa,Trott:2020szf}, HERA \citep[][]{DeBoer:2016tnn}, or SKA \citep[][]{Koopmans:2015sua} are expected to detect for the first time the strongly redshifted 21-cm clustering signal during and prior to the epoch of reionisation.
%Current and upcoming radio interferometer telescopes such as the Low Frequency Array \citep[LOFAR,][]{Haarlem:2006aaa, Mertens:2020llj}, the Murchison Widefield Array \citep[MWA,][]{Tingay:2013aaa,Trott:2020szf}, the Hydrogen Epoch of Reionisation Array \citep[HERA,][]{DeBoer:2016tnn}, or the Square Kilometre Array \citep[SKA,][]{Koopmans:2015sua} are expected to detect for the first time the strongly redshifted 21-cm clustering signal during and prior to the epoch of reionisation.
These measurements will open up a new window on the universe \citep{Madau:1996cs,Shaver:1999gb}, providing insights into both astrophysics \citep{Ewall-Wice:2018bzf,Mirocha:2019gob,Park:2018ljd} and cosmology \citep{McQuinn:2005hk,Liu:2019srd}. Next to the very first stars and galaxies \citep{Barkana:2004vb,Greig:2020hty}, the 21-cm signal has the potential to find evidence for exotic sources of radiation \citep{Liu:2018uzy,Mena:2019nhm}, new signs from the dark matter sector \citep{Sitwell:2013fpa,Lopez-Honorez:2016sur,Schneider:2018xba,Lidz:2018fqo,Nebrin:2018vqt} or other deviations from the standard $\Lambda$CDM cosmological model \citep[][]{Hill:2018lfx,Yang:2019nhz,Lopez-Honorez:2020lno}.

Predicting the 21-cm signal of the cosmic dawn, however, remains a challenging task. On the one hand, the complicated physics of radiation-hydrodynamics combined with the enormous ranges of relevant scales make brute-force simulations extremely difficult. On the other hand, the poorly known characteristics of early star-forming sources as well as the complicated interplay between gas cooling and feedback add important uncertainties that need to be either understood or properly parametrised.

First attempts to predict the 21-cm clustering before reionisation were based on analytical techniques, using a combination of cosmological perturbation theory and excursion-set modelling prescriptions for the sources \citep{Furlanetto:2004nh,Barkana:2004vb,Pritchard:2006sq} (for a more recent attempt see also Ref.~\citep{Raste:2017yse}). While these calculations were able to predict many important features of the 21-cm power spectrum -- such as for example the characteristic double-peaked shape due to the Lyman-$\alpha$ coupling and heating epochs -- it remains unclear how well they agree with more detailed calculations \citep{Santos:2007aaa}.

A further important step towards more realistic predictions of the 21-cm signal at cosmic dawn was taken with the development of semi-numerical methods such as {\tt 21cmFAST} \citep{Mesinger:2007pd,Mesinger:2010aaa} or {\tt simfast21} \citep{Santos:2009aaa} (see also Refs.~\citep{Fialkov:2013uwm,Mutch:2016aaa,Hutter:2018qxa} for other semi-numerical methods). These models numerically evolve the matter perturbations, Lyman-$\alpha$ coupling, and temperature fluctuations on a grid, where the source distributions are either obtained via a halo catalogue from N-body simulations or via an excursion-set recipe. Semi-numerical techniques consist of a major improvement with respect to analytical approaches, mainly because they follow the evolution of the spin temperature in configuration space and are therefore able to produce maps of the 21-cm signal.

In principle, more accurate predictions can be obtained by post-processing numerical $N$-body simulations using radiative-transfer (RT) calculations. This is the strategy followed e.g. by {\tt Grizzly} \citep{Thomas:2008uq,Ghara:2014yfa}, {\tt C$^2$-Ray} \citep{Mellema:2005ht,Friedrich:2012aaa}, and {\tt CRASH} \citep{Graziani:2013aaa}. While the former code is based on approximate but faster one-dimensional RT calculations, the latter two follow a full three dimensional ray-tracing approach. First results for the epoch of pre-reionisation can be found in Refs.~\citep{Ghara:2014yfa,Ghara:2017vby} as well as Ref.~\citep{Ross:2018uhh} and \citep{Eide:2020xyi}.

Full radiation-hydrodynamical simulations, including self-consistent formation of sources, are becoming an increasingly common tool to study the process of reionisation \citep[e.g. Refs.~][]{Ocvirk:2015xzu,Doussot:2017cee,Ocvirk:2018pqh,Rosdahl:2018aaa,Katz:2019due}. However, for the pre-reionisation epoch of cosmic dawn, such simulations remain rare \citep[][]{Semelin:2017xgv}. This is because they require narrowly binned multi-frequency RT calculations in a simulation box that resolves the mini-haloes hosting the first sources and that accounts for the large distances travelled by Lyman-$\alpha$ and X-ray radiation.

In this paper we propose an analytical approach to predict the 21-cm global signal and power spectrum at cosmic dawn. While being potentially less accurate than the different numerical techniques mentioned above, such a model has the advantage of providing fast predictions based on a well defined framework. Different parametrisations of the source modelling or effects from varying cosmology can be readily implemented and tested. An analytical method is also particularly well suited for parameter inference, which is an important aspect of 21-cm cosmology due to the large uncertainties related to early-universe galaxy formation that have to be parametrised and marginalised over. Comparing multi-parameter models with observations requires a very large number of calculations that can only be performed with fast prediction routines.

Our method is inspired by earlier work of \citet{Holzbauer:2011aaa} who used the halo model to predict the clustering from the Lyman-$\alpha$ coupling between the gas and the first star-light at cosmic dawn. We extend this approach adding a description for the temperature fluctuations as well as an improved modelling of the halo accretion and the star formation rates. With this at hand, we obtain a complete prediction of the 21-cm power spectrum at cosmic dawn.

The paper is structured as follows. In Sec.~\ref{sec:sources} we discuss the source modelling including halo mass function, bias, gas accretion, and star-formation efficiency. Sec.~\ref{sec:dTb} summarises the derivation of the differential brightness temperature. In Sec.~\ref{sec:hm} we present the halo model of flux profiles and we show results for the 21-cm power spectrum assuming different star-formation efficiencies. We then go on and compare our model to other analytical and semi-numerical methods in Sec.~\ref{sec:comparisons} before concluding in Sec.~\ref{sec:conclusions}.

\section{\label{sec:sources}Source modelling}
Accurately quantifying the abundance, distribution, and emission of sources is a crucial step for any method predicting the 21-cm signal. The source distribution, for example, can be readily quantified using prescriptions for the halo mass function and the halo bias. Regarding the source emission, analytical and semi-numerical models usually rely on estimates for the halo accretion rates combined with a parametrisation of the star-formation efficiency. While the former describes the accretion of primordial gas onto the parent halo, the latter generically parametrises all star formation and feedback processes.

In the present section, we will first discuss the halo mass function and biasing model, before we describe and compare different estimates for halo mass accretion rates. At the end we present our parametrisation for the star formation efficiency parameter.

\subsection{Halo mass function and bias}\label{sec:massfctbias}
Both the halo mass function and the halo bias are central ingredients of the halo model which our method is based on. For the halo mass function, we assume the standard form motivated by the extended Press-Schechter (PS) approach \citep{Press:1973iz,Bond:1990iw}, i.e., 
\be\label{massfct}
\frac{dn}{d\ln M} = \frac{1}{2}\frac{\bar\rho}{M}f(\nu)\frac{d\ln \nu}{d\ln M}
\ee
where the peak-height of perturbations is defined as $\nu=\delta_{c}(z)^2/\sigma^2$ with $\delta_{c}(z)=1.686/D(z)$, $D(z)$ being the cosmological growth factor. The variance of the density field ($\sigma^2$) is defined as
\begin{equation}
\sigma^2(M) = \int\frac{d^3\mathbf{k}}{(2\pi)^3}P(k)W^2(k|M),
\end{equation}
where $P(k)$ is the linear power spectrum at redshift zero and $W(k|M)$ is the Fourier transform of the top-hat window function. The first crossing distribution ($f$) us given by
\be\label{fnu}
f(\nu)=A\sqrt{2\frac{q\nu}{\pi}}\left[1+(q\nu)^{-p}\right]\exp(-q\nu/2),
\ee
which comes with two free model parameters $q$ and $p$ \citep[see Ref.][]{Sheth:2001dp}. The factor A is fixed via the relation $A=(1 + 2^{-p}\Gamma(1/2-p)/\sqrt{\pi})^{-1}$.

Using the peak-background split model introduced in Ref.~\citep{Sheth:1999mn}, the halo bias can be described within the same extended PS formalism, leading to the relation \citep[see e.g Ref.][]{Cooray:2002dia}
\be\label{bias}
b(M)=1+\frac{q\nu-1}{\delta_{c}(z)} + \frac{2p}{\delta_{c}(z)[1+(q\nu)^p]}.
\ee

In the following, we use the values $q=0.85$ and $p=0.3$ unless stated otherwise. We have verified that this setup provides a  better fit to the halo mass function of the high-redshift $N$-body simulations from Refs.~\citep{Iliev:2011aaa,Schneider:2018xba} than the original Press-Schechter \citep{Press:1973iz} or Sheth-Tormen \citep{Sheth:2001dp} mass functions (which are characterised by $q=1$, $p=0$ and $q=0.707$, $p=0.3$, respectively).

\begin{figure*}[tbp]
\centering
\includegraphics[width=0.32\textwidth,trim=0.0cm 0.0cm 1.0cm 0.0cm,clip]{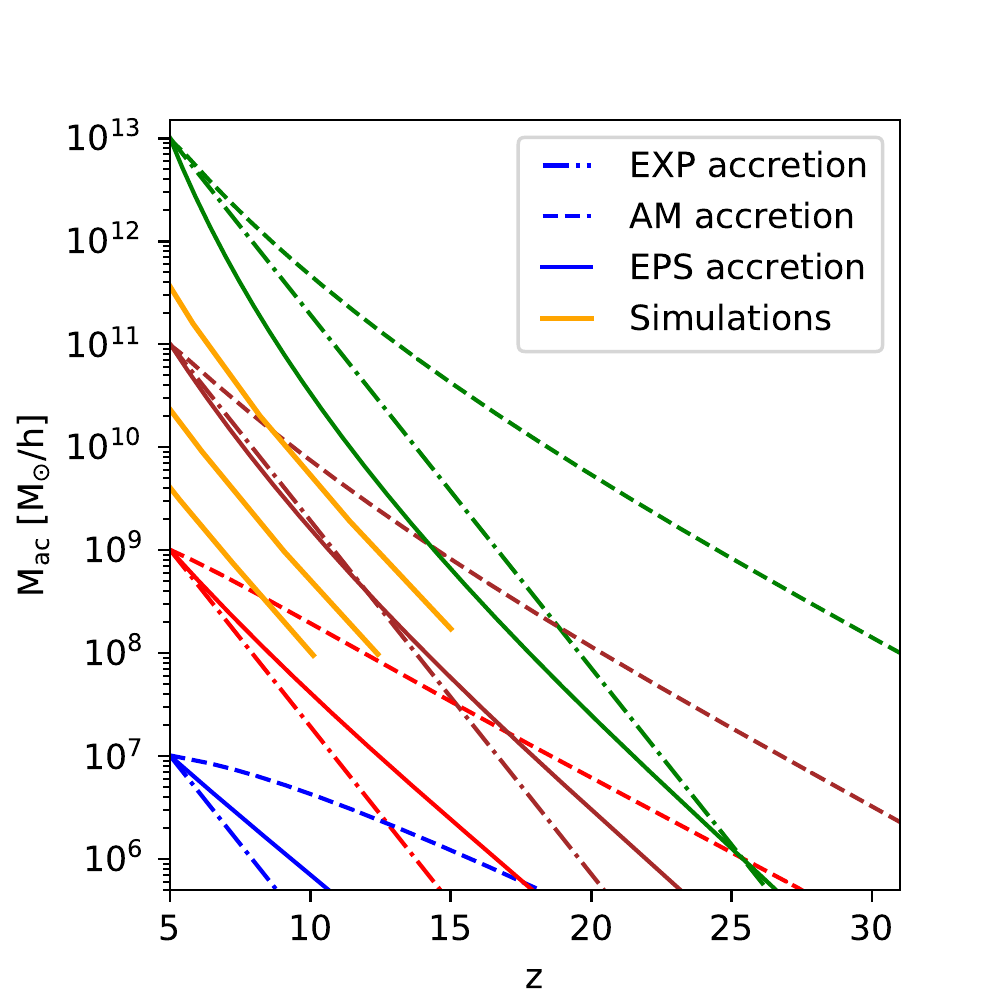}
\includegraphics[width=0.32\textwidth,trim=0.0cm 0.0cm 1.0cm 0.0cm,clip]{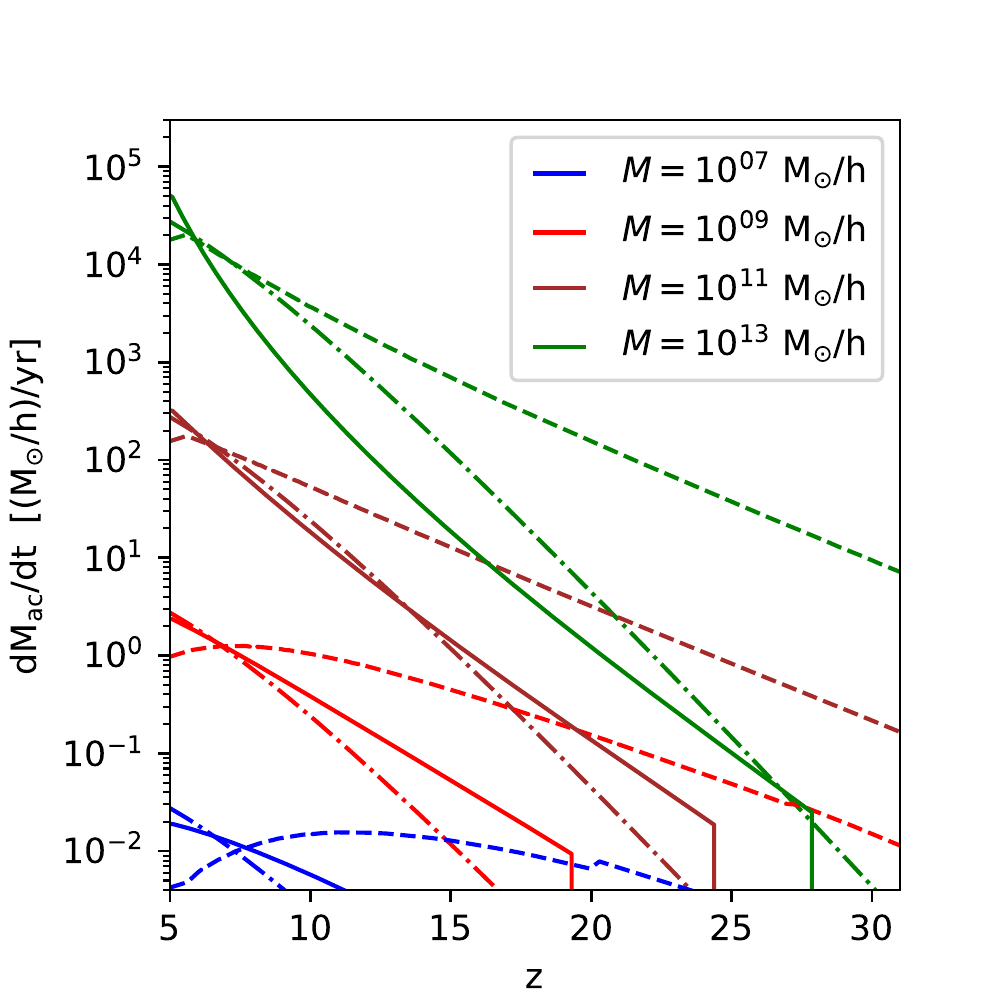}
\includegraphics[width=0.32\textwidth,trim=0.0cm 0.0cm 1.0cm 0.0cm,clip]{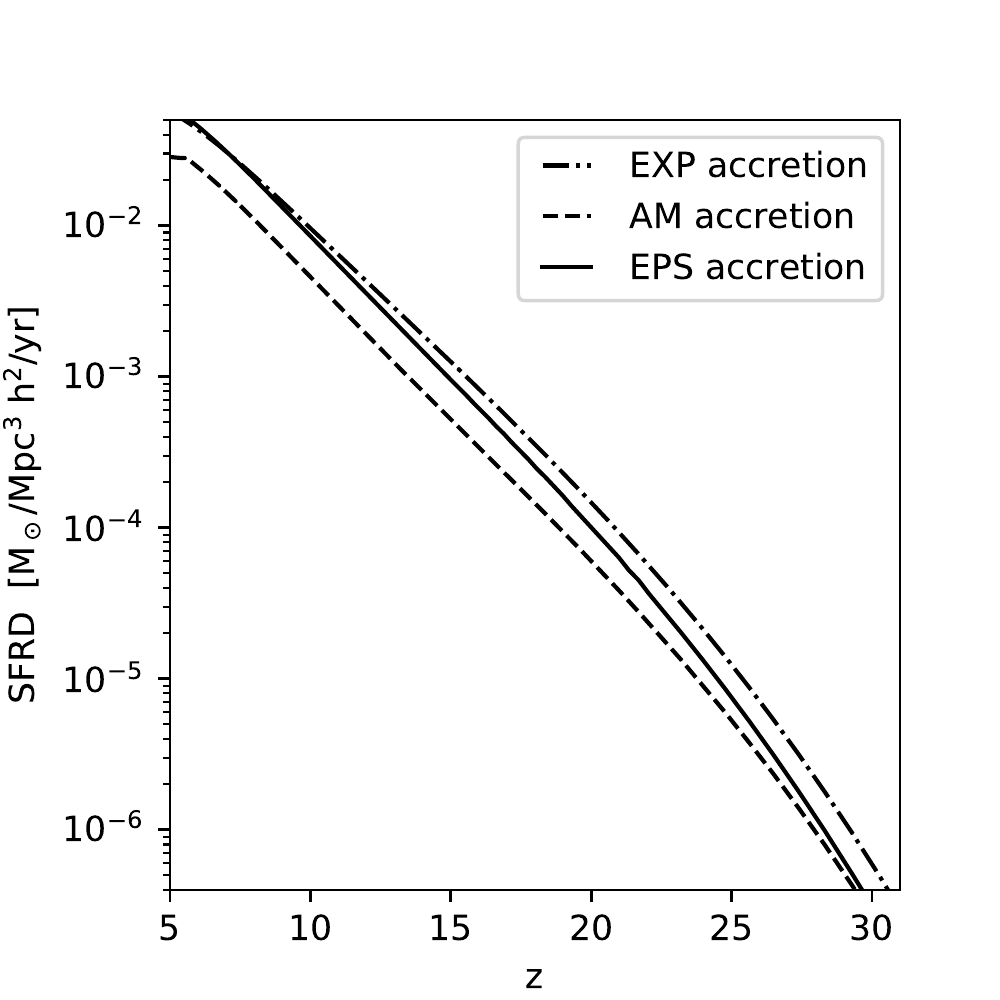}
\caption{Halo growth, mass accretion rate, and star-formation rate density (SFRD) for the three different models discussed in Sec.~\ref{sec:MA}. \emph{Left:} Halo mass as a function of redshift for four different final masses $M$ (green, brown, red, and blue lines). The orange lines correspond to the mean halo growth from the B20 \citep{Behroozi:2020jhj} simulations. \emph{Centre:} Mass accretion rates for the same models and halo masses. \emph{Right:} Resulting star-formation rate densities (SFRD) for the three different mass-accretion rate models (see Eq.~\ref{sfrd_Maccr})}
\label{fig:MA}
\end{figure*}

\subsection{Halo mass accretion}\label{sec:MA}
There exists different ways to analytically quantify the growth of haloes. Here we discuss three models that are used in the literature, and we compare them to results from $N$-body simulations.

\subsubsection{Exponential (EXP) halo growth}
The simplest way to obtain a halo accretion history is to assume that haloes grow exponentially. This assumption has been shown to provide a good match to simulations, especially at high redshifts \citep[see e.g. Refs][]{McBride:2009aaa,Dekel:2013uaa,Trac:2015rva}. We consider the following equation
\be\label{AMexp}
M_{\rm ac}(M,z) = M \exp\left[\alpha (z_0-z)\right]
\ee
where $M$ is the halo mass at the final redshift $z_0$. The factor $\alpha$ can be determined using simulations and has been shown to be only weakly mass dependent at high redshift. Following Ref.~\citep{Dekel:2013uaa} we therefore assume $\alpha=0.79$ independent of the halo mass $M$.

In the left-hand and centre panels of Fig.~\ref{fig:MA} we show the redshift evolution of the halo mass ($M_{\rm ac}$) and the accretion rate ($\dot M_{\rm ac}$). The dash-dotted lines correspond to the exponential (EXP) growth model of Eq.~(\ref{AMexp}). The model is in very good agreement with the numerical $N$-body simulations from \citet[B20]{Behroozi:2020jhj} (orange lines). Note, however, that the simulations do not cover all redshifts and mass ranges of interest to the present study.

\subsubsection{Extended Press-Schechter (EPS) method}
Halo accretion rates can also be obtained by means of the extended Press-Schechter (EPS) formalism. For example, \citet{Neistein:2006ak} calculated an ensemble of EPS merger trees and determined the average growth of their main branches. Based on this, they proposed the equation
\be\label{MAeps}
\frac{dM_{\rm ac}}{dz} = -\sqrt{\frac{2}{\pi}}\frac{M_{\rm ac}}{\sqrt{S(QM_{\rm ac})-S(M_{\rm ac})}}\frac{d\delta_c(z)}{dz},
\ee
where $S\equiv \sigma^2$ at redshift zero \citep[see also Ref.][]{Correa:2014xma}. This differential equation can be solved assuming $M_{\rm ac}(M,z_0)\equiv M$, where $z_0$ designates the final redshift of interest. The value of $Q$ has to be selected empirically, and we use $Q=0.6$. This number is larger than the range $Q=0.43-0.5$ proposed by \citet{Neistein:2006ak} but it provides a better agreement with the simulations from B20 (see orange lines in Fig.~\ref{fig:MA}).

The predicted EPS halo growth and accretion rates are shown as solid lines in the left-hand and central panels of Fig.~\ref{fig:MA}. The evolution is similar to the exponential model except at small masses, where the halo growth is less pronounced. The model is in good agreement with the B20 simulations, which should not come as a surprise since we have recalibrated the $Q$-parameter according to these simulations.

The advantage of the EPS model compared to the much simpler exponential model lies in the fact that Eq.~(\ref{MAeps}) is, in principle, sensitive to changes in cosmology. Whether the true cosmology dependence of the mass accretion rate can be accurately reproduced by the EPS model, remains, however, to be tested.

\subsubsection{Abundance matching (AM) method}
The final method we are investigating here is inspired by the abundance matching (AM) technique \citep{Vale:2004yt} and was first applied in Ref.~\citep{Furlanetto:2016aaa} as a measure of high-redshift galaxy growth. The method aims to connect haloes between different redshift bins ($z_n$) by matching
\be
\int_{M_n}^{\infty} dM \frac{dn}{dM}(M,z_n) = \int_{M_{n-1}}^{\infty} dM \frac{dn}{d\ln M}(M,z_{n-1}),
\ee
where the halo mass function $dn/d M$ is obtained via Eq.~(\ref{massfct}).
%\be
%\int_{M_n}^{\infty} dM n(M,z_n) = \int_{M_{n-1}}^{\infty} dMn(M,z_{n-1}),
%\ee
%where $n(M,z)$ is the cumulative halo mass function defined by
%\be
%n(M,z)=\int_M^{\infty} \frac{dn}{d\ln M'} \left(\frac{dM'}{M'}\right).
%\ee
Connecting all masses $M_n$ at different redshifts $z_n$ allows us to estimate the accretion rate of haloes. Note that the AM method connects haloes in a strictly hierarchical way, which means that the $i$-largest halo at the final redshift $z_{0}$ will be assumed to stay the $i$ largest halo at all higher redshifts. As a consequence, the model implicitly assumes haloes to exclusively grow via smooth accretion, since halo mergers cannot be accounted for.

In Fig.~\ref{fig:MA} the halo growth and accretion rates of the AM model are shown as dashed lines. Compared to the other methods, the AM model predicts significantly slower halo growth over all redshifts and mass ranges. The AM results do not match the B20 simulations very well. Note that our findings are in qualitative agreement with Ref.~\citep{Membane:2017aaa}, where the AM model was compared to a fitting function of Ref.~\citep{Trac:2015rva}.

\subsection{Star-formation efficiency and stellar-to-halo mass ratio}\label{sec:sfe}
The halo growth rate discussed above provides an estimate of the amount of total matter accreted onto a halo. However, we still need to parametrise the star-formation efficiency, i.e., how much of that accreted matter will be transformed into stars that emit radiation. Following Refs.~\citep{Mason:2015cna,Sun:2016aaa,Mirocha:2016aaa}, we define 
\be\label{deffstar}
f_*(M)\equiv\dot M_*/\dot M_{\rm ac},
\ee
where $\dot M_*$ and $\dot M_{\rm ac}$ are the stellar and halo accretion rates. Note that Eq.~(\ref{deffstar}) is different from the stellar-to-halo mass ratio
\be
\tilde{f}_*(M)\equiv M_*/M_{\rm ac}
\ee
often used in the literature \citep[see e.g. Ref.][]{Pritchard:2011xb,Park:2018ljd}. The star-formation efficiency and the stellar-to-halo mass ratio can be connected by the integral
\be\label{ftilde}
\tilde{f}_*(M)= \frac{1}{M_{\rm ac}}\int f_* \dot M_{\rm ac} dt.
\ee
Note that for the simple case of a redshift-independent star-formation efficiency, Eq.~(\ref{ftilde}) leads to $\tilde{f}_*=f_*$. A more general study of the relation between star-formation efficiency and stellar-to-halo mass ratio, including explicit redshift dependences motivated by feedback processes, can be found in Refs.~\citep{Furlanetto:2016aaa,Mirocha:2020sid}.

In this paper, we use the functional form
\be\label{fstar}
f_*(M)= \frac{2 (\Omega_b/\Omega_m)f_{*,0}}{\left(M/M_p\right)^{\gamma_1}+\left(M/M_p\right)^{\gamma_2}}\times S(M)
\ee
as a parametrisation for the star-formation efficiency. Eq.~(\ref{fstar}) consists of a double-power law, multiplied with a small-scale function
\be
S(M)=\left[1+(M_t/M)^{\gamma_3}\right]^{\gamma_4},
\ee
that may provide either a suppression or a boost at the truncation mass scale $M_t$. A suppression at small scales could naturally occur at scales where atomic cooling processes become inefficient. A small-scale boost, on the other hand, could emerge due to the presence of population III stars in mini-haloes.

Note that Eq.~(\ref{fstar}) does not depend on redshift, and we can therefore set $f_*=\tilde f_*$ in this paper. We will nonetheless formally distinguish between $f_*$ and $\tilde f_*$ in order to avoid confusion and to acknowledge the fact that a more realistic description of the star-formation efficiency may well include an explicit redshift dependence.

\subsection{Star-formation rate density and collapse fraction}
With the halo mass function, accretion rate, and star formation efficiency at hand, it is possible to calculate the mean star-formation-rate density (SFRD), which is an important ingredient for calculating the global 21-cm signal (see Sec.~\ref{sec:dTb}). We define the SFRD as the integral over the halo mass function weighted by the star-formation efficiency and the accretion rate, i.e.,
\be\label{sfrd_Maccr}
\dot{\rho}_{*}(z) = \int dM \frac{dn}{dM} f_{*}(M) \dot{M}_{\rm ac}(M,z).
\ee
The definition above differs from the relation $\dot{\tilde\rho}_{*}(z) = \bar\rho df_{\rm coll}/dt$, which is based on the global collapse fraction ($f_{\rm coll}$) and is often used in the literature  \citep[e.g. Refs.][]{Barkana:2004vb,Pritchard:2005an}. The collapse fraction is given by the integral
\be\label{fcoll}
f_{\rm coll}(z)=\frac{1}{\bar\rho}\int dM \tilde{f}_{*}(M)\frac{dn}{d M}M,
\ee
providing the total ratio of stars to matter in the universe. Note that Eq~(\ref{fcoll}) includes the stellar-to-halo mass ratio inside of the integral and is therefore slightly different from the standard definition in the literature \citep[see e.g. Refs.][]{Barkana:2003qk,Mirocha:2014faa}.

The star-formation rate densities based on the three different mass accretion models introduced above are shown as black lines in the right-hand panel of Fig.~\ref{fig:MA}. While they have similar general trends, the exponential growth (EXP) model is about a factor of two larger than the method based on abundance matching (AM). The extended Press-Schechter (EPS) model lies in between, being closer to the AM method at very high and closer to the EXP model at lower redshifts.

The differences between the star-formation rate densities shown in Fig.~\ref{fig:MA} directly affect the modelling of the Lyman-$\alpha$ coupling, gas heating, and ionisation. This is one of the main reasons why comparisons between different methods to calculate the 21-cm signal are hard to interpret when they are using different implementations for the star-formation rate (as we will see in Sec.~\ref{sec:comparisons}).

\section{\label{sec:dTb}21-cm brightness temperature}
The 21-cm differential brightness temperature ($T_{21}$) is a function of the background radiation ($T_{\gamma}$), the spin temperature of the gas ($T_s$), the neutral hydrogen fraction ($x_{\rm HI}$) and the gas density field ($\delta_b$) which all depend on redshift $z$ and position $\mathbf{x}$. Following e.g. Ref.~\citep{Furlanetto:2006tf}, the brightness temperature can be written as
\begin{multline}\label{T21}
T_{21}(\mathbf{x},z) = 27x_{\rm HI}(1+\delta_b) \\
\times\left(\frac{\Omega_bh^2}{0.023}\right)^{}\left(\frac{0.15}{\Omega_mh^2}\frac{(1+z)}{10}\right)^{\frac{1}{2}}\left(1-\frac{T_{\gamma}}{T_s}\right)%\,\, {\rm mK}
\end{multline}
in milli-Kelvin [mK]. Assuming a standard $\Lambda$CDM model without exotic radio sources, the background temperature is dominated by the cosmic microwave background (CMB) radiation. The rightmost expression in Eq.~(\ref{T21}) can be written as
\be\label{Tratio}
\left(1-\frac{T_{\gamma}}{T_s}\right)\simeq \frac{x_{\rm tot}}{1+x_{\rm tot}}\left(1-\frac{T_{\gamma}}{T_{k}}\right),
\ee
where $x_{\rm tot}\equiv x_{\alpha}+x_{c}$ ($x_{\alpha}$ and $x_{c}$ denoting the radiative and collisional coupling coefficients). Throughout this paper, we set the collisional coupling to zero, as it is only important at very high redshifts beyond $z\sim 30$.

The gas temperature ($T_{k}$) is obtained via the differential equation
\be\label{Tgas}
\frac{3}{2}\frac{dT_{k}(\mathbf{x},z)}{dz}=\frac{T_{k}(\mathbf{x},z)}{\rho(\mathbf{x},z)}\frac{d\rho(\mathbf{x},z)}{dz}-\frac{\Gamma_{h}(\mathbf{x},z)}{k_B(1+z)H},
\ee
where $\rho$ is the matter density and $\Gamma_{h}$ the heating source term. The latter is given by the sum
\begin{multline}\label{Gammaheat}
\Gamma_{h}(\mathbf{x},z)=4\pi\sum_i f_if_{X,h}\\
\times \int_{\nu_{\rm th}^i}^{\infty} d\nu \left(\nu-\nu_{\rm th}^i\right)h_P\sigma_{i}(\nu)J_{X,\nu}(\mathbf{x},z)
\end{multline}
with $i=\{\rm H,He\}$, i.e. the hydrogen and helium components with fractions $f_i=n_{i}^{0}/n_b^{0}$, threshold energies $\nu_{\rm th}^i h_P$=\{13.6, 26.5\} eV, and cross sections $\sigma_i(\nu)$. For the fraction of X-ray energy deposited as heat, we assume $f_{X,h}=\bar x_e^{0.225}$, where $\bar x_e$ is the free electron fraction \citep[see Fig. 4 in Ref.][]{Furlanetto:2010aaa} that is calculated according to Eqs.~12 and 13 in Ref.~\citep{Mirocha:2014faa}.

The radiation coupling coefficient, induced by the Wouthuysen-Field effect \citep{Wouthuysen:1952aaa,Field:1958aaa}, can be written as
\be\label{xalpha}
x_{\alpha}(\mathbf{x},z)=\frac{1.81\times10^{11}}{(1+z)} S_{\alpha}J_{\alpha}(\mathbf{x},z),
\ee
where $S_{\alpha}$ is given by Eq.~(55) in Ref.~\citep{Furlanetto:2006jb}. 
%where we set $S_{\alpha}=1$ for simplicity \citep[following e.g. Refs.][]{Chen:2003gc,Hirata:2005mz}.

The gas temperature and Lyman-$\alpha$ coupling depend on the flux terms $J_{X}$ and $J_{\alpha}$. While the former is dominated by X-ray radiation between a few hundred to a few thousands eV, the latter stems from the narrow spectral range between the Lyman-$\alpha$ and the Lyman-limit frequencies.

We parametrise the spectral energy distributions of the Lyman-$\alpha$ and X-ray flux as simple power laws
\be\label{SED}
I_s(\nu) = A_s \nu^{-\alpha_s},
\ee
with $s=\{\alpha,X\}$. The normalisation $A_s$ is defined so that integrating $I_s(\nu)$ over the corresponding energy range becomes unity \citep[see Ref.][]{Mirocha:2015jra}. The number emissivity of UV photons between the Lyman-$\alpha$ and Lyman-limit range is given by
\be\label{lyalemission}
\varepsilon_{\alpha}(\nu)=\frac{N_{\alpha}}{m_p}I_{\alpha}(\nu),
\ee
where $N_{\alpha}$ is the number of photons per baryon emitted in the range between the Lyman-$\alpha$ ($\nu_{\alpha}$) and the Lyman-limit ($\nu_{LL}$) frequencies. The energy emissivity of X-ray photons is
\be\label{xrayemission}
\varepsilon_{X}(\nu)=f_X c_X \frac{I_{X}(\nu)}{\nu h_P}
\ee
where $f_X$ is a free parameter of order unity and $c_X$ is a normalisation factor constrained by observations. It is set to $c_X=3.4\times10^{40}$ erg yr s$^{-1}$M$_{\odot}^{-1}$ based on the findings of Ref.~\citep{Gilfanov:2003bd}.
%$A_s=(1-\alpha_s)\left[\nu_{\rm s, max}^{(1-\alpha_s)}-\nu_{\rm s, min}^{(1-\alpha_s)}\right]^{-1}$ where $s=\{\alpha,X\}$

\subsection{\label{sec:gs}Global Signal}
The global differential brightness temperature is directly obtained by averaging Eqs.~(\ref{T21}) and (\ref{Tratio}). We thereby set $\delta_b$ to zero and assume spatially averaged values for the gas temperature and coupling coefficient. The Lyman-$\alpha$ coupling is obtained via Eq.~(\ref{xalpha}), where the mean Lyman-$\alpha$ flux is given by
\be\label{meanJa}
{\bar J}_{\alpha}(z)=\frac{(1+z)^2}{4\pi}\sum_{n=2}^{n_m}f_n\int_{z}^{z_{\rm max}^{(n)}}dz'\frac{c\varepsilon_{\alpha}(\nu')}{H(z')}{\dot\rho}_{*}(z'),
\ee
with $\nu'=\nu(1+z')/(1+z)$. The recycling fractions $f_n$ are taken from Ref.~\citep{Pritchard:2005an} with the sum truncated at $n_m=23$. The integration limit is given by
\be
z_{\rm max}^{(n)}=(1+z)[1-(n+1)^{-2}]/(1-n^{-2})-1,
\ee
designating the maximum redshift from which photons can Doppler-shift into the Lyman resonances.

The global X-ray number flux per frequency is given by the relation  \citep{Pritchard:2011xb}
\be\label{meanJx}
{\bar J}_{X,\nu}(z)=\frac{(1+z)^2}{4\pi}\int_{z}^{\infty} dz'\frac{c\varepsilon_{X}(\nu')}{H(z')}{\rm e}^{-\tau_{\nu}}{\dot\rho}_{*}(z').
\ee
Here we have introduced the optical depth parameter $\tau$ defined as
\be\label{tau}
\tau_{\nu}(z,z')=\int_z^{z'}dz''\frac{dl}{dz''}\sum_i  n_{i}\sigma_{i}(\nu''),
\ee
where $i=\{{\rm HI, HeI}\}$ and where $\nu''$ is the frequency redshifted from the source at $z'$ to $z''$. Plugging Eq.~(\ref{meanJx}) into Eq.~(\ref{Gammaheat}) leads to the mean heating rate $\bar\Gamma_{h}(z)$. The global temperature evolution is finally obtained by solving
\be\label{meanTgas}
\frac{3}{2}\frac{d\bar T_k}{dz}= \frac{3\bar T_k}{(1+z)}-\frac{\bar\Gamma_{h}(z)}{k_B (1+z)H},
\ee
which corresponds to Eq.~(\ref{Tgas}) at order zero in density perturbations.

The solution of Eq.~(\ref{meanTgas}) together with $\bar x_{\alpha}$ from Eq.~(\ref{meanJa}) allows us to obtain the global differential brightness temperature ($\bar T_{21}$). Examples of the $\bar T_{21}$ signal are shown in the bottom-left panel of Fig.~\ref{fig:results}.

\subsection{\label{sec:ps}Power spectrum}
The 21-cm brightness temperature defined in Eq.~(\ref{T21}) is a function of both redshift and position. At linear order, the 21-cm perturbations are given by
\be\label{delta21}
\delta_{21}(\mathbf{x},z)=\beta_b\delta_b + \beta_{\alpha}\delta_{\alpha}+\beta_{h}\delta_{h} + \beta_{p}\delta_{p} - \delta_{dv},
\ee
This expansion is identical to the one proposed in Ref.~\citep{Barkana:2003qk}, except that, for reasons that will become evident later on (see Sec.~\ref{sec:Pterms}), we furthermore separate the temperature fluctuations into a heating ($\delta_h$) and a primordial ($\delta_p$) term:
\be\label{deltaT}
\delta_{T} = f_{\rm T}\delta_{h} + (1-f_{\rm T})\delta_{p},
\ee
where $f_{\rm T}\equiv(\bar T_{k}-\bar T_{p})/\bar T_{k}$. In this context \emph{primordial} means prior to any heating from sources, i.e., the regime where the gas cools adiabatically and the temperature fluctuations are seeded by the matter perturbations. Note furthermore that Eq.~(\ref{deltaT}) is a direct consequence of the assumption $T_{k}=T_{h}+T_{p}$.

The other terms of Eq.~(\ref{delta21}) designate the baryon perturbations ($\delta_b$), the Lyman-$\alpha$ coupling perturbations ($\delta_{\alpha}$), and the perturbations due to the line-of-sight velocity gradient ($\delta_{dv}$) caused by redshift-space distortion effects. At linear order and in Fourier space, $\delta_{dv}$ is simply given by $ \delta_{dv}=\mu \delta_m$, where $\mu$ is the cosine of the angle between the wave vector $\mathbf{k}$ and the line of sight \citep{Bharadwaj:2004nr}.

The pre-factors of the individual perturbations in Eq.~(\ref{delta21}) are given by
\be
\beta_b &\simeq& 1,\\
\beta_{\alpha} &=& \frac{\bar x_{\alpha}}{\bar x_{\rm tot}(1+\bar x_{\rm tot})},\\
\beta_{h}&\simeq&f_{\rm T}\frac{\bar T_{\gamma}}{(\bar T_{k}-\bar T_{\gamma})},\\
\beta_{p}&\simeq&(1-f_{\rm T})\frac{\bar T_{\gamma}}{(\bar T_{k}-\bar T_{\gamma})},
\ee
and only depend on redshift $z$ but not on the position vector $\mathbf{x}$. 

Based on Eqs.~(\ref{delta21}) and (\ref{deltaT}), it is straight-forward to calculate the power spectrum and to sort all components with respect to their power of $\mu$. Taking the average over the angle then leads to \citep{Barkana:2004zy}
\begin{multline}\label{P21}
P_{\rm 21}= P_{\rm \alpha\alpha} +  P_{hh} + P_{pp} + P_{bb}\\
+ 2  \left(\frac{}{} P_{\alpha h} + P_{\alpha p} + P_{\alpha b} + P_{hp} + P_{hb} + P_{pb}\right) \\
+\frac{2}{3}\left(\frac{}{} P_{\alpha m}+ P_{hm} + P_{pm} + P_{bm}\right) + \frac{1}{5} P_{mm}.
\end{multline}
In the next section, we propose a method to calculate all individual auto and cross power spectra of Eq.~(\ref{P21}) in order to obtain a fast, analytical estimate for the power spectrum of the 21-cm brightness fluctuations. Note furthermore, that when talking about the 21-cm power spectrum, we will mean either $P_{21}$ or the expression $\bar T^2 \Delta_{21}^2$ depending on the context. The dimnesionless quantity $\Delta_{21}^2$ is defined as $\Delta_{21}^2\equiv k^3P_{21}/(2\pi^2$).

\section{\label{sec:hm}Halo model}
In the framework of the original halo model haloes are considered as the building blocks of the universe \citep[see e.g. Ref.][]{Cooray:2002dia}. The halo abundance, distribution, and internal profiles are used to calculate the matter power spectrum at both linear and nonlinear scales. In the context of 21-cm clustering, however, the focus is not so much on haloes per se, but rather on the sources inhabiting haloes, which emit radiation, thereby affecting the gas cells around them. Instead of halo profiles, we therefore consider radiation profiles around sources that extend into the intergalactic space far beyond the halo limits. This means that there will typically be several overlapping radiation profiles from different sources affecting any single gas volume.

Although the picture of overlapping radiation profiles is rather different from the original halo model, it turns out that the 21-cm power spectrum can be described in a very similar way. In this section, we first introduce the formalism, before going into the details of the source profiles and the description of temperature fluctuations. At the end we show the resulting 21-cm power spectrum assuming three benchmark models with different astrophysical parameters.

\begin{figure*}[tbp]
\centering
\includegraphics[width=0.32\textwidth,trim=0.0cm 0.0cm 1.0cm 0.95cm,clip]{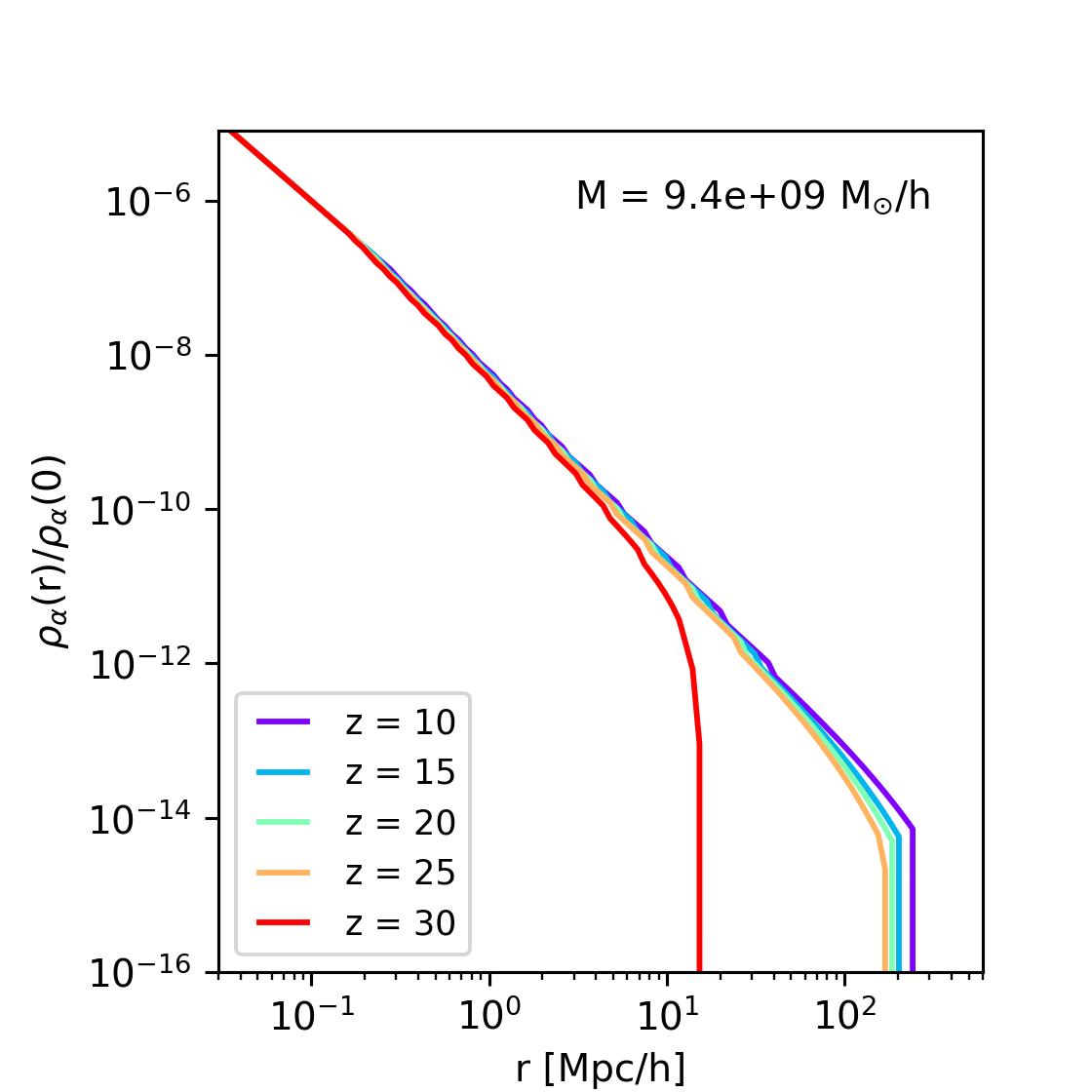}
\includegraphics[width=0.32\textwidth,trim=0.0cm 0.0cm 1.0cm 0.95cm,clip]{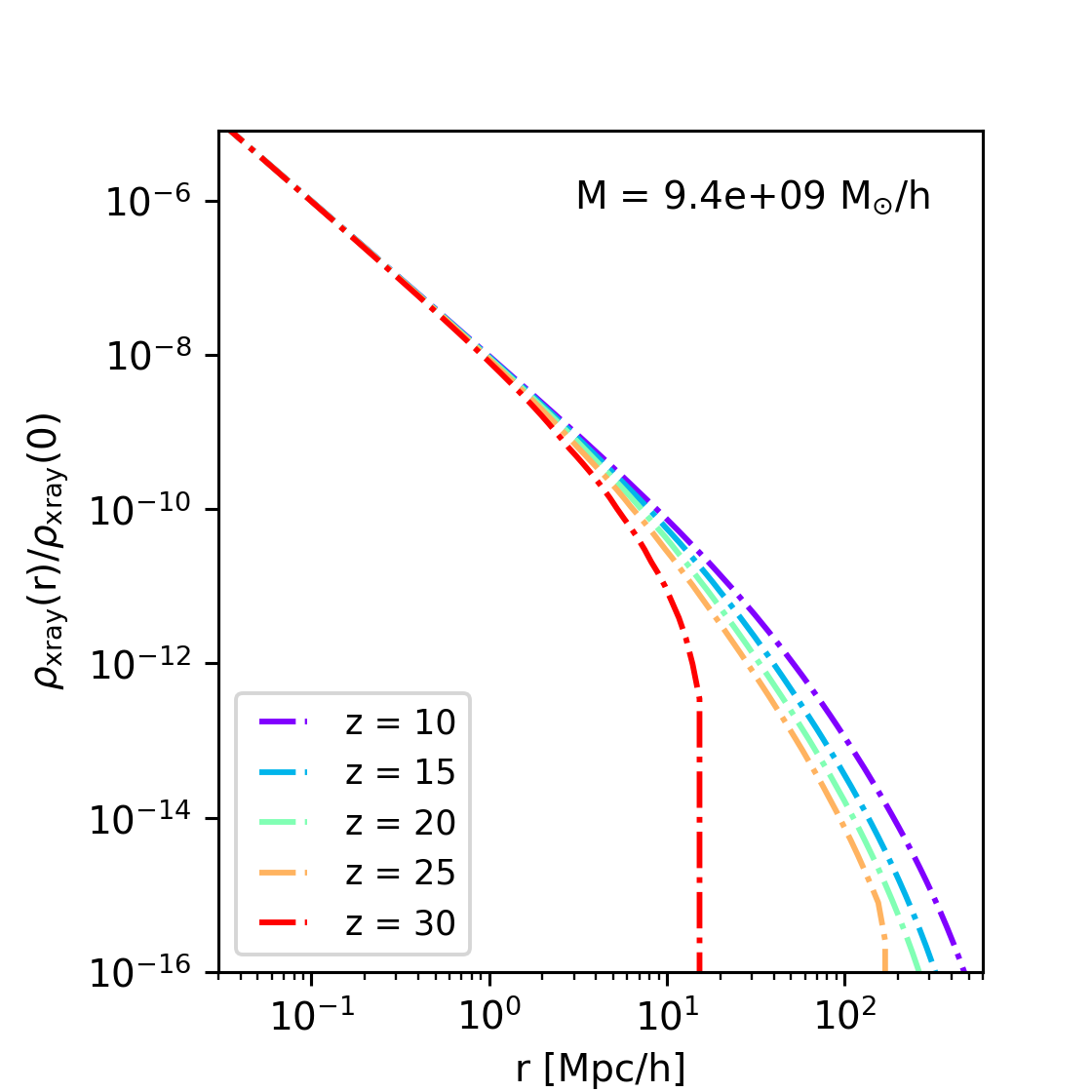}
\includegraphics[width=0.32\textwidth,trim=0.0cm 0.0cm 1.0cm 0.95cm,clip]{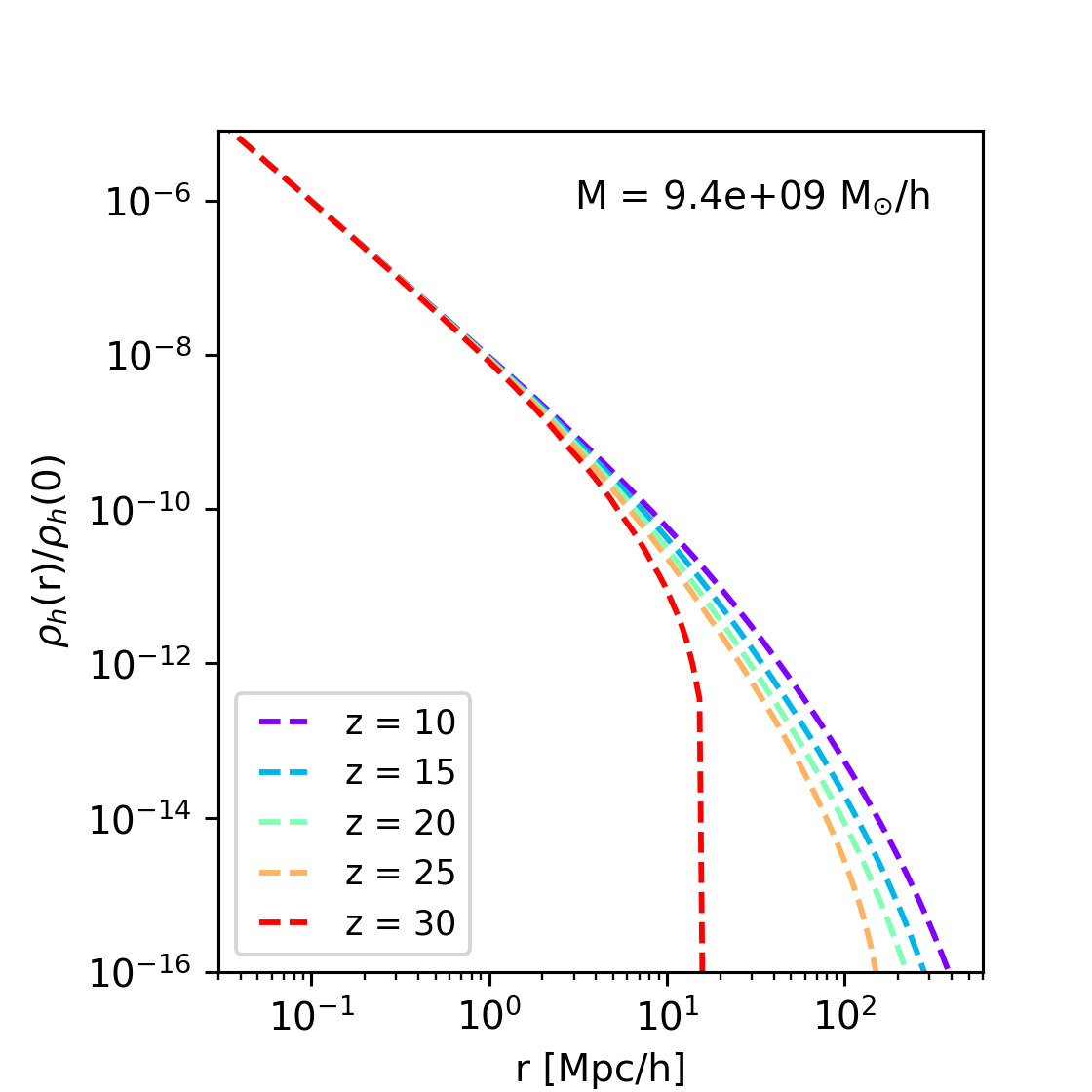}\\
\includegraphics[width=0.32\textwidth,trim=0.0cm 0.0cm 1.0cm 0.95cm,clip]{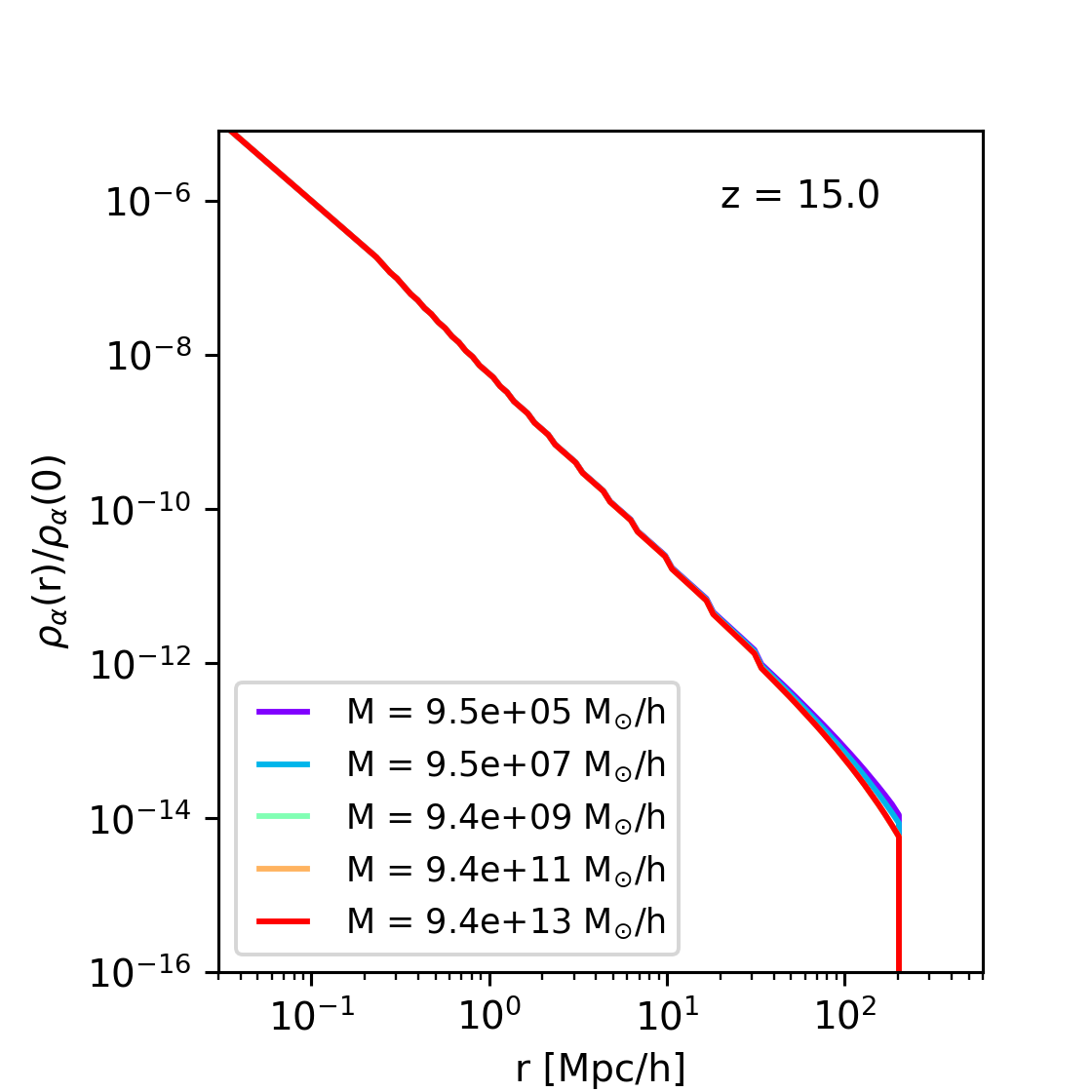}
\includegraphics[width=0.32\textwidth,trim=0.0cm 0.0cm 1.0cm 0.95cm,clip]{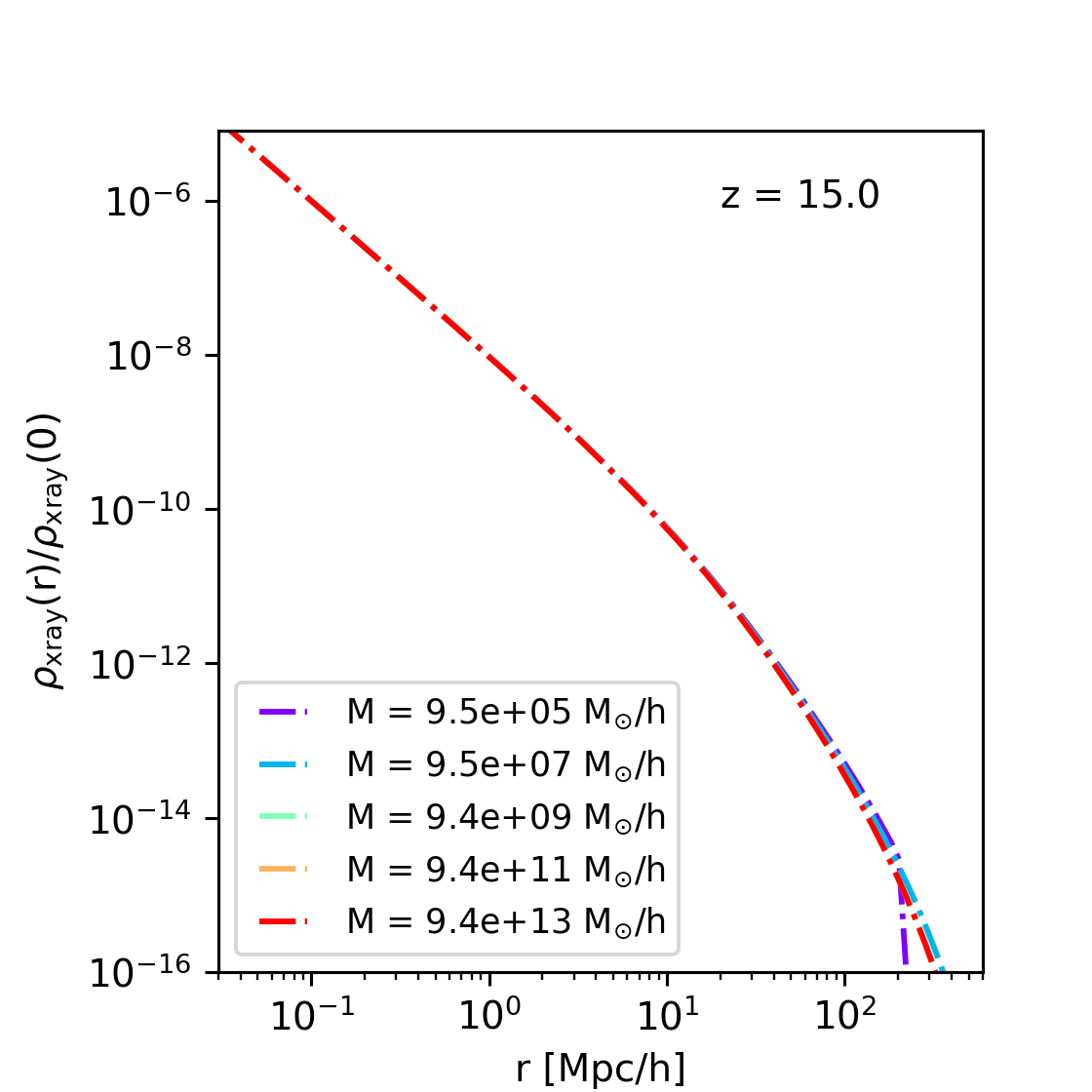}
\includegraphics[width=0.32\textwidth,trim=0.0cm 0.0cm 1.0cm 0.95cm,clip]{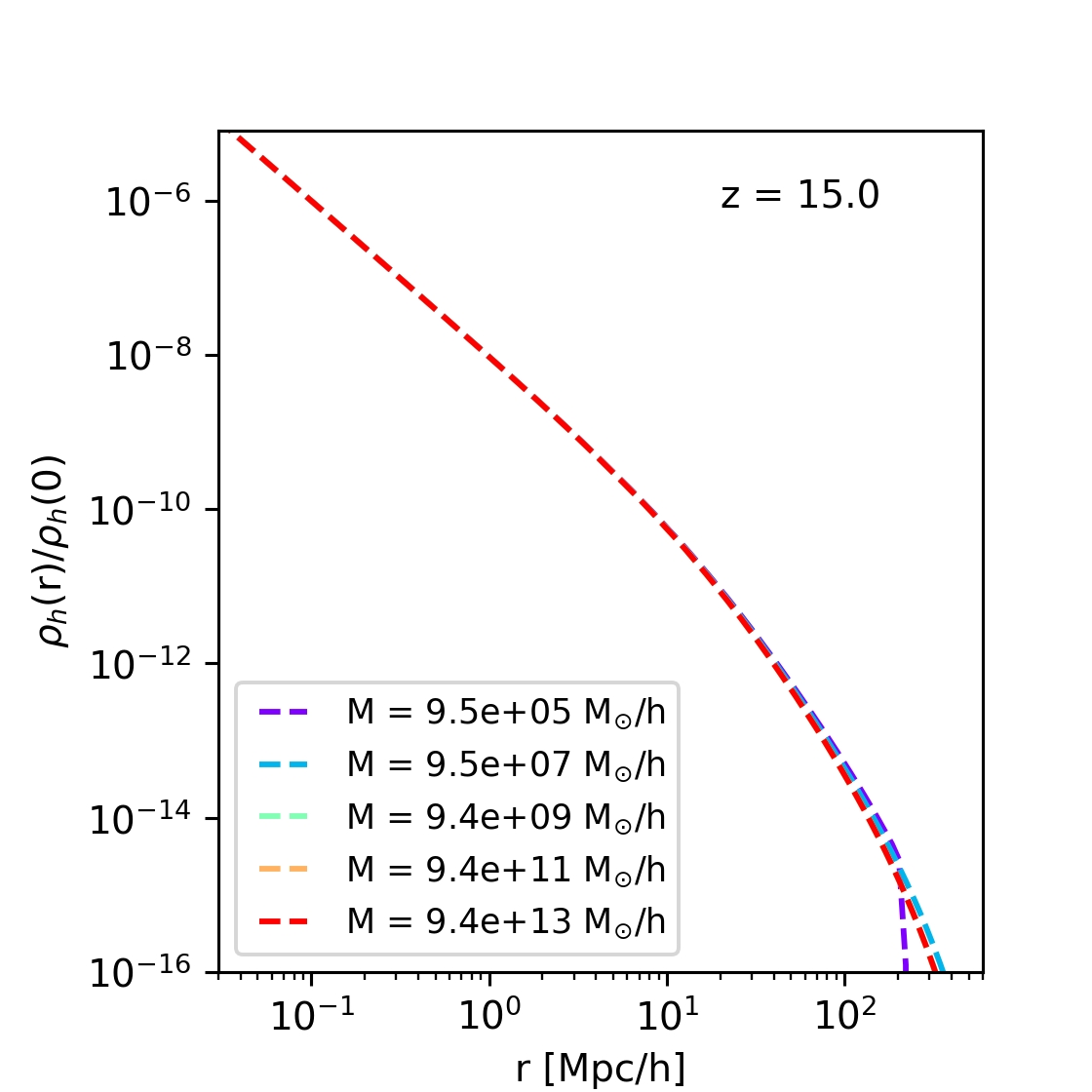}
\caption{Normalised Lyman-$\alpha$ coupling (left), X-ray energy deposition (centre), and heating (right) profiles. In the top-row we vary the redshift at fixed halo mass and in the bottom-row we vary the halo mass at fixed redshift.}
\label{fig:profiles}
\end{figure*}

\subsection{Power spectrum description}
In the context of the 21-cm halo model, the power spectra of different components can be calculated in the following way:
\begin{align}\label{Phm}
P_{XY}^{\rm 1h}(k,z)=&\frac{\beta_X\beta_Y}{({\bar\rho} f_{\rm coll})^2}\int dM \frac{dn}{dM}\tilde f_{*}^2M^2 |u_{X}||u_{Y}|,\nonumber\\
P_{XY}^{\rm 2h}(k,z)=&\frac{\beta_X}{{(\bar\rho} f_{\rm coll})}\int dM \frac{dn}{dM}\tilde f_{*}M |u_{X}| b_{X}\\
\times&\frac{\beta_Y}{({\bar\rho} f_{\rm coll})}\int dM \frac{dn}{dM}\tilde f_{*}M |u_{\rm Y}| b_{Y} \times P_{\rm lin},\nonumber\\
P_{XY}(k,z) =&P_{XY}^{\rm 1h}(k,z) + P_{XY}^{\rm 2h}(k,z),\nonumber
\end{align}
where $P_{\rm lin}(k)$ the linear matter power spectrum, $u_{X,Y}(k,M,z)$ the Fourier transformed flux profile, and $b_{X,Y}(M,z)$ the halo bias. The subscripts $X$ and $Y$ refer to the mass ($m$), baryon ($b$), Lyman-$\alpha$ ($\alpha$), and heating ($h$) components. Note furthermore that Eq.~(\ref{Phm}) describes both auto and cross spectra depending on whether $X=Y$ or $X\neq Y$. This means that the halo model provides all components of Eq.~(\ref{P21}) except the ones that include the primordial gas temperature ($p$). We will derive these in Sec.~\ref{sec:Pterms}.

In the halo model framework, the one-halo term ($P_{XY}^{\rm 1h}$) describes the clustering within one single source profile, while the two-halo term ($P_{XY}^{\rm 2h}$) accounts for the signal induced by different sources. As a consequence $P_{XY}^{\rm 1h}$ and $P_{XY}^{\rm 2h}$ dominate at small and large scales, respectively, with a transition region corresponding to the typical size of the source profiles.  Not surprisingly, only the two-halo term carries information about the spatial distribution of sources via the components $b_{X,Y}$ and $P_{\rm lin}$. The one-halo term, on the other hand, carries information about the shot-noise of sources, and its shape is only controlled by the Fourier transformed radiation profiles.

\subsection{Flux profiles}
The key components of the halo model are the radiation profiles around sources. Following \citet{Holzbauer:2011aaa}, the profile of the Ly-$\alpha$ radiation can be written as
\be\label{rho_alpha}
\rho_{\alpha}(r|M,z)=\frac{1}{4\pi r^2}\sum_{n=2}^{n_m}f_n\varepsilon_{\alpha}(\nu')f_{*}\dot{M}_{\rm ac}(z'|M,z)
\ee
with $\nu'=\nu(1+z')/(1+z)$. The profile is proportional to the halo mass accretion $\dot{M}_{\rm ac}$ and exhibits the characteristic $r^2$ decrease with radius $r$. It furthermore depends on the look-back redshift $z'=z'(r)$ that corresponds to the redshift when a photon at radius $r$ has been emitted at the source. At the emission redshift $z'$, the source is in an earlier stage of evolution (compared to the redshift of the signal $z$), which means that its accretion rate is smaller as well. The look-back redshift is obtained by inverting the co-moving distance
\be
r(z'|z) = \int_z^{z'}\frac{c}{H(z'')}dz''
\ee
which has to be done numerically.

\begin{figure*}[tbp]
\centering
\includegraphics[width=0.49\textwidth,trim=0.2cm 0.0cm 1.5cm 0.95cm,clip]{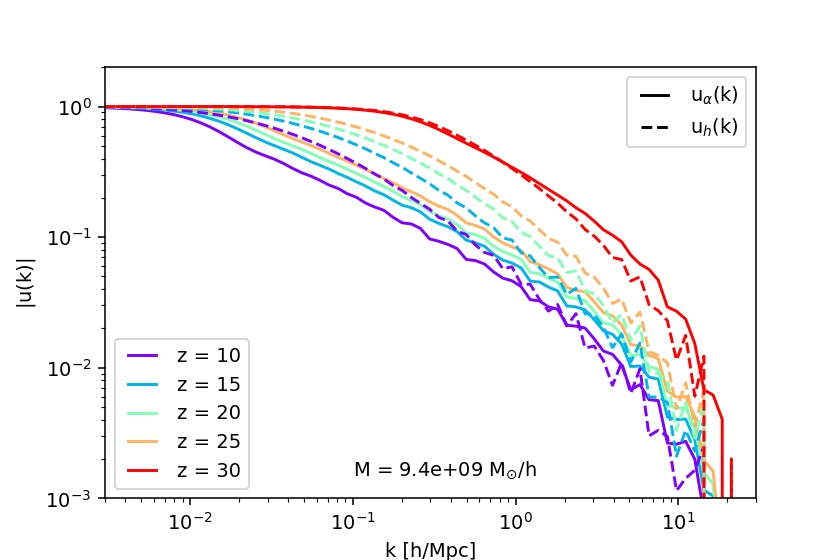}
\includegraphics[width=0.49\textwidth,trim=0.2cm 0.0cm 1.5cm 0.95cm,clip]{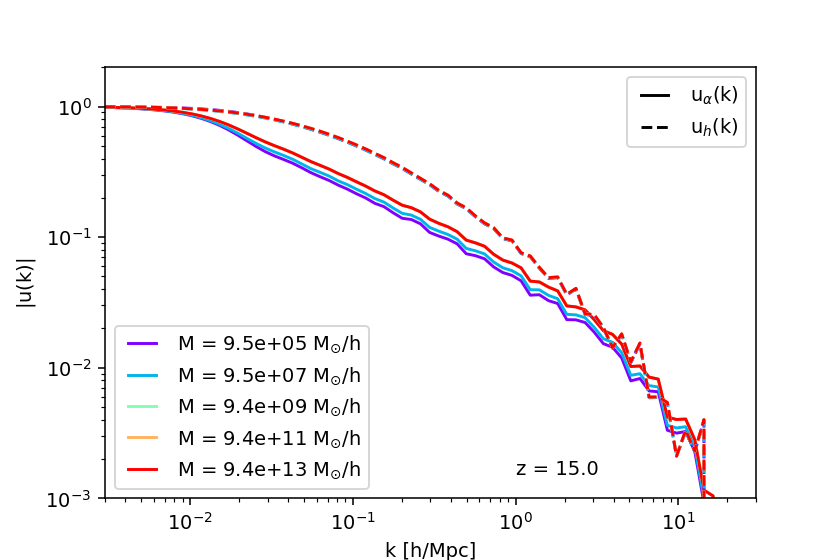}
\caption{Normalised Fourier transforms of the Lyman-$\alpha$ (solid) and the heating profiles (dashed). In the left- and right-hand panels we vary redshift and halo mass while keeping the other one constant.}
\label{fig:FTprofiles}
\end{figure*}

The shape of the Lyman-$\alpha$ flux profile is plotted in the left-hand panels of Fig.~\ref{fig:profiles}, where the top and bottom panels specifically highlight the dependencies on redshift ($z$) and halo mass ($M$). Here we have used the EPS mass accretion model, but the plots look very similar if another model is applied instead. All Lyman-$\alpha$ profiles are characterised by a $1/r^2$ decrease close to the source, which becomes gradually steeper towards the outer parts. The steepening is a direct result of the radiation originating from the source at a higher redshift ($z'$) when the accretion rate onto the source was smaller. At around 200 Mpc/h, the profiles exhibit a steep drop that is due to the photons having redshifted out of the Lyman-$\alpha$ series. At very high redshift beyond z=25, the drop-off shifts towards much smaller radii below 100 Mpc/h because the source is so young that the radiation did not have time to expand further. Note that this effect is not limited to the Lyman-$\alpha$ radiation, but is also visible in the X-ray and heating profiles discussed below.

A closer inspection of the Lyman-$\alpha$ flux profiles in Fig.~\ref{fig:profiles} reveals small discontinuities in the form of step-like features between $r=0.1-100$ Mpc/h. These small steps are a consequence of the sum in Eq.~\ref{rho_alpha} and have been predicted in earlier work \citep[see Ref.][]{Holzbauer:2011aaa}.

The heating of the gas depends on the flux of X-ray radiation and can be defined similarly to the Lyman-$\alpha$ profile, i.e.,
\begin{multline}\label{rho_heat}
\rho_{\rm xray}(r|M,z)=\frac{1}{r^2}\sum_i f_i f_{X,h} \\
 \times \int_{\nu_{\rm th}^i}^{\infty} d\nu\left(\nu-\nu_{\rm th}^i\right) h_P\sigma_i(\nu)\varepsilon_{\rm X}(\nu')e^{-\tau_{\nu}}f_{*}\dot{M}_{\rm ac}(z'|M,z),
\end{multline}
where $i$=$\{\rm H,He\}$ and $\nu_{\rm th}^i h_P$=\{13.6, 26.5\} eV. Compared to Eq.~(\ref{rho_alpha}) there is an additional attenuation term due the optical depth defined in Eq.~(\ref{tau}).

The shape of the energy deposition profile ($\rho_{\rm xray}$) is shown in the middle panels of Fig.~\ref{fig:profiles}, the top and bottom panels again highlighting its redshift and mass dependencies. As a general trend, the profile extends further in radius (to around 500 Mpc/h) and is more gradually suppressed than the Lyman-$\alpha$ coupling profile. This is due to the fact that hard X-ray radiation can travel large distances until it is deposited as heat, and X-ray photons do not redshift out of a well defined spectral range. Since the large distances traveled by X-ray photons also lead to increased look-back redshifts (the difference between $z$ and $z'$), the energy deposition profile shows a stronger attenuation towards large radii.
 
The energy deposition from X-ray emission leads to a strong increase of the temperature fluctuations around sources. The corresponding profile can be obtained via the differential equation
\begin{multline}\label{rhoT}
\frac{3}{2}\frac{d\rho_{h}(r|M,z)}{dz} = \frac{3\rho_{h}(r|M,z)}{(1+z)} - \frac{\rho_{\rm xray}(r|M,z)}{k_B (1+z)H}.
\end{multline}
The first term on the right-hand-side of this equation describes the cooling due to the expansion of space, the second term corresponds to the energy deposition from X-ray radiation. The solution of Eq.~(\ref{rhoT}) is called the heating profile ($\rho_{h}$) in agreement with the notation of Sec.~\ref{sec:ps}, where we separated the temperature fluctuations into a primordial and a heating term (see Eq.~\ref{deltaT}).

The heating profile is illustrated on the right-hand-side of Fig.~\ref{fig:profiles}, where the top and bottom panels again show the dependency with redshift and halo mass. The shape of the heating profile ($\rho_{h}$) is similar to the energy deposition profile ($\rho_{\rm xray}$), which is not a surprise, since X-ray radiation is assumed to be the only source of heating in Eq.~(\ref{rhoT}).

The source profiles of Eqs.~(\ref{rho_alpha}) and (\ref{rhoT}) are key ingredients of the halo model. However, in order to substitute them into Eq.~(\ref{Phm}), we need to first calculate their normalised Fourier transforms. They are given by
\be\label{uk}
u_i (k|M,z) = \frac{\int dr r^2\rho_i(r|M,z)j_0(kr)}{\int dr r^2\rho_i(r|M,z)}
\ee
where $j_0(x)=\sin(x)/x$ is the spherical Bessel function of order 0. The subscript $i$ either stands for $\alpha$ or $h$, denoting the Lyman-$\alpha$ emission and heating profiles. 

In Fig.~\ref{fig:FTprofiles} we show the Fourier transformed profiles with their redshift and halo mass dependencies. All profiles go to unity at low $k$-modes, a key characteristic that is guaranteed by the normalisation of Eq.~(\ref{uk}). Towards higher values of $k$, the profiles become strongly suppressed. In general, the Lyman-$\alpha$ profiles are more suppressed than the heating profiles, which is a consequence of the fact that soft X-ray photons do not travel far before being absorbed by the gas. This early absorption leads to an excess of small-scale clustering compared to the more freely emitted Lyman-$\alpha$ flux. One exception to this behaviour becomes visible at $z=30$ (see red lines in the left-hand panel), where both profiles have a very similar shape. This is not surprising because at very high redshifts, the shape of the profiles is driven by the emission of a very young source with photons that did not have time to travel far.

Fig.~\ref{fig:FTprofiles} shows a clear redshift dependence of the normalised Fourier profiles with more small scale clustering towards higher redshifts. The mass dependence, on the other hand, is very weak. This means that, in principle, the halo model of Eq.~(\ref{Phm}) could be simplified considerably by assuming the profiles $u_i$ not to depend on halo mass. Note, however, that for the sake of completeness we keep the full mass dependence in our model.

\subsection{Temperature fluctuations}\label{sec:Pterms}
In Sec.~\ref{sec:ps}, we have separated the temperature fluctuations into a heating ($\delta_{h}$) and a primordial ($\delta_{p}$) term. While the heating term is sourced by the X-ray flux emission and can therefore be readily described by the halo model, the primordial fluctuations are driven by the matter fluctuations and can be directly solved via Eq.~(\ref{Tgas}) at the linear level. Since we have separated out the contribution from the sources, we can set $\Gamma=0$. After linearising $T=\bar T (1+\delta_{p})$ as well as $\rho=\bar\rho(1+\delta)$ we obtain the solution
\be
\delta_{p}=(1+\delta)^{2/3}-1.
\ee
for the primordial (prior to sources) temperature fluctuations. Note that in Fourier space, the matter perturbations are readily obtained by setting $\delta=\sqrt{P_{\rm mm}}$. With this at hand, we can now derive the reminding auto and cross power spectra of Eq.~(\ref{P21}) which include the primordial heating term and are thus not covered by Eq.~(\ref{Phm}). They are given by
\be
P_{pp}(k,z)&=&\beta_{p}(z)^2\delta_{p}(k,z)^2,\\
P_{Xp}(k,z)&=&\beta_{p}(z)\sqrt{P_{XX}(k,z)}\delta_{p}(k,z),
\ee
where $X$ again stands for $\{ \alpha, h,b,m\}$.

%\begin{figure*}[tbp]
%\centering
%\includegraphics[width=0.42\textwidth,trim=0.2cm 0.0cm 1.0cm 0.8cm,clip]{fstar3M.png}
%\includegraphics[width=0.42\textwidth,trim=0.2cm 0.0cm 1.0cm 0.8cm,clip]{Pk_B3M.png}\\
%\includegraphics[width=0.42\textwidth,trim=0.2cm 0.0cm 1.0cm 0.8cm,clip]{gs3M.png}
%\includegraphics[width=0.42\textwidth,trim=0.2cm 0.0cm 1.0cm 0.8cm,clip]{Pz_B3M.png}
%\caption{\emph{Leftmost panels:} Star-formation efficiency (top) and resulting global signal (bottom) for the three benchmark models with suppressed, unchanged and boosted small-scale behaviour (blue, cyan, and magenta). \emph{Remaining panels:} Power spectra as a function of $k$-modes (top) and redshift (bottom) for the same models.}
%\label{fig:results}
%\end{figure*}

\begin{figure*}[tbp]
\centering
\includegraphics[width=0.242\textwidth,trim=0.2cm 0.0cm 1.0cm 0.8cm,clip]{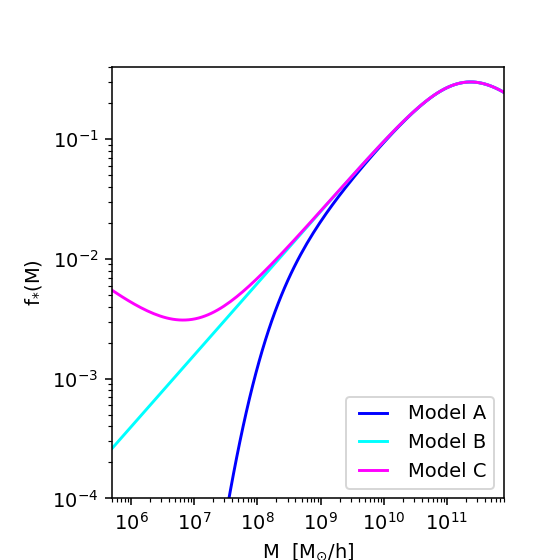}
\includegraphics[width=0.242\textwidth,trim=0.2cm 0.0cm 1.0cm 0.8cm,clip]{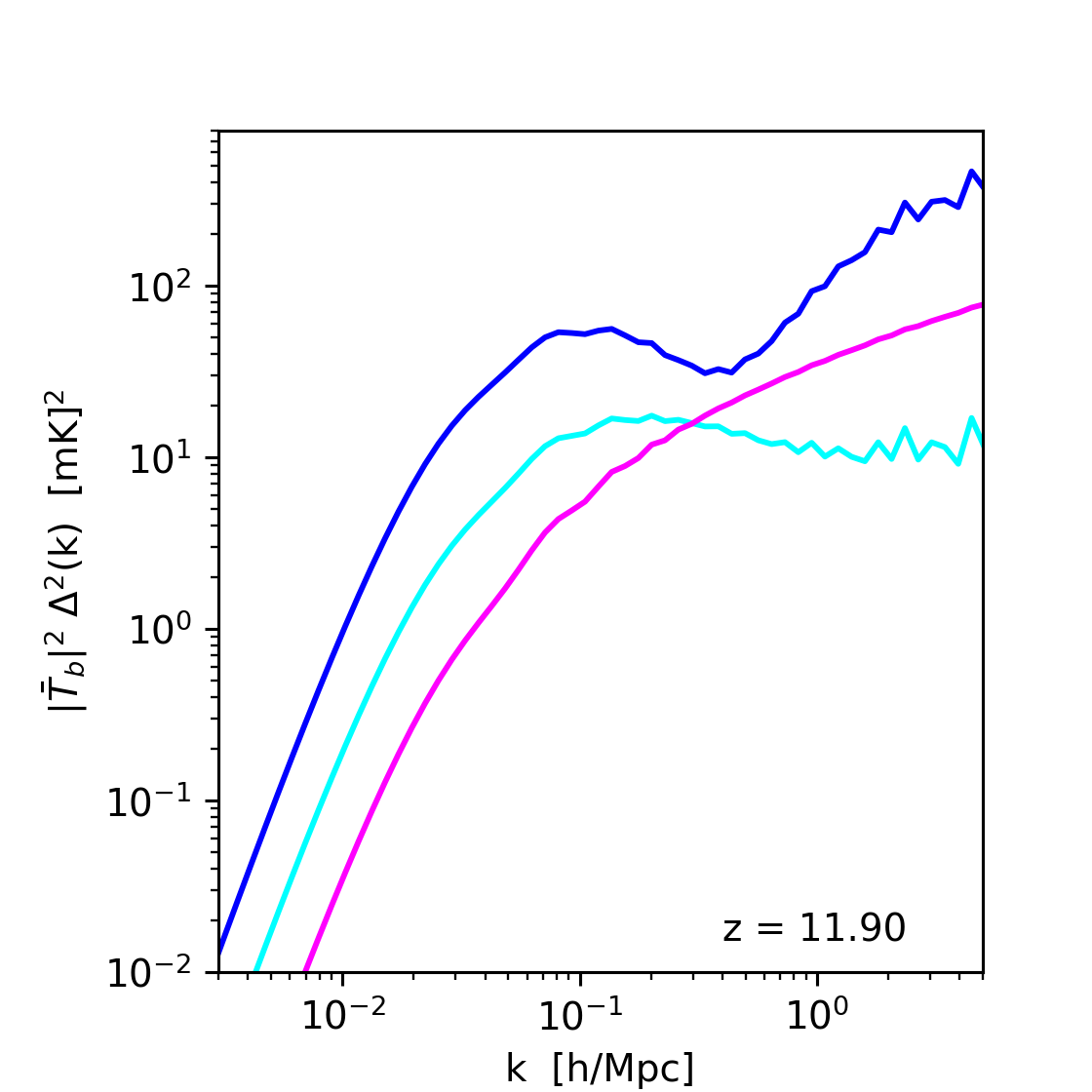}
\includegraphics[width=0.242\textwidth,trim=0.2cm 0.0cm 1.0cm 0.8cm,clip]{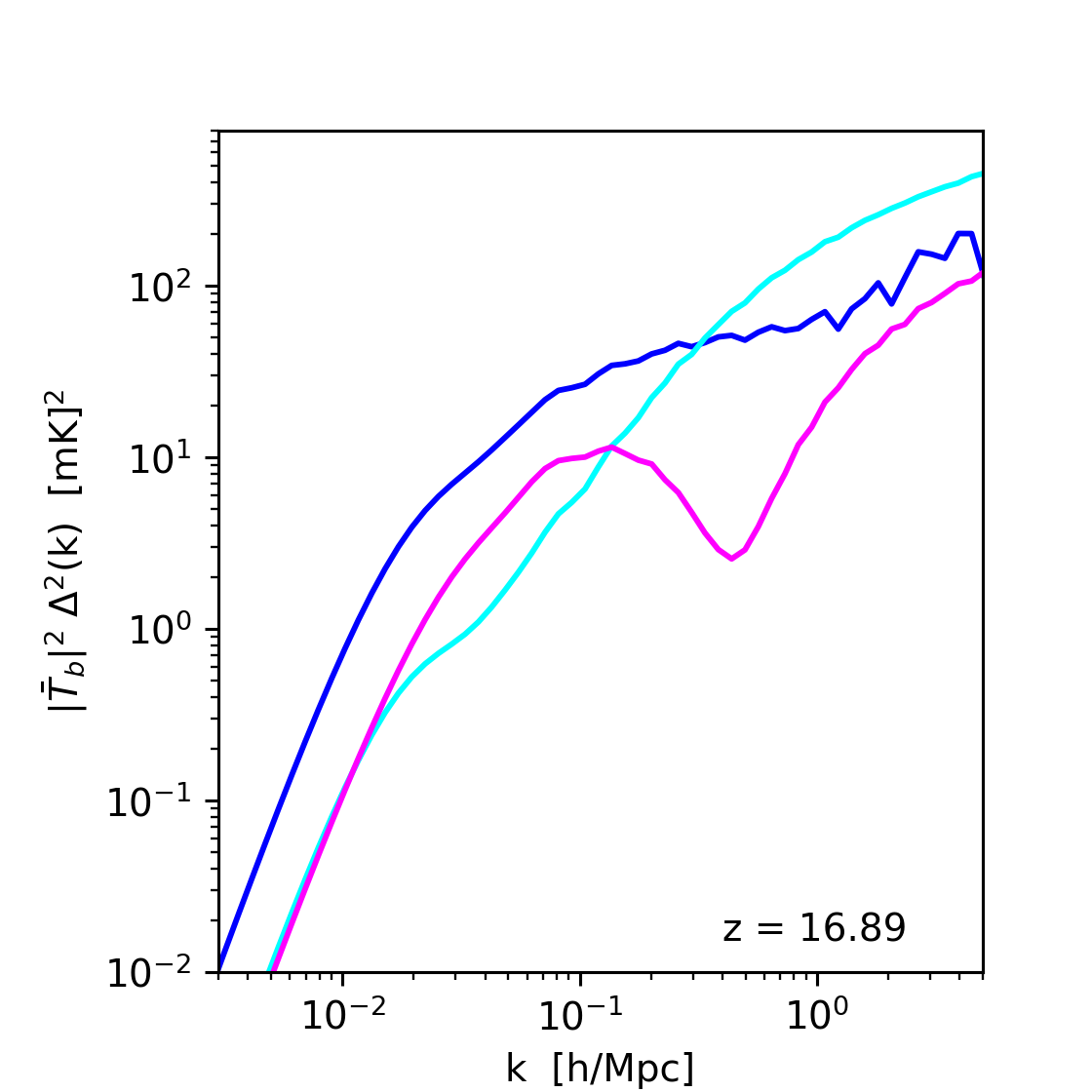}
\includegraphics[width=0.242\textwidth,trim=0.2cm 0.0cm 1.0cm 0.8cm,clip]{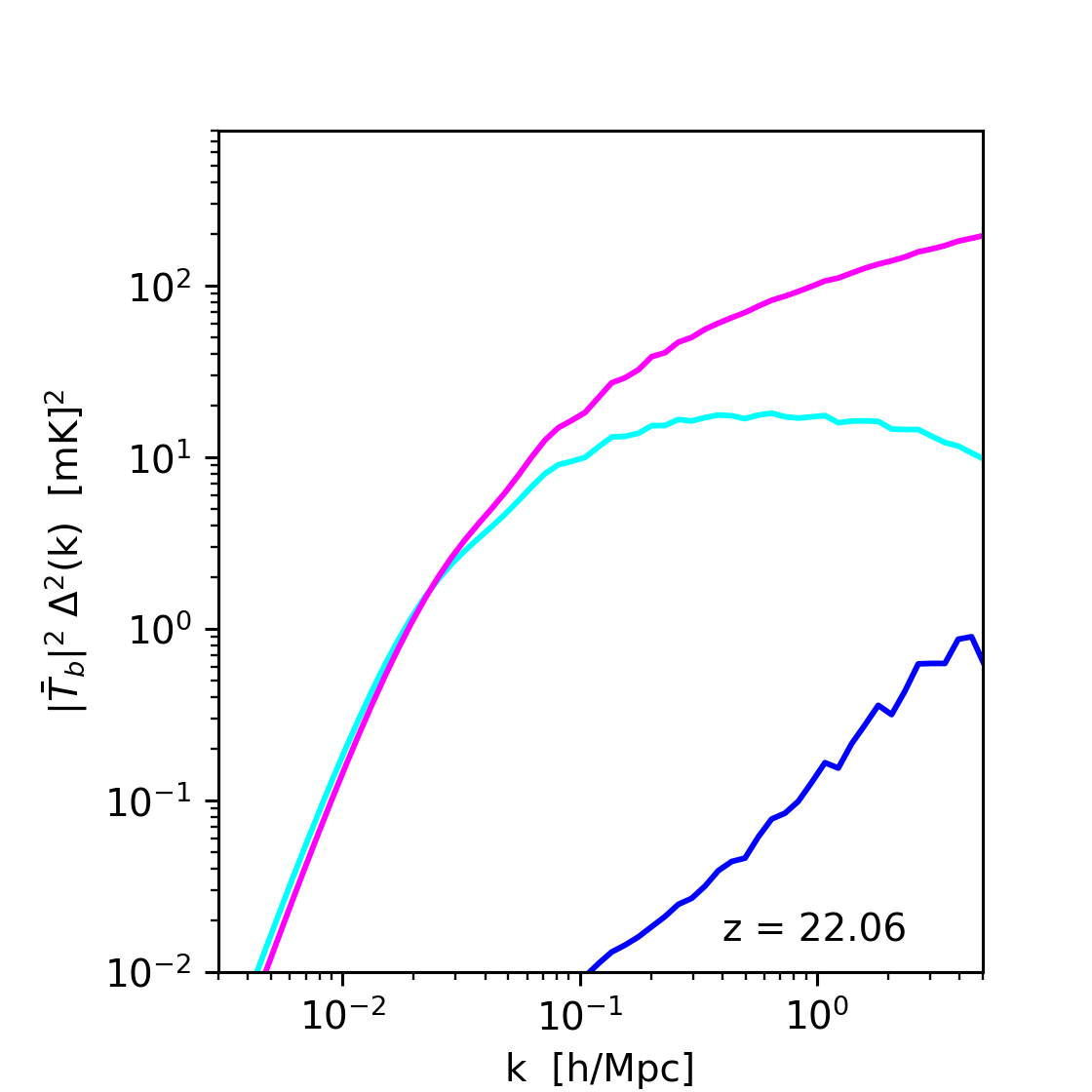}\\
\includegraphics[width=0.242\textwidth,trim=0.2cm 0.0cm 1.0cm 0.8cm,clip]{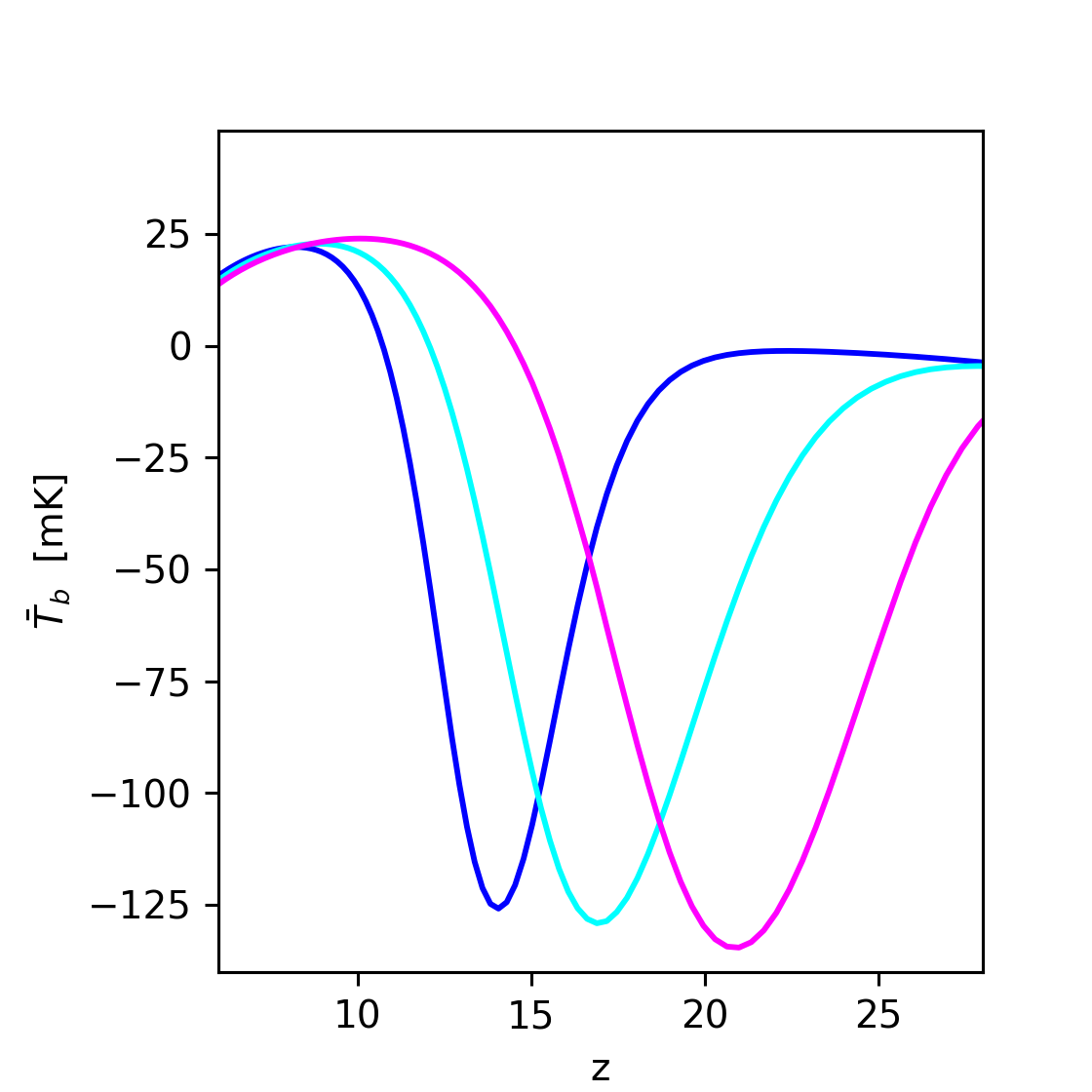}
\includegraphics[width=0.242\textwidth,trim=0.2cm 0.0cm 1.0cm 0.8cm,clip]{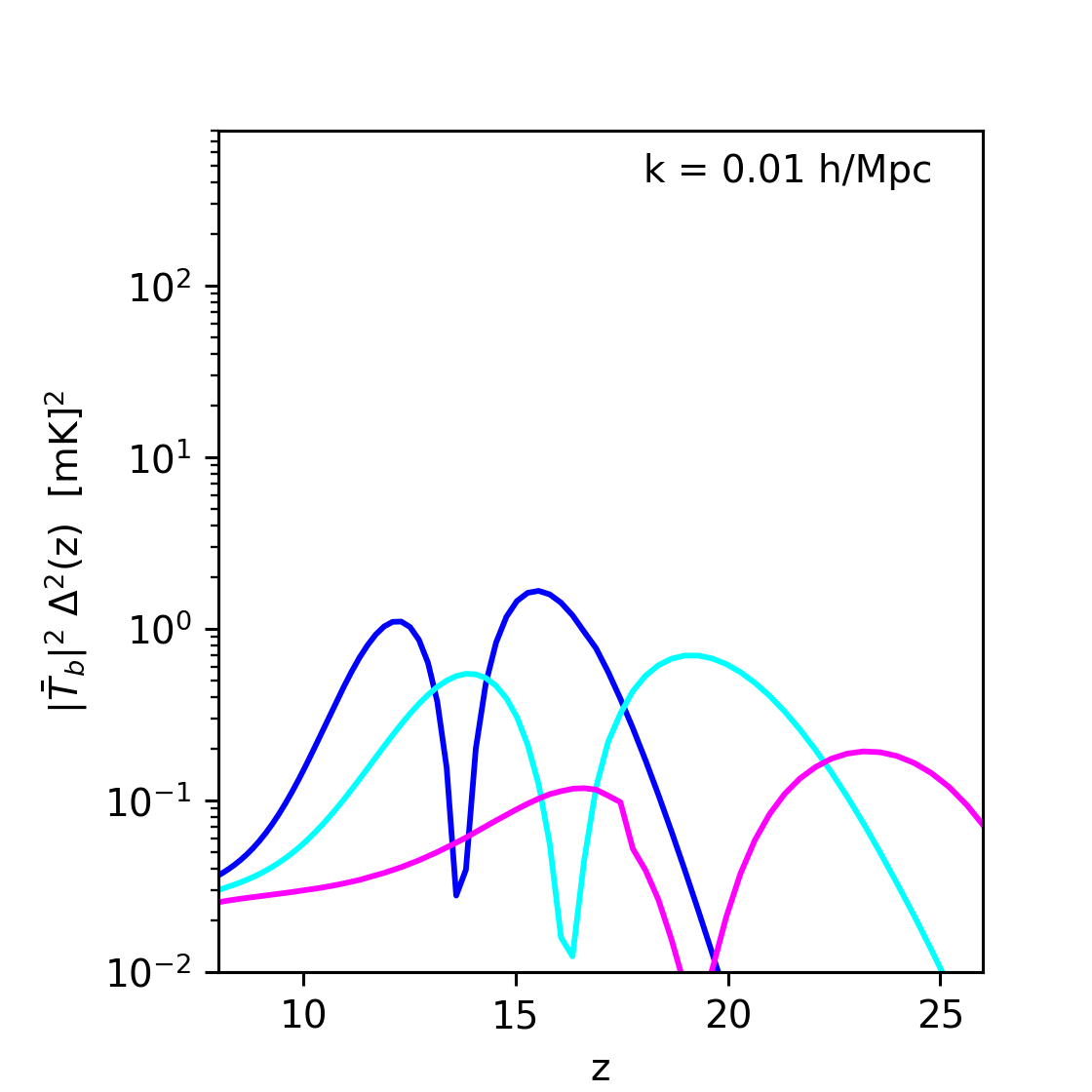}
\includegraphics[width=0.242\textwidth,trim=0.2cm 0.0cm 1.0cm 0.8cm,clip]{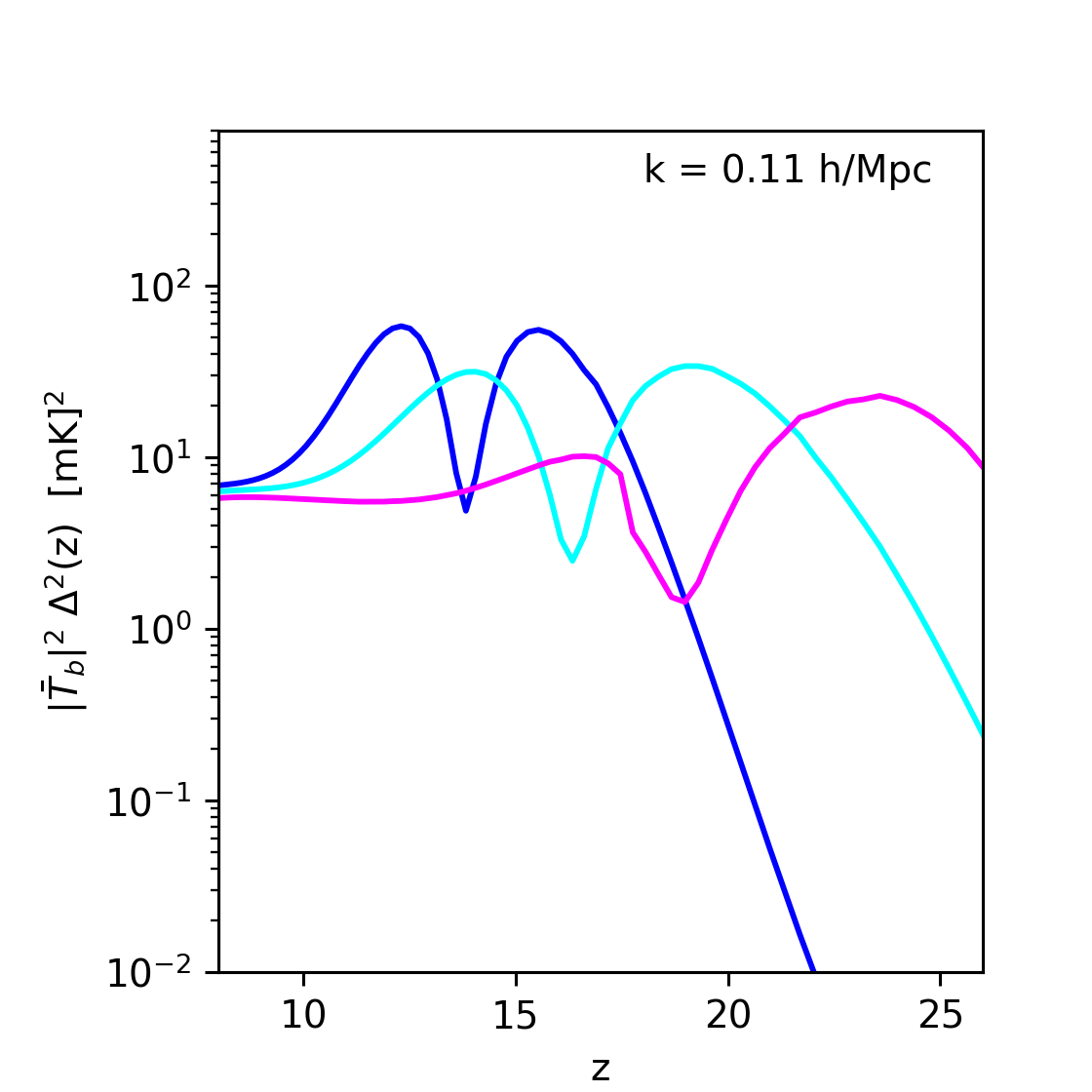}
\includegraphics[width=0.242\textwidth,trim=0.2cm 0.0cm 1.0cm 0.8cm,clip]{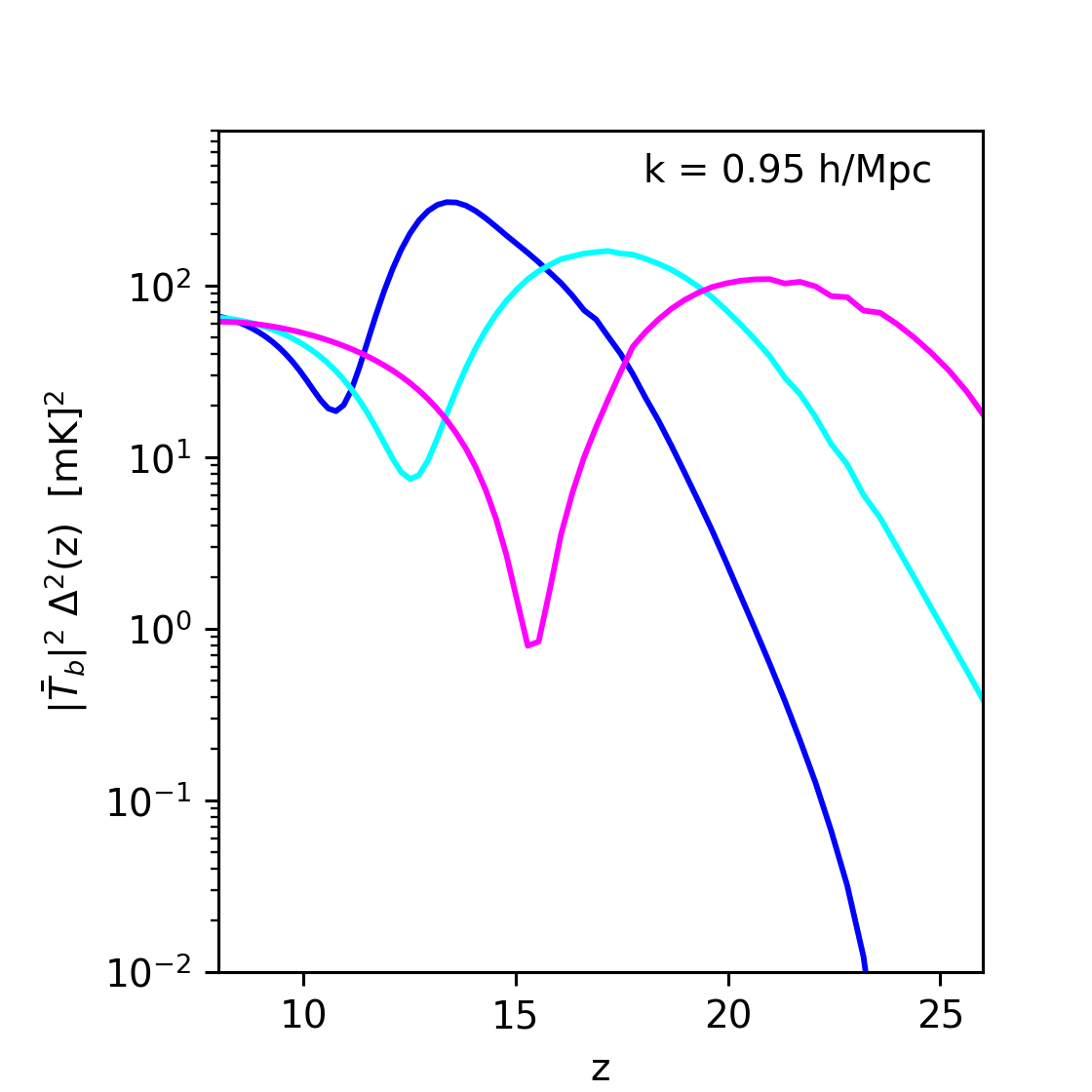}
\caption{\emph{Leftmost panels:} Star-formation efficiency (top) and resulting global signal (bottom) for the three benchmark models with suppressed, unchanged and boosted small-scale behaviour (blue, cyan, and magenta). \emph{Remaining panels:} Power spectra as a function of $k$-modes (top) and redshift (bottom) for the same models.}
\label{fig:results}
\end{figure*}

\subsection{Results}\label{results3M}
Based on the formalism derived above, it is now possible to calculate the 21-cm power spectrum for a given set of model parameters. In this section, we will first present our choices of parametrisation before showing the power spectrum for a selected set of redshift and $k$-modes.

For the Lyman-$\alpha$ sources, we assume $N_{\alpha}=10000$ evenly distributed ($\alpha_{\alpha}=0$) over the energy range between the Lyman-$\alpha$ and Lyman-limit frequencies (see Eq.~\ref{lyalemission}). The X-ray energy emission is defined by $f_X=1$ plus a power-law spectral energy distribution with $\alpha_X=1.5$ over the range $E=0.5-5$ keV (see Eq.~\ref{xrayemission}). Regarding the halo mass function, we assume the \citet{Sheth:2001dp} prescription (Eq.~\ref{massfct}) with modified parameter $q=0.85$. The halo bias is modelled based the peak-background split approach (Eq.~\ref{bias}) with the same parameters than what is used for the halo mass function.
 
In order to highlight the sensitivity of the results to the source parametrisation, we focus on three benchmark models (A, B, and C) that are characterised by different star formation efficiencies ($f_*$). All models have the same large-scale behaviour given by the double power law of Eq.~(\ref{fstar}) with $f_{*,0}=0.3$, $\gamma_1=0.49$, $\gamma_2=-0.61$, and $M_p=2\times10^{11}$ M$_{\odot}$/h \citep[see Ref.][]{Mirocha:2016aaa}. At small mass scales, however, Model A assumes a strong additional suppression (with $M_t=5\times10^7$ M$_{\odot}$/h, $\gamma_3=1$, $\gamma_4=-4$) mimicking the effects of inefficient cooling processes. Model C, on the other hand, is characterised by a boost of $f_*$ towards very small masses (with $M_t=10^7$ M$_{\odot}$/h, $\gamma_3=1$, $\gamma_4=1$). Such a behaviour can be motivated by the presence of Population-III stars in mini-haloes. Model B finally shows neither additional suppression nor boost of $f_*$ but a continuation of the power-law decrease down to the smallest masses (i.e. $\gamma_4=0$). All models are truncated at $M_{\rm min}=5\times10^5$ M$_{\odot}$/h (which roughly corresponds to the smallest halo mass where stars can form via molecular gas cooling).

The star-formation efficiencies of the three benchmark models are plotted in the top-left panel of Fig.~\ref{fig:results}. The typical shape from the double-power law prescription is visible at large masses above $\sim10^9$  M$_{\odot}$/h. At smaller halo masses the models diverge showing the characteristic suppression, power-law continuation, and boost of the benchmark models A, B, and C described above. 

The effect of the different star-formation efficiencies on the global differential brightness signal is illustrated in the bottom-left panel of Fig.~\ref{fig:results}. While model A leads to a narrow absorption signal at a rather low redshift of $z\sim 14$, model B and C show wider troughs shifted towards $z\sim17$ and $z\sim21$, respectively \footnote{Note that although our parametrisation is similar to that of \citet{Mirocha:2016aaa}, the global signal is shifted to slightly higher redshift due to differing assumptions about the star-formation efficiency normalisation and the minimum halo mass.}.

The 21-cm power spectra as a function of $k$-values are plotted in the remaining top-panels of Fig.~\ref{fig:results} with increasing redshift from left to right. Many of the lines have a wave like feature with a local flattening or minimum at $k\sim 0.1-1$ h/Mpc. This feature is located at the transition where the 1-halo starts to dominate over the 2-halo term.

The remaining panels at the bottom of Fig.~\ref{fig:results} show the power spectra as a function of redshift for the three specific $k$-modes. At $k=$ 0.01 h/Mpc (left) and 0.1 h/Mpc (middle) the power spectra show the characterised double-peak feature, which indicate the two characteristic epochs where the Lyman-$\alpha$ coupling and the X-ray heating dominate. At $k=1$ h/Mpc the two peaks have merged into one single, broader peak in qualitative agreement with other work from the literature \citep[see e.g. Refs.][]{Santos:2007aaa,Ross:2018uhh}.

The results presented in Fig.~\ref{fig:results} are based on the EPS mass accretion model plus the halo mass function and bias parameters described above. However, it is important to notice that these modelling choices introduce significant uncertainties regarding the 21-cm power spectrum. In the Appendices~\ref{app:MAC} and \ref{app:MFB}, we quantify the effects due to the choice of the mass accretion model, the halo mass function and halo bias. The bottom line of this analysis is that modelling choices, such as switching from Sheth-Tormen to Press-Schechter halo prescription, or using the AM instead of the EPS accretion rate modelling, has an effect on the 21-cm power spectrum that can in some cases be as large as a factor of $\sim10$. The difference can be significantly reduced if the flux parameters $N_{\alpha}$ are re-adapted so that the global signals are forced to match, but a remaining difference of the power spectrum of about a factor of $\sim2$ remains. %This means that there may remain significant uncertainties not only present in our model but in any semi-analytical modelling approach.

\section{Comparison with other approaches}\label{sec:comparisons}
In this section we compare the 21 cm halo model with the analytical model from Refs.~\citep{Barkana:2004vb,Pritchard:2006sq} (abbreviated to BLPF model) and with the semi-numerical code {\tt 21cmFAST} \citep{Mesinger:2007pd,Mesinger:2010aaa}. While the former is more of a general consistency check, the latter consist of a true test for our model. Note, however, that the comparison with {\tt 21cmFAST} is not straight-forward, mainly because there are subtle differences in the parametrisation of source properties which cannot be fully accounted for without changing the code itself, something we postpone to future work. %We therefore do not expect a better agreement than leave a more careful, apple-to-apple comparison for future work.

%However, we want to emphasise that comparing different methods is not always straight-forward, mainly because there are subtle differences in the parametrisation of source properties. Accounting for all these differences would require substantial efforts including changes in the codes which we leave for future work. As a consequence, we do not always expect a precise match between the models.

\begin{figure*}[tbp]
\centering
\includegraphics[width=0.32\textwidth,trim=0.35cm 0.0cm 1.23cm 0.95cm,clip]{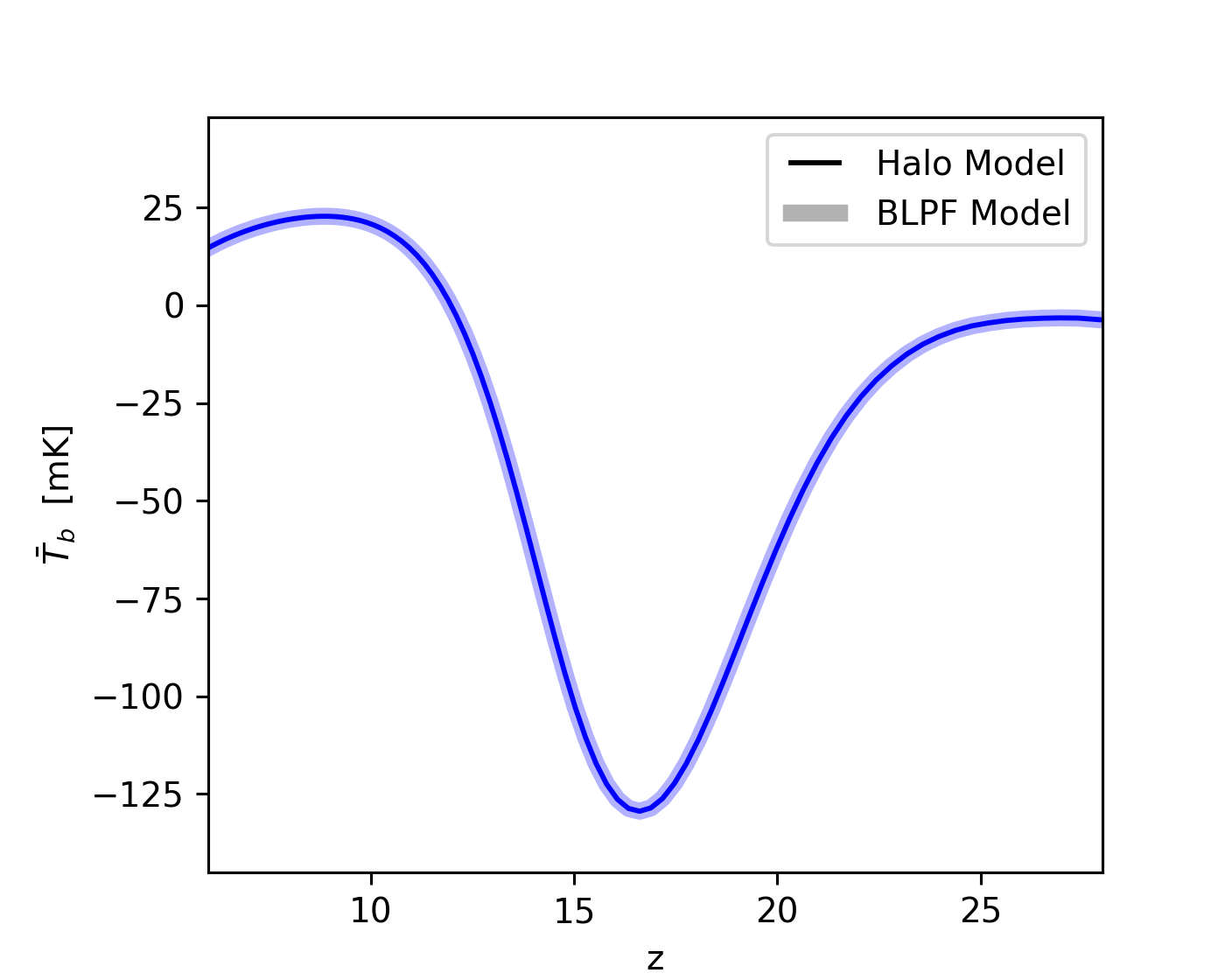}
\includegraphics[width=0.32\textwidth,trim=0.35cm 0.0cm 1.23cm 0.95cm,clip]{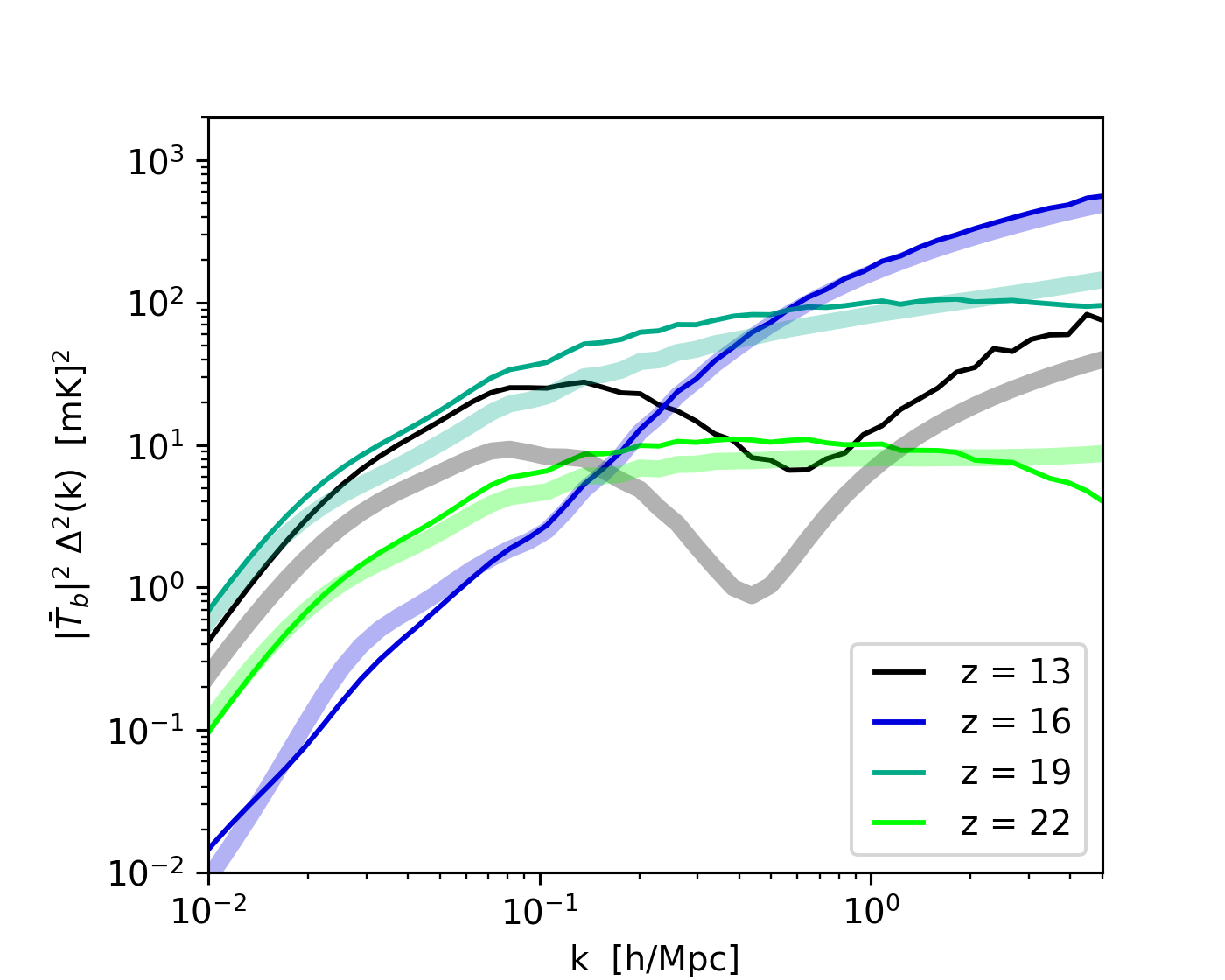}
\includegraphics[width=0.32\textwidth,trim=0.35cm 0.0cm 1.23cm 0.95cm,clip]{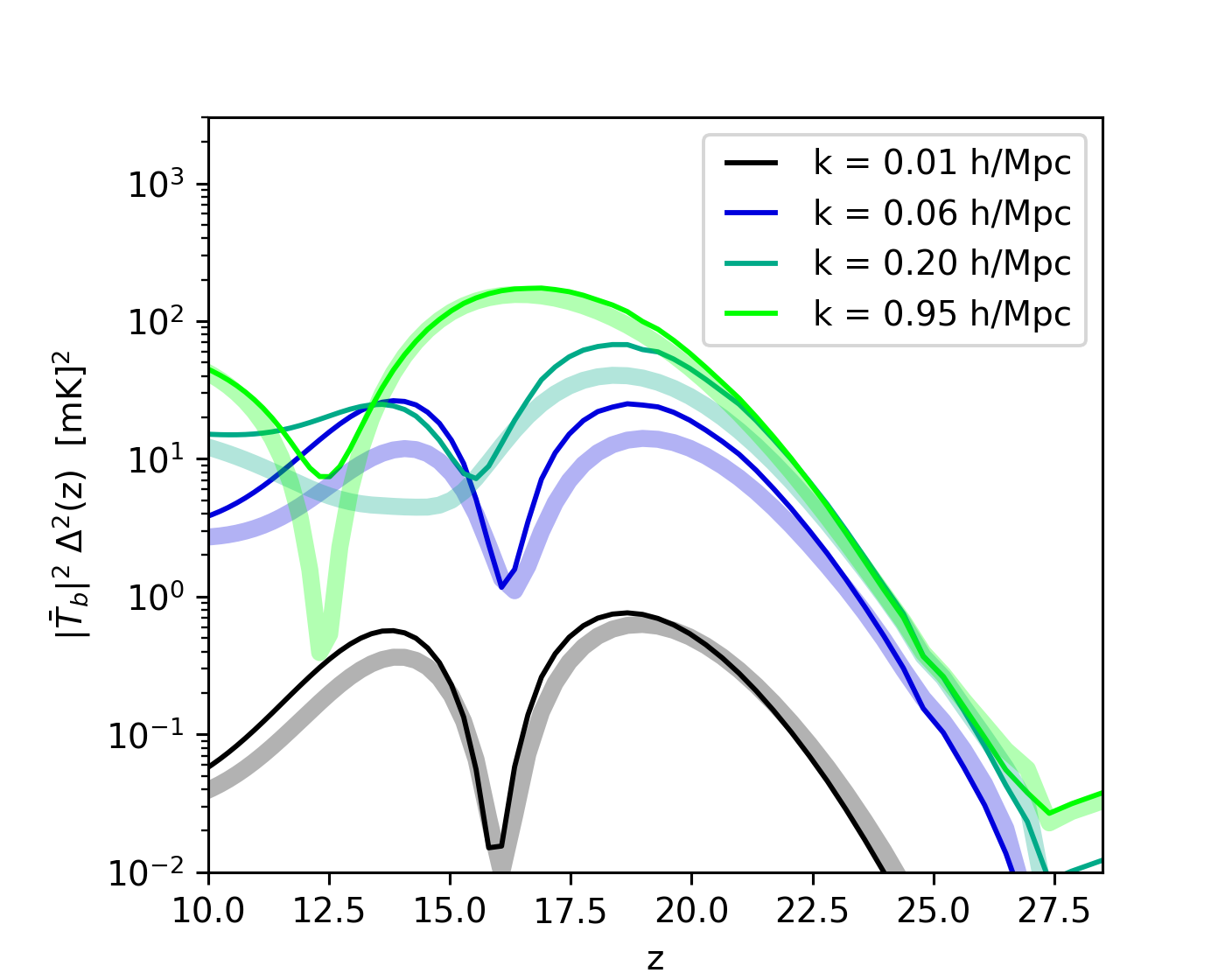}\\
\includegraphics[width=0.32\textwidth,trim=0.35cm 0.0cm 1.23cm 0.95cm,clip]{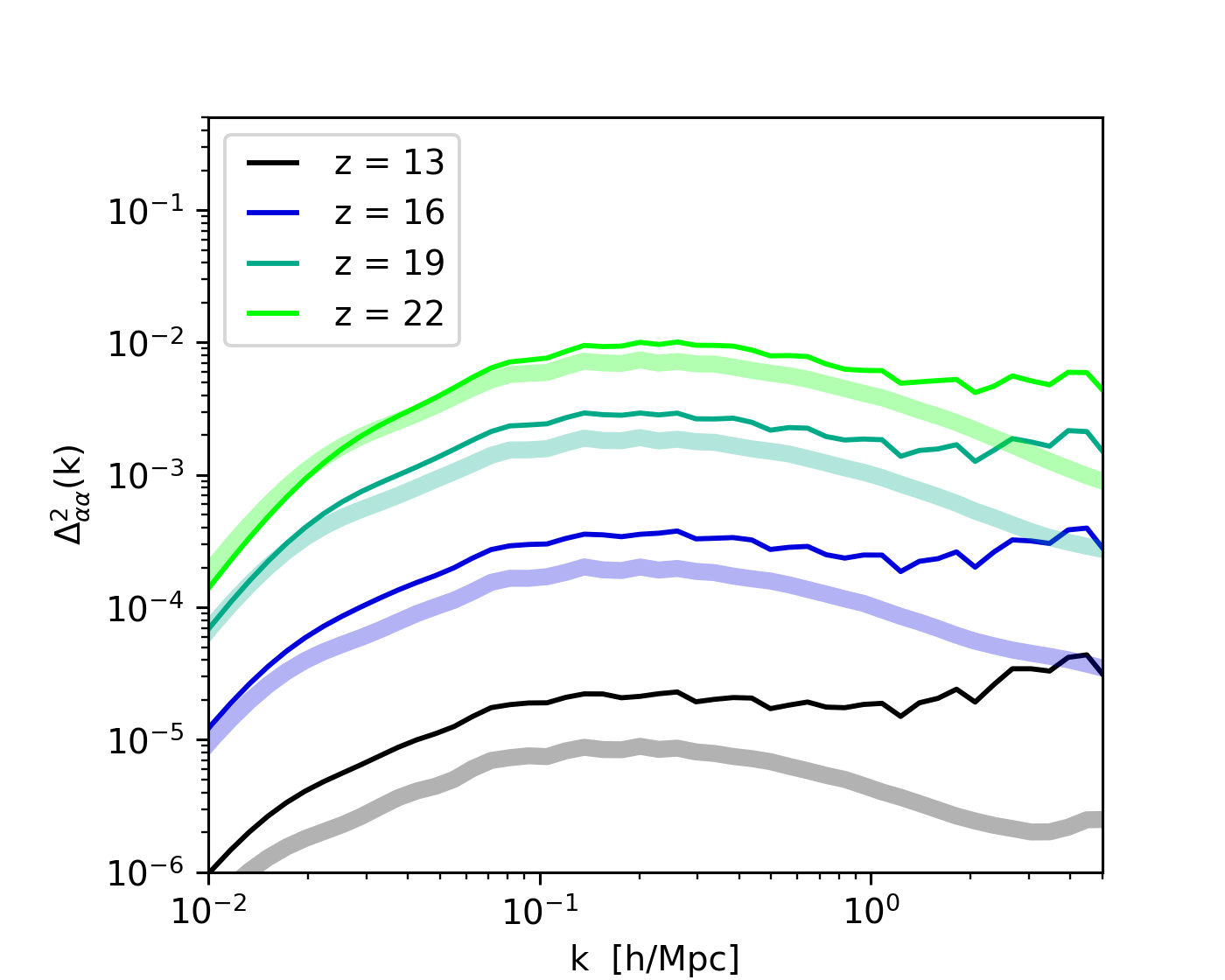}
\includegraphics[width=0.32\textwidth,trim=0.35cm 0.0cm 1.23cm 0.95cm,clip]{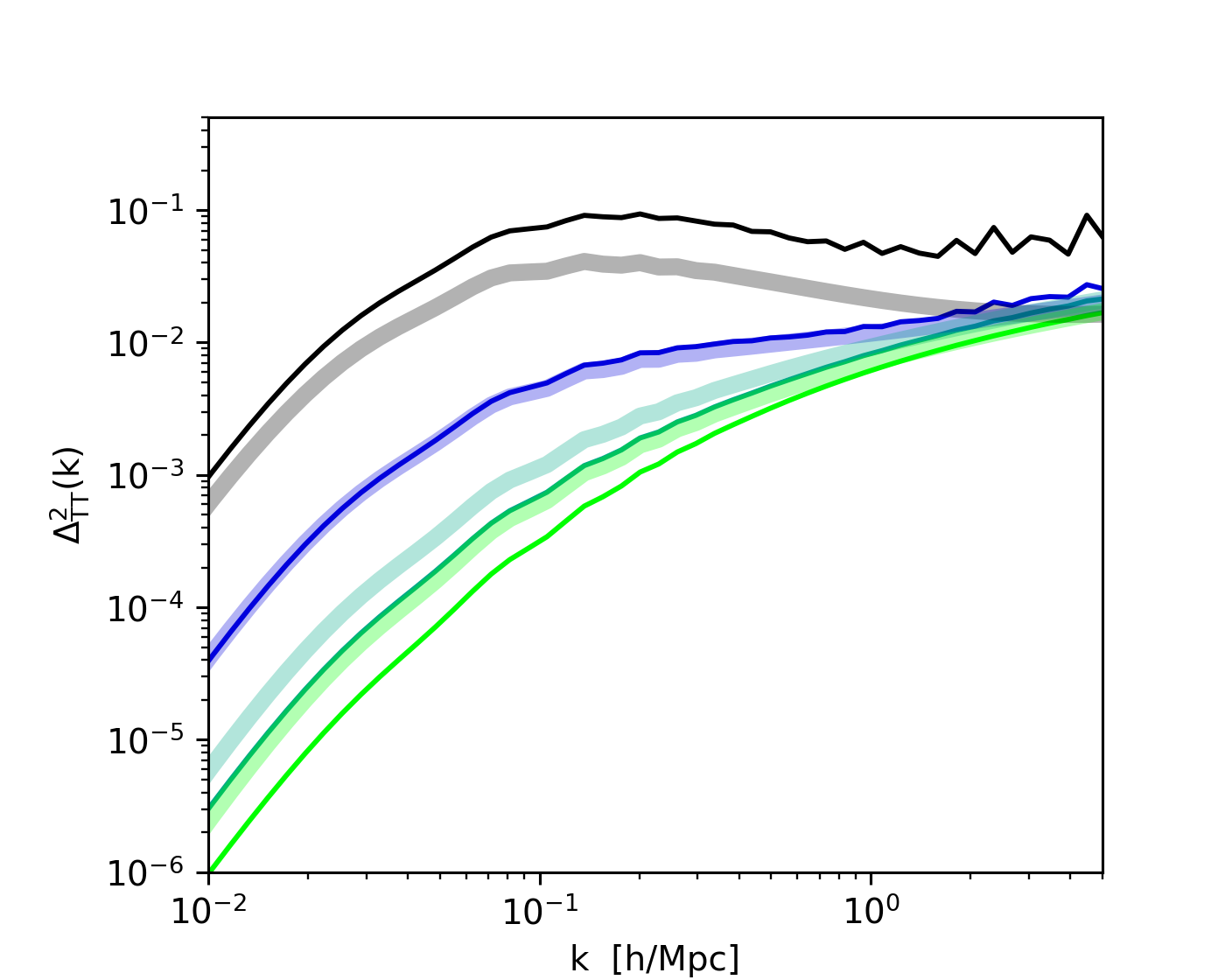}
\includegraphics[width=0.32\textwidth,trim=0.35cm 0.0cm 1.23cm 0.95cm,clip]{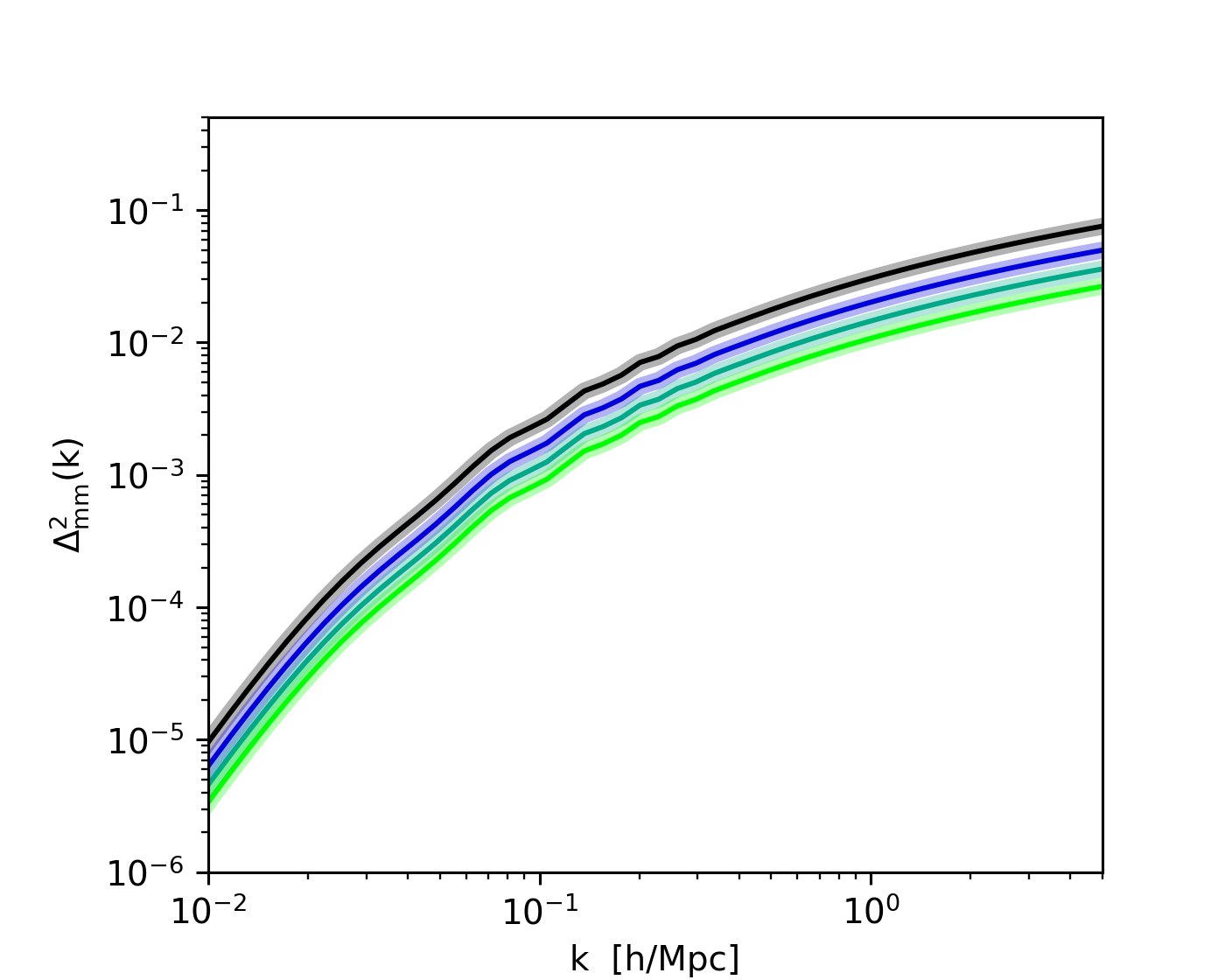}\\
\includegraphics[width=0.32\textwidth,trim=0.35cm 0.0cm 1.23cm 0.95cm,clip]{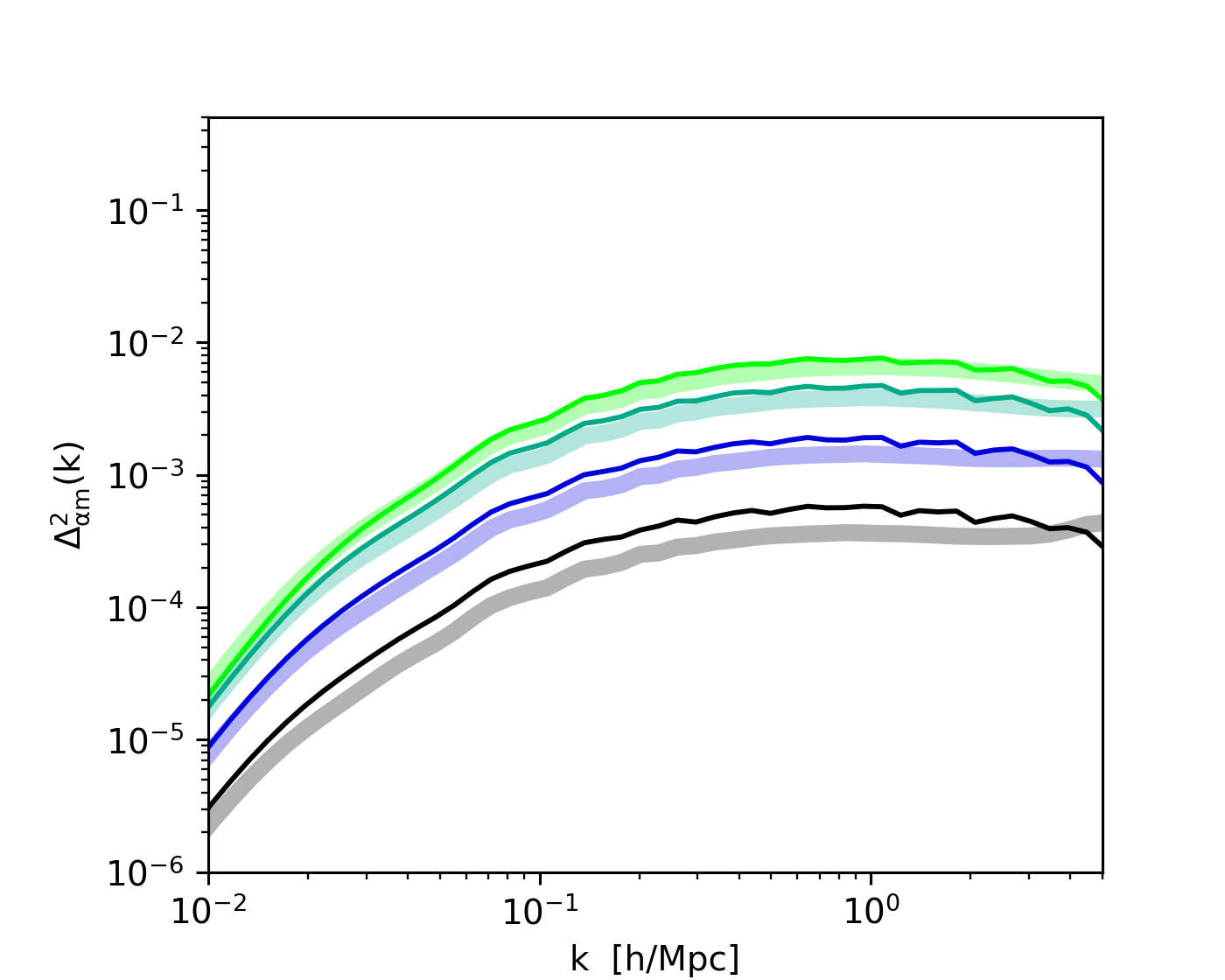}
\includegraphics[width=0.32\textwidth,trim=0.35cm 0.0cm 1.23cm 0.95cm,clip]{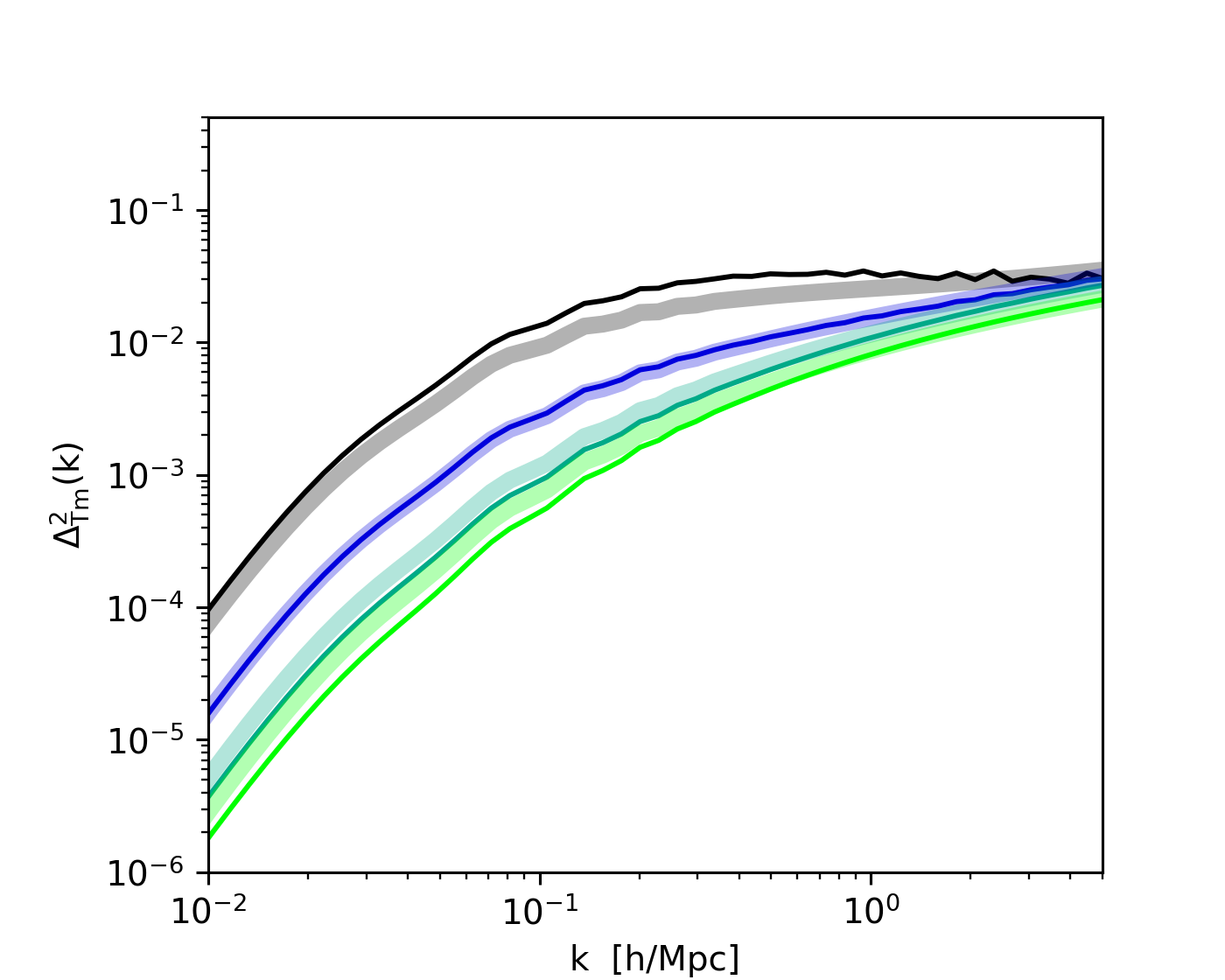}
\includegraphics[width=0.32\textwidth,trim=0.35cm 0.0cm 1.23cm 0.95cm,clip]{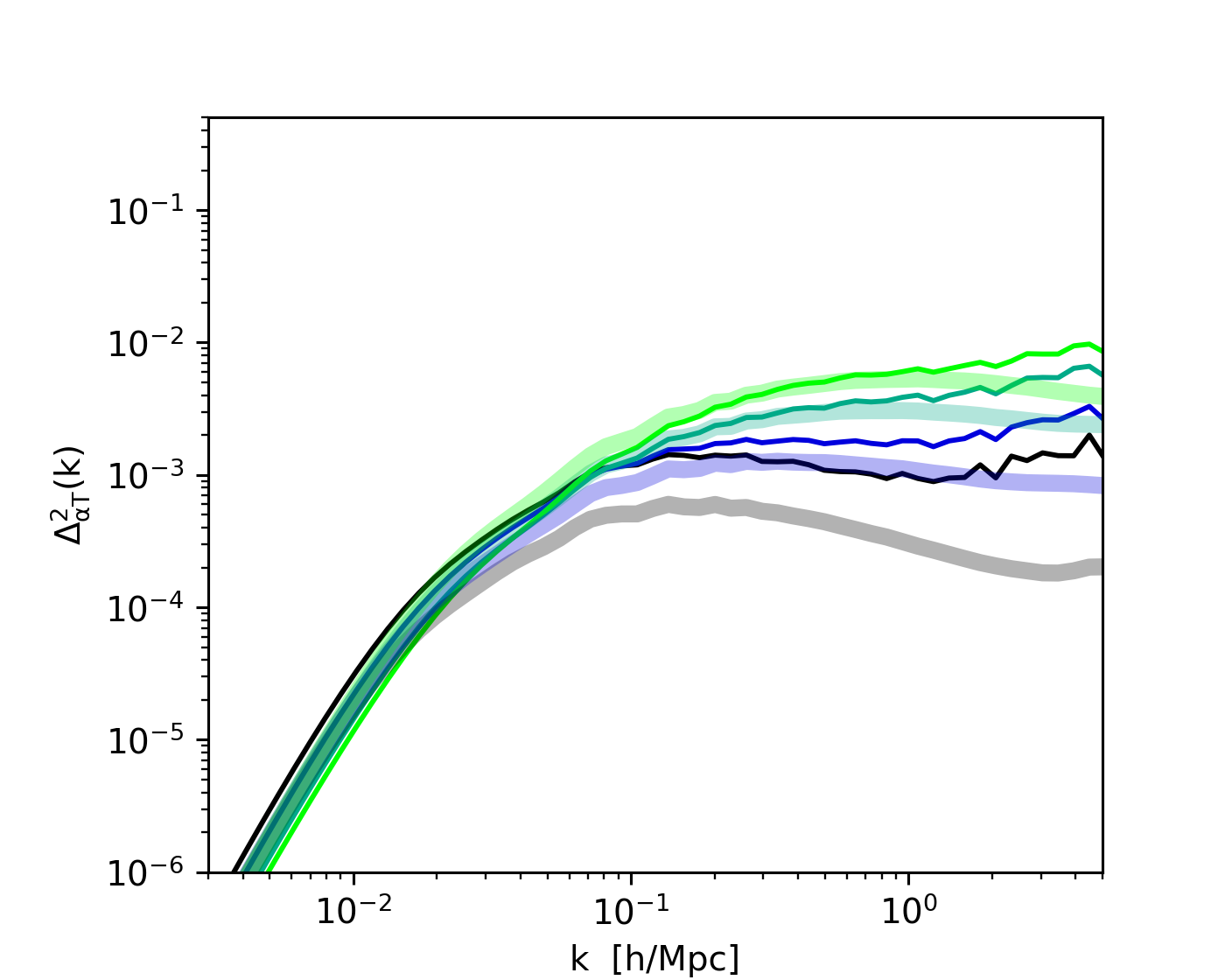}
\caption{Comparison between the 21-cm halo model developed in this paper (solid lines) and the BLPF method from Refs.~\citep{Barkana:2004vb,Pritchard:2006sq} (coloured bands) assuming astrophysical parameters from the benchmark model B (see Fig~\ref{fig:results}). \emph{Top panels:} global signal (left) and the total 21-cm power spectrum with respect to $k$-modes (centre) and redshift (right). \emph{Middle panels:} Auto power spectra of the Lyman-$\alpha$ coupling (left), temperature (centre), and mass (right) components. \emph{Bottom panels:} Corresponding cross power spectra of the individual components.}
\label{fig:PFcomparison}
\end{figure*}

\subsection{Comparing to the analytical approach of BLPF}
A first analytical calculation of the 21-cm power spectrum at cosmic dawn has been performed by \citet{Barkana:2004vb}, focusing on the perturbations induced by the Lyman-$\alpha$ coupling. Their model has been extended to include temperature fluctuations by \citet{Pritchard:2006sq} which is why we abbreviate it as BLPF. In contrast to the halo model approach, which is centred on the radiation sources as building blocks, the BLPF approach focuses on the gas, calculating the light-cone effects from the surrounding sources. A good summary of the model can be found in Ref.~\citep{Pritchard:2011xb}.

For this comparison we have implemented the BLPF model, mostly following the descriptions in Ref.~\citep{Pritchard:2006sq}. The source parametrisation, however, which includes the halo mass function, the bias, and the spectral energy distribution, has been adapted to the description presented in this paper. We furthermore omit any shot-noise corrections, since there is no fully worked-out model that includes both shot-noise of the Lyman-$\alpha$ and temperature fluctuations \citep{Pritchard:2006sq}. Finally, the star-formation rate density is based on the EPS accretion rate modelling (see Eq.~\ref{sfrd_Maccr}) and not on the time derivative of the collapse fraction. These changes with respect to the original work of Refs.~\citep{Barkana:2004vb,Pritchard:2006sq} allow us to carry out a fair comparison, where any resulting discrepancies can be fully attributed to differences between the methods and are not the result of different source descriptions.

The comparison between the BLPF and the halo model is performed using the source parametrisation of benchmark model B. This means we assume the radiation flux parameters $N_{\alpha}=10000$ and $f_X=1$, a non-truncated double-power law for the star formation efficiency, as well as power-law spectra with indices $\alpha_{\alpha}=0$ and $\alpha_X=1.5$ for the UV and X-ray radiation. See Sec.~\ref{results3M} for more details about the parametrisation.

The global signal is shown in the top-left panel of Fig.~\ref{fig:PFcomparison}. Since we use the exact same source modelling and the same calculation for the SFRD, it is not surprising that both approaches yield the exact same result. This perfect agreement is very convenient because it guarantees that any changes at the level of the power spectrum are not induced by different amplitudes of the global differential brightness temperature.

The central-top panel of Fig.~\ref{fig:PFcomparison} shows the 21-cm power spectrum as a function of $k$-modes for a selection of four different redshifts. The redshift values are chosen to lie in the heating dominated regime ($z=13$), in the transition regime of maximum absorption ($z=16$), in the Lyman-$\alpha$ regime ($z=19$), and at the epoch of the very first stars ($z=22$). The solid lines correspond to the 21-cm halo model, while the coloured bands show the results from the BLPF model. In general, the BLPF model predicts less power, especially during the heating (black) and, to a lesser extend, during the Lyman-$\alpha$ epochs (dark green). The differences between the models are typically of the order of a few, but they can grow to about an order of magnitude for specific redshifts and $k$-ranges.

The right-hand panel in the top row of Fig.~\ref{fig:PFcomparison} illustrates the 21-cm power spectrum, this time as a function of redshift for the selected modes $k\sim 0.01$, $0.06$, $0.2$, and $0.95$ h/Mpc. As before, we observe a good qualitative agreement between the models, both of them showing a characteristic double-peak feature for low $k$-modes merging into one single peak at $k\sim 1$ h/Mpc. A closer look reveals, however, that the double-peak feature survives to higher $k$-values in the halo model compared to the BLPF model. In general, we conclude that the two models differ by no more than a factor of a few in their redshift evolution, with some exceptions where the difference can grow to about an order of magnitude at most.

In the middle and bottom rows of Fig.~\ref{fig:PFcomparison} we show the auto and cross power spectrum for the individual components $\alpha$, $T$, and $m$. We observe as a general rule that the halo model and BLPF power spectra are well converged at the largest scales (lowest $k$ values) before they start to slowly diverge towards higher $k$-values. Beyond  $k\sim 1$ h/Mpc, the divergence is accentuated which is due to the dominance of the one-halo term. The largest differences are visible in the $P_{\alpha\alpha}$ and $P_{TT}$ auto spectra as well as the $P_{\alpha T}$ cross spectrum.

We conclude that the 21-cm halo model predictions are in qualitative agreement with the results from the BLPF model. In general, the halo model power spectrum is larger by a factor of a few, and, for exceptional $k$ and $z$-values the differences can grow to about an order of magnitude. We want to emphasise, however, that at this point we do not know which of the two models is more accurate. Although former findings by Ref.~\citep{Santos:2007aaa} suggest that the BLPF model lacks power with respect to semi-numerical calculations as well, we want to remind that these findings were based on a different source parametrisation and are therefore not directly comparable.

\begin{figure*}[tbp]
\centering
\includegraphics[width=0.41\textwidth,trim=-0.02cm 0.4cm 1.41cm 0.4cm,clip]{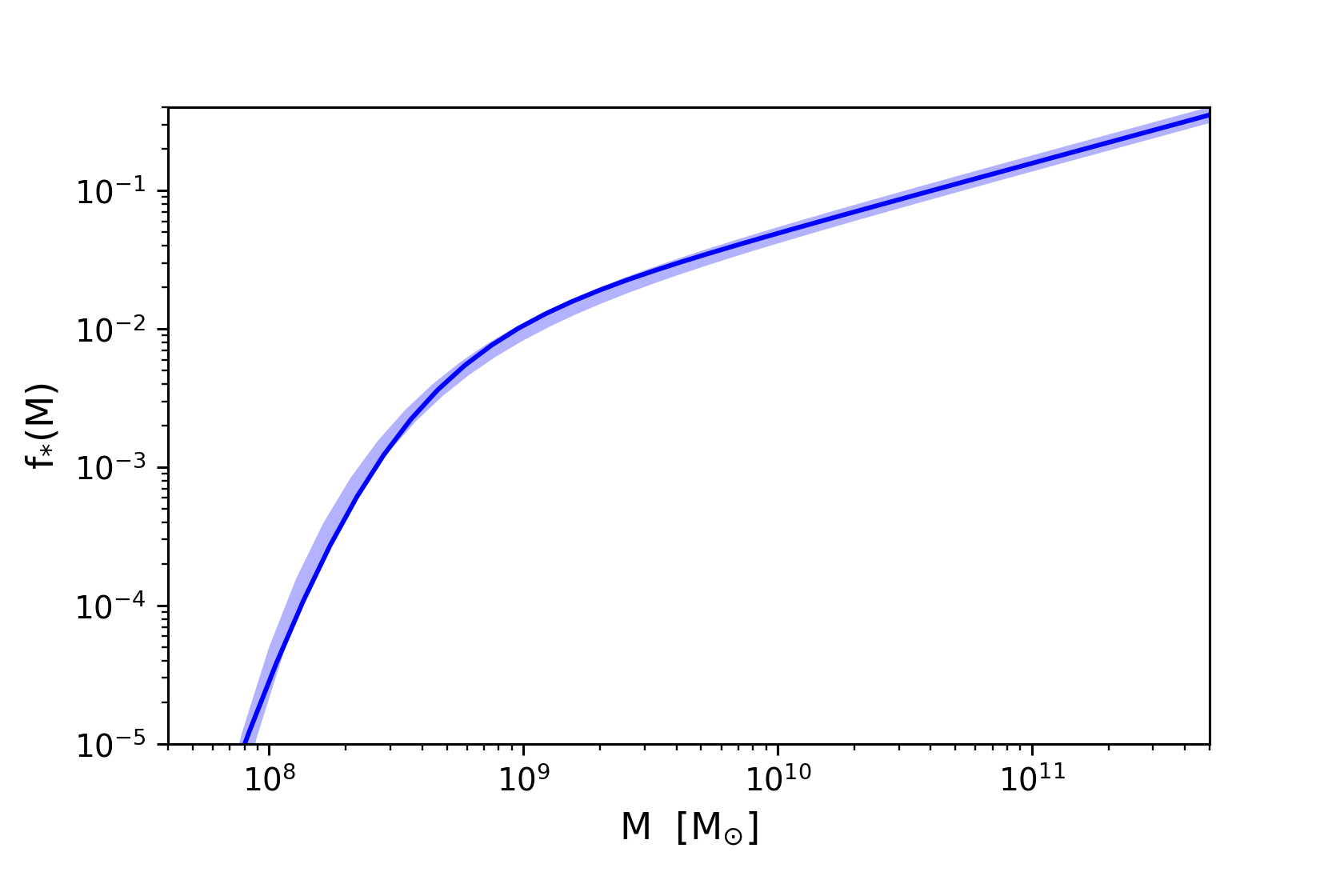}
\includegraphics[width=0.41\textwidth,trim=-0.02cm 0.4cm 1.41cm 0.4cm,clip]{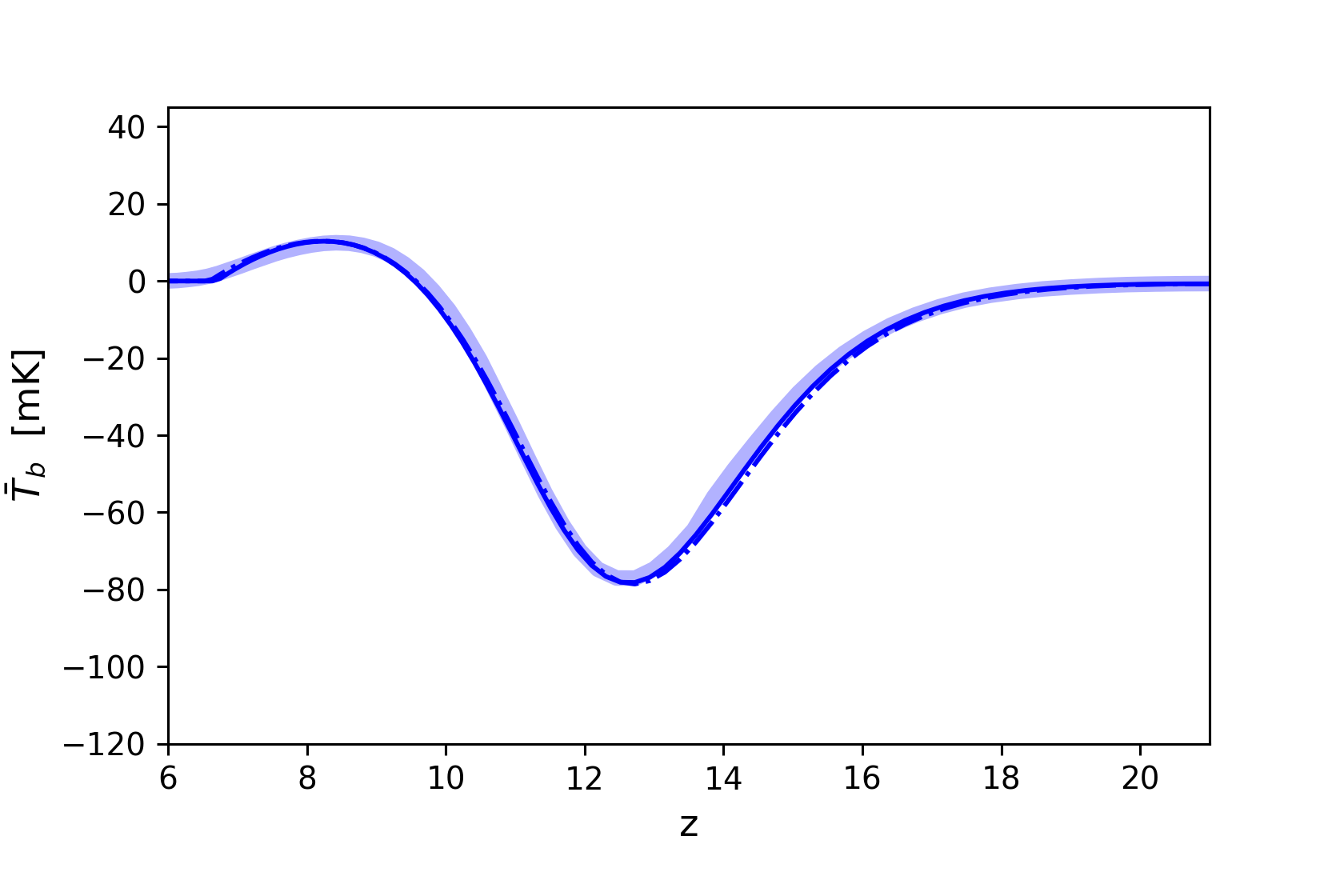}\\
\includegraphics[width=0.87\textwidth,trim=0.9cm 0.0cm 1.5cm 0.6cm,clip]{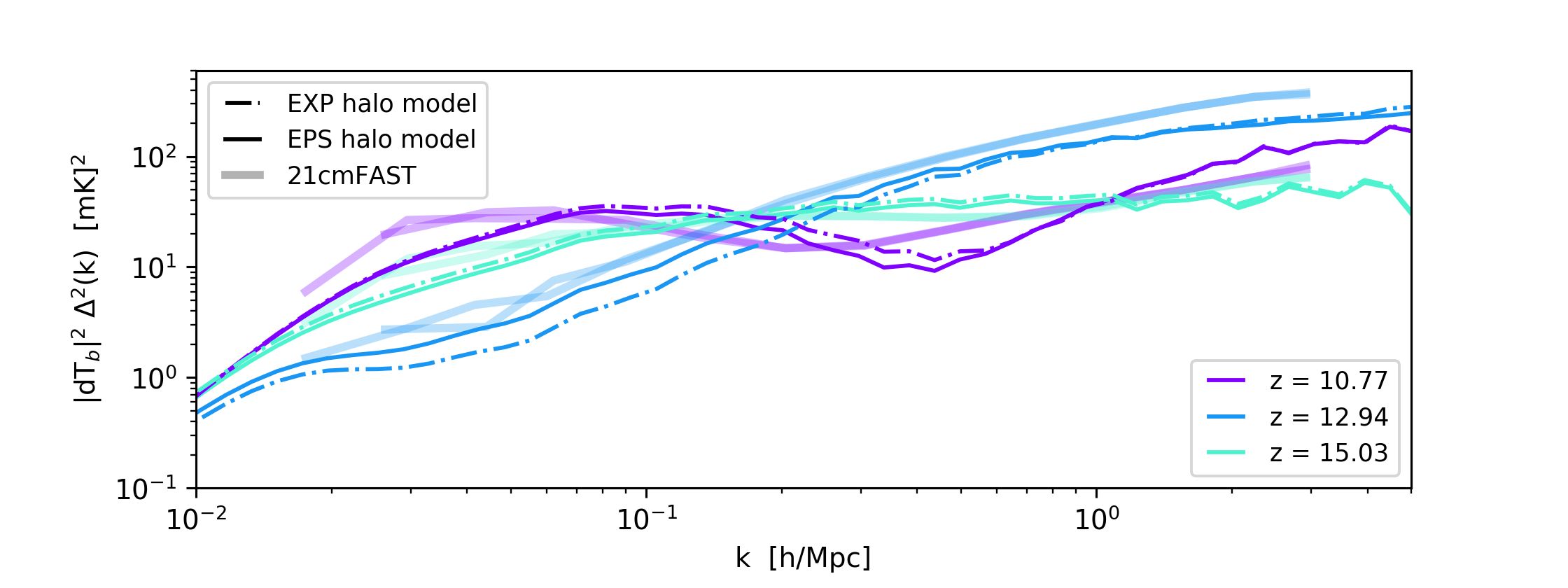}\\
\includegraphics[width=0.87\textwidth,trim=0.9cm 0.3cm 1.5cm 0.9cm,clip]{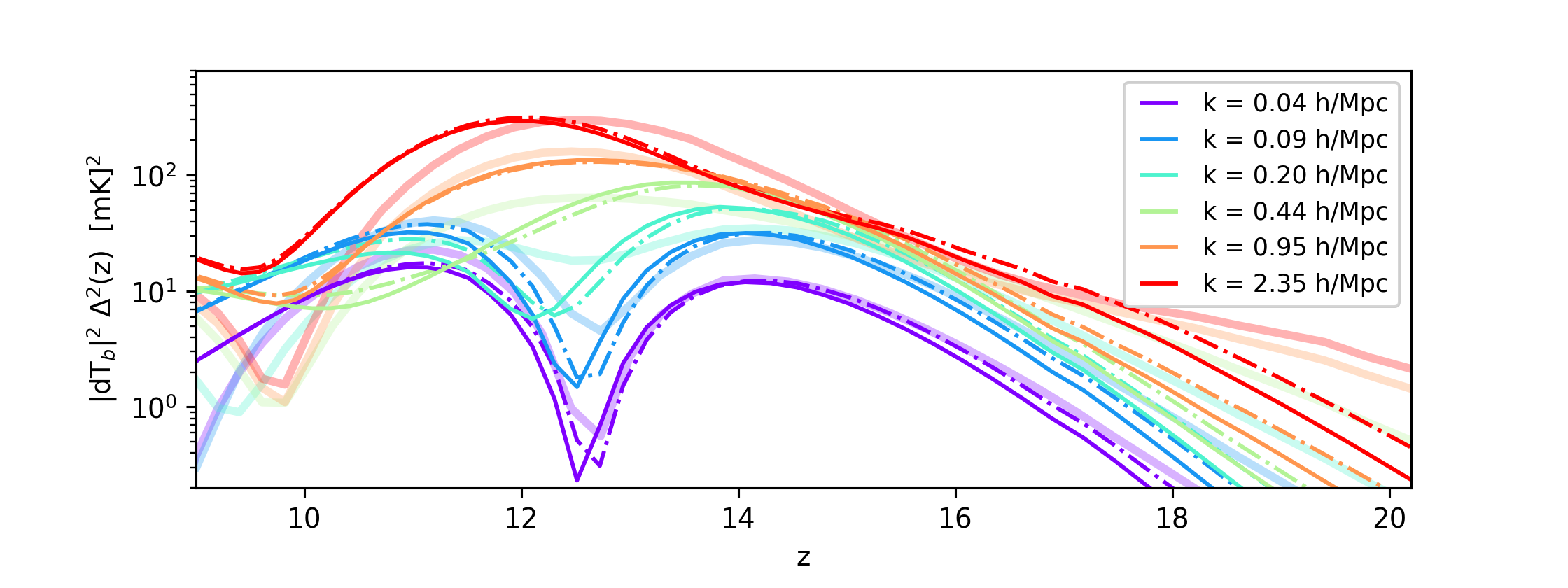}
\caption{Comparison between the halo model and 21cmFAST. \emph{Top-left:} Star formation efficiency of Eq.~\ref{fstar} fitted to the one from 21cmFAST (blue line and band, respectively). \emph{Top-right:} Global differential brightness temperature obtained by assuming the EXP and EPS accretion rates (dash-dotted and solid lines) and matched to the 21cmFast results (band) by refitting the total Lyman-$\alpha$ and X-ray photon numbers. \emph{Centre and Bottom:} Corresponding power spectra as a function of $k$-modes and redshift. Note that redshifts below $z\sim 10$ should not be trusted since effects related to reionisation become important.}
\label{fig:comparison21CMFAST}
\end{figure*}

\subsection{Comparing to {\tt 21cmFAST}}
The code {\tt 21cmFAST} \citep{Mesinger:2010aaa,Murray:2020trn} is based on a semi-numerical approach to predict 21-cm maps by solving the spin temperature evolution and the reionization process on a three-dimensional grid. The matter field is evolved according to a first-order Zel'Dovich displacement \citep{Zeldovich:1970aaa}. The sources are not resolved individually, but their distribution is calculated using an excursion-set method directly applied to the matter field.

A one-to-one comparison between our model and {\tt 21cmFAST} is not straight-forward because of subtle differences in the parametrisation and implementation that may significantly affect the results. For example, we do not exactly know which halo bias and halo mass function agrees best with the excursion set implementation used in {\tt 21cmFAST}. Furthermore, there are small differences in the parametrisations of the spectral energy distributions and the star-formation efficiency, that might be of relevance.

For the {\tt 21cmFAST} run, we assume the fiducial model from Ref.~\citep{Park:2018ljd} whose astrophysical parameters guarantee an agreement with the observed luminosity function at $z=6-10$. At the same time, it leads to a pre-reionisation signal that is shifted to somewhat lower redshifts compared to the benchmark models introduced above. The stellar-to-halo mass ratio assumed by Ref.~\citep{Park:2018ljd} corresponds to a power law, followed by an exponential cutoff towards small scales (with parameters $f_{*,10}=0.05$ for the amplitude, $\alpha_*=0.5$ for the power-law index, and $M_{\rm turn}=5\times10^{8}$ M$_{\odot}$ for the mass scale of the exponential downturn, see Eq.~(2) in Ref.~\citep{Park:2018ljd}). We attempt to reproduce the same functional form without changing the parametrisation described in Eq.~(\ref{fstar}). The best agreement is found with the parameters $f_{*,0}=0.05$, $M_p=10^{10}$ M$_{\odot}$, $\gamma_1=-0.5$, $\gamma_2=-0.5$, $M_t=2\times10^{8}$ M$_{\odot}$, $\gamma_3=1.4$, and $\gamma_4=-4$. A comparison of the two functions is shown in the top-left panel of Fig.~\ref{fig:comparison21CMFAST}.

Regarding the halo mass function, we use the Sheth-Tormen model with $q=0.707$ and $p=0.3$. Although this mass function is not such a good fit to high-redshift $N$-body simulations (see Ref.~\citep{Iliev:2011aaa,Schneider:2018xba} and discussions in Sec.~\ref{sec:massfctbias}), it is used as the reference for the collapse fraction calculated in {\tt 21cmFAST} \citep[see e.g. Eq.~14 in ][]{Mesinger:2010aaa}. For the halo bias, on the other hand, we apply Eq.~(\ref{bias}) with the Press-Schechter (PS) parameters $q=1$ and $p=0$. This is because in {\tt 21cmFAST}, sources are populated with respect to the PS conditional mass function. Finally, we do not know which prescription for the halo-accretion rate is supposed to match best with the algorithm of {\tt 21cmFAST}. For this reason we decide to show the EPS as well as the EXP approach, since both are in good agreement with high-redshift $N$-body simulations (see Sec.~\ref{sec:MA}).

In order to allow for a meaningful comparison, we select the flux parameters $N_{\alpha}=1400$ and $f_X=0.26$ so that we obtain a good match to the {\tt 21cmFAST} global signal. The spectral energy range and power-law index of the X-ray radiation, on the other hand, is kept fix at the default values of {\tt 21cmFAST} (which are $E_{\rm min}=0.5$ keV, $E_{\rm max}=2$ keV, and $\alpha_X=1$).

In the top-right panel of Fig.~\ref{fig:comparison21CMFAST}, we plot the global differential brightness temperature. The results from the 21-cm halo model with EPS and EXP accretion are shown as solid and dash-dotted lines, while the global signal from {\tt 21cmFAST} is plotted as broad blue band. Although the agreement between the lines and the band is very good (which is not so surprising, since we have selected the flux parameters $N_{\alpha}$ and $f_{\rm X}$ to obtain the best fit to the {\tt 21cmFAST} global signal), the curves are not identical. The  differences are of order 10 percent or less, and could either stem from unaccounted deviations in the source parametrisation or from the different ways the global signal is calculated.

The central panel of Fig.~\ref{fig:comparison21CMFAST} shows the power spectrum as a function of $k$-modes for three different redshifts, representing the regime dominated by heating ($z\sim10.8$, purple), the transition regime where the global absorption signal is at its maximum ($z\sim12.9$, blue), and the regime dominated by the Lyman-$\alpha$ coupling ($z\sim15.0$, cyan). The results from {\tt 21cmFAST} are again shown as broad coloured bands. Note that we plot two runs with box-size $L=300$ and $450$ Mpc, respectively (where the number of low- and high-resolution cells are kept constant at $300^3$ and $1200^3$). The resulting power spectra from the 21-cm halo model are plotted as solid and dashed lines. %They newer differ agreement with {\tt 21cmFAST} is always better than a factor of three

The bottom panel of Fig.~\ref{fig:comparison21CMFAST} illustrates the power spectrum, this time as a function of redshift. At large scales (low $k$-values) the two characteristic peaks due to the Lyman-$\alpha$ coupling and the heating epochs are clearly visible at  $z\sim11$ and 14. Arounf $k\sim0.3$ h/Mpc, a continuous transition from two to one prominent peak is visible in both models. The agreement between {\tt 21cmFAST} and the 21-cm halo model is best at very large and very small scales. In between, at the transition scale (see $k=0.2$ and $0.44$ h/Mpc), there is some visible discrepancies between the two models, with {\tt 21cmFAST} predicting more power by a factor of a few between $z\sim11$ and $13$. Furthermore, there are increasing differences towards very low and very high redshifts. Note, however, that at redshifts below $z\sim10$ the halo model results cannot be trusted because effects from reionisation are ignored. At very high redshifts, on the other hand, both predictions from {\tt 21cmFAST} and the halo model become increasingly uncertain due to the poor knowledge of source numbers and distributions.

As a summary, let us emphasise that we find good qualitative agreement between {\tt 21cmFAST} and the 21-cm halo model. At the quantitative level, the differences do not exceed a factor of $\sim3$, except at very low and very high redshifts, where the models cannot be fully trusted. Note that this level of disagreement is of the same order than the expected systematic effects due to modelling choices regarding the halo accretion, the halo abundance, and the halo bias (see Appendix~\ref{app:MAC} and \ref{app:MFB} for more details). We therefore conclude that, at the current stage, it is impossible to know if the observed differences between the 21-cm halo model and {\tt 21cmFAST} are a result of different modelling choices or if they hint towards more fundamental issues regarding the halo model approach.

\section{Conclusions}\label{sec:conclusions}
%The recent publications of strongly improved upper limits for the 21-cm power spectrum \citep{} nurture hope for a first detection in the near future.
Fast and sufficiently accurate predictions for the 21-cm clustering signal are important in order to develop a better understanding of the cosmic dawn, the epoch of the high-redshift universe right before the phase transition from neutral to ionised hydrogen. Many important aspects of the prediction pipeline, especially related to the source modelling, remain unknown, and fast models may help to explore the vast parameter space of possibilities.

In this paper we present a new analytical method based on the framework of the halo model, where Lyman-$\alpha$ coupling and temperature fluctuations are described with the help of overlying flux profiles that properly include red-shifting and source attenuation due to the expansion of the universe and the finite speed of light. The model provides a natural framework to predict all auto and cross power spectra of the Lyman-$\alpha$ coupling, the temperature fluctuations, and the matter perturbations. The temperature fluctuations are separated in a primordial component sourced by the matter perturbations and a heating component induced by the first sources. The effects form the process of reionisation could, in principle, be added to the framework, but this is left as future work.

A distinctive advantage compared to other analytical methods is that the halo model approach naturally includes shot-noise effects induced by the low number of sources in the early universe. This is important for high redshifts and small scales, especially when investigating models with very luminous sources. The influence of shot-noise on the power spectrum is model-dependent and can only be quantified in combination with the assumed astrophysical parameters.

Next to presenting the framework of a 21-cm halo model, we investigate the effects of the stellar accretion, the source abundance, and the halo bias. We show that different choices regarding these model ingredients can strongly affect both the global signal absorption trough and the 21-cm power spectrum. If the flux parameters are re-fitted to correct for the shift in the global signal, the effect on the power spectrum becomes smaller, but the difference can still be as large as a factor of a few. This means that analytical and semi-numerical methods require a precise understanding of structure formation at very early times in order to avoid significant systematic errors.

In order to check the general validity of the 21-cm halo model, we compare it to the earlier analytical model by Barkana, Loeb, Pritchard, and Furlanetto (BLPF) introduced in Refs.~\citep{Barkana:2004vb,Pritchard:2006sq}. In general, we find good qualitative agreement between the two approaches regarding both the $k$-mode and redshift evolution. At the quantitative level, the halo model predicts a somewhat larger clustering signal compared to the BLPF model. However, the difference is no more than a factor of a few, except for specific redshifts and $k$-values where it can grow to about an order of magnitude.

Furthermore, we compare the 21-cm halo model to the semi-numerical code {\tt 21cmFAST} \citep{Mesinger:2010aaa,Murray:2020trn}. Although an exact comparison between the two methods is currently unfeasible due to small differences in the source parametrisations, we nevertheless find a very encouraging agreement. At redshifts before the onset of reionisation, the difference between the halo model and {\tt 21cmFAST} stays below a factor of $\sim 3$ (being considerably better than this at most $k$-modes and redshifts). Whether the remaining differences between the models are due to the specifics of the source modelling, or whether they are a sign for a more fundamental failure of the halo model remains to be investigated in the future.

In general, a detailed apple-to-apple comparison including analytical, semi-numerical, and full simulation-based calculations of the 21-cm signal would be extremely beneficial for the community. Only a combined use of accurate but very expensive simulations together with much faster approximate methods will lead to a more complete understanding of the complex and rich signal from the epoch of the cosmic dawn.

\begin{acknowledgments}
We thank Romain Teyssier and Kai Hofmann for very helpful suggestions related to the model. This work is supported by the Swiss National Science Foundation via the grant \tt{PCEFP2\_181157}. 
\end{acknowledgments}

\appendix

\begin{figure*}[tbp]
\centering
\includegraphics[width=0.32\textwidth,trim=0.0cm 0.0cm 1.0cm 0.8cm,clip]{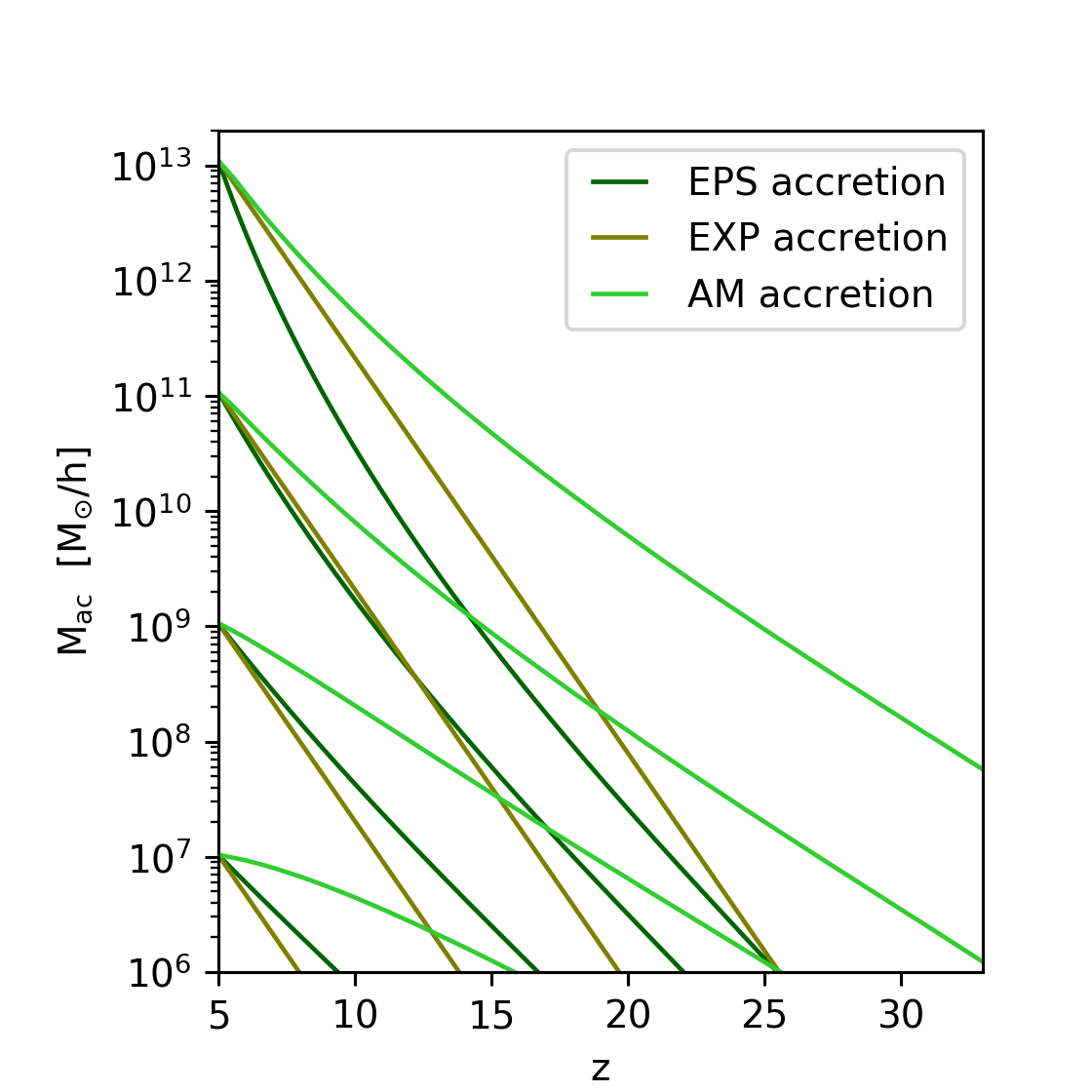}
\includegraphics[width=0.32\textwidth,trim=0.0cm 0.0cm 1.0cm 0.8cm,clip]{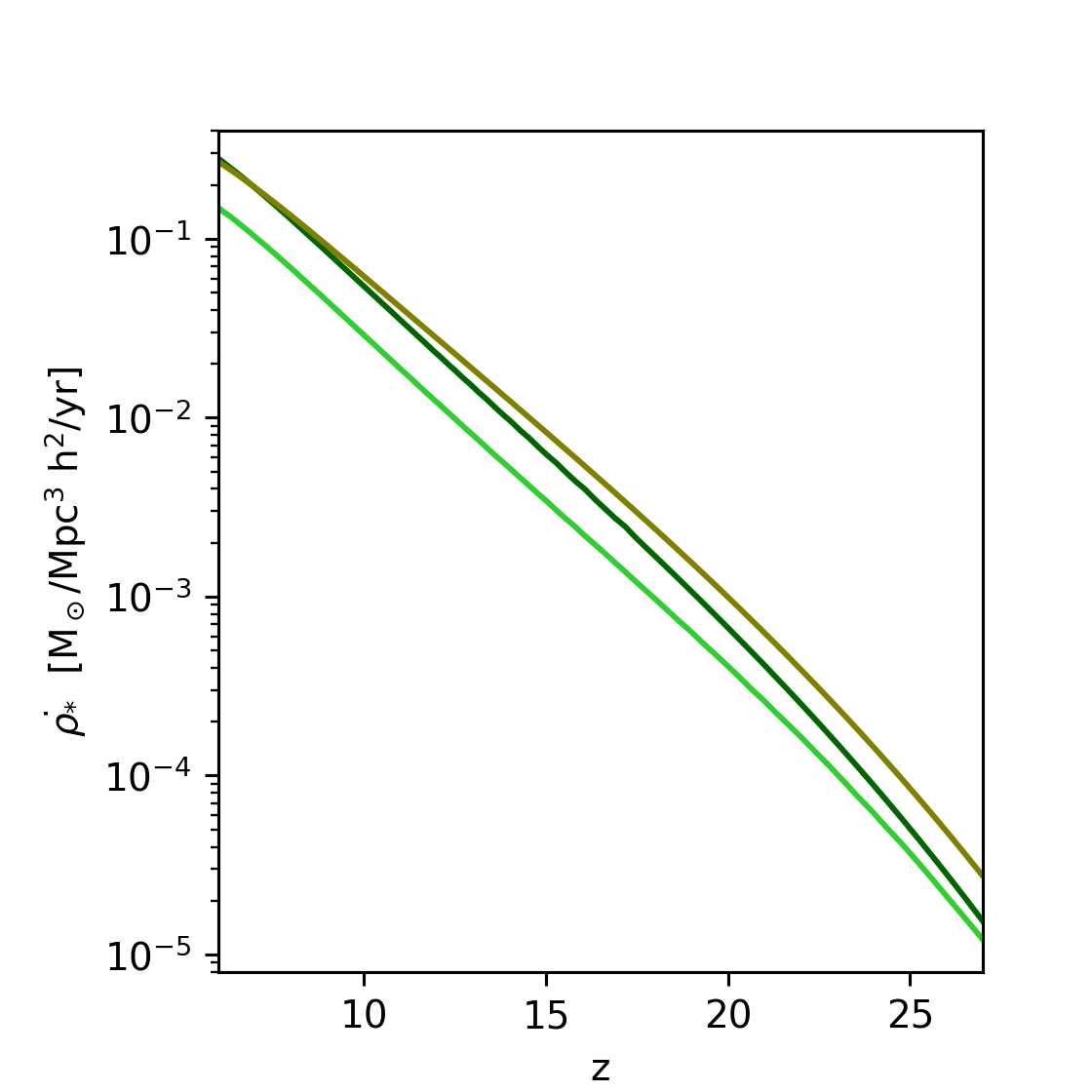}
\includegraphics[width=0.32\textwidth,trim=0.0cm 0.0cm 1.0cm 0.8cm,clip]{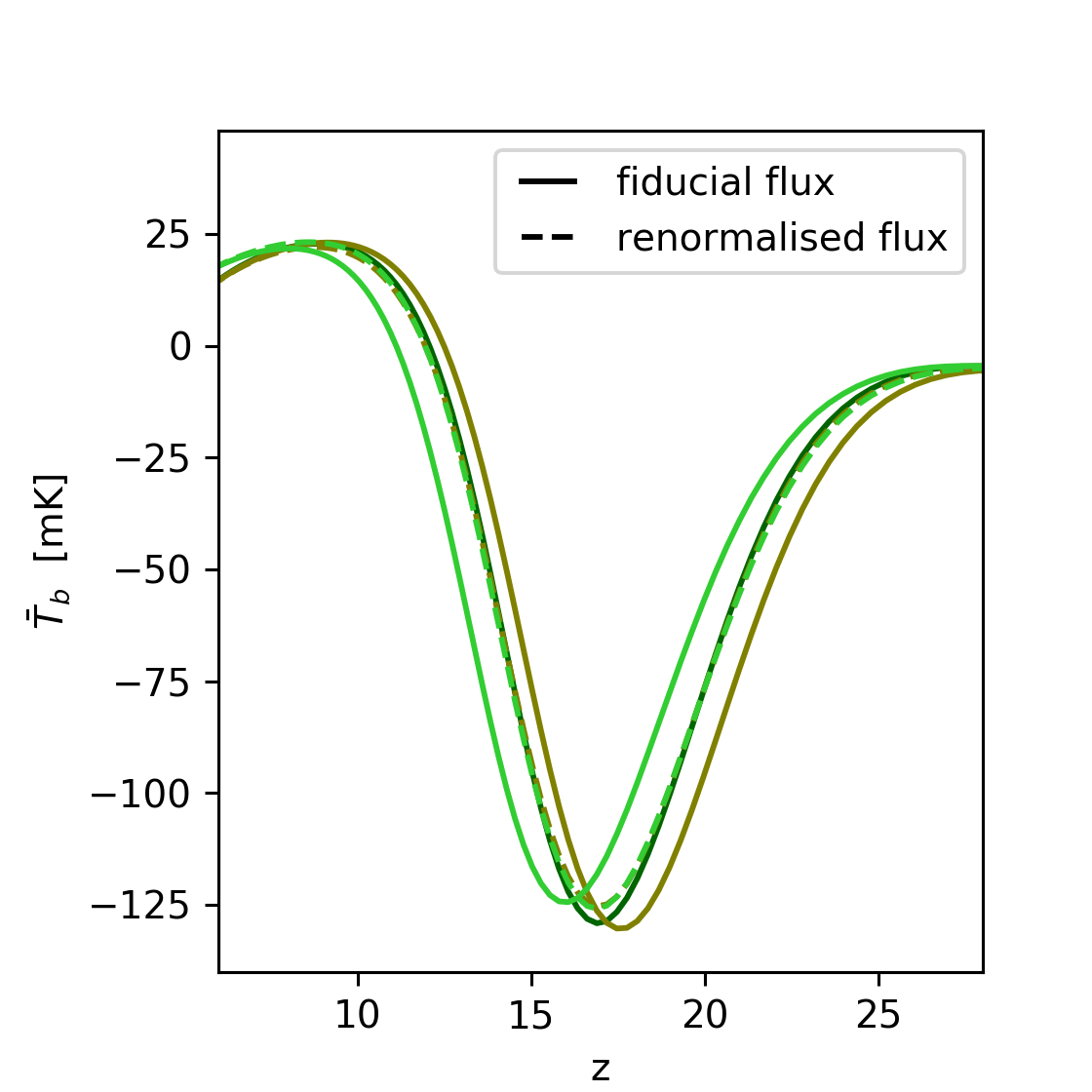}\\
\includegraphics[width=0.32\textwidth,trim=0.0cm 0.0cm 1.0cm 0.8cm,clip]{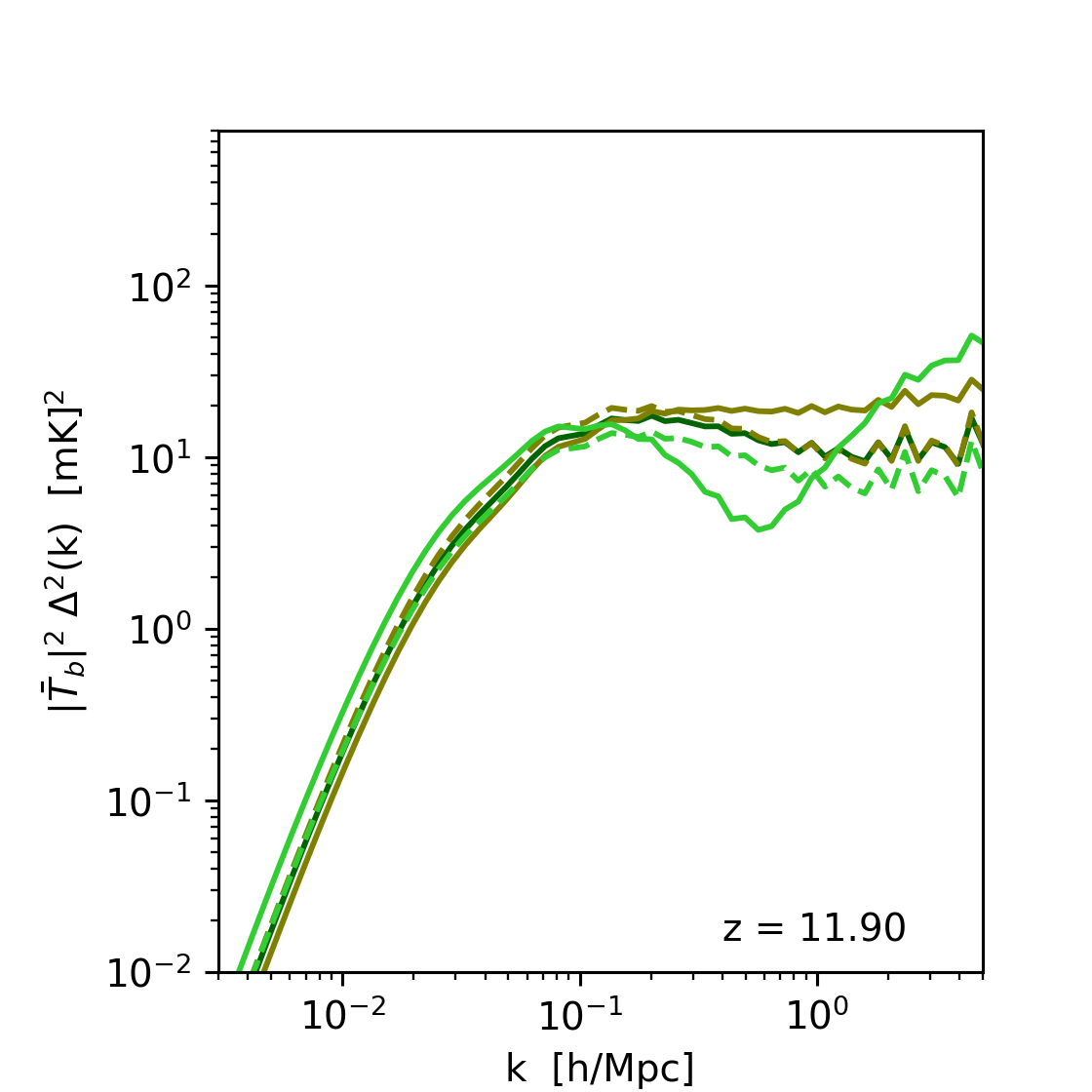}
\includegraphics[width=0.32\textwidth,trim=0.0cm 0.0cm 1.0cm 0.8cm,clip]{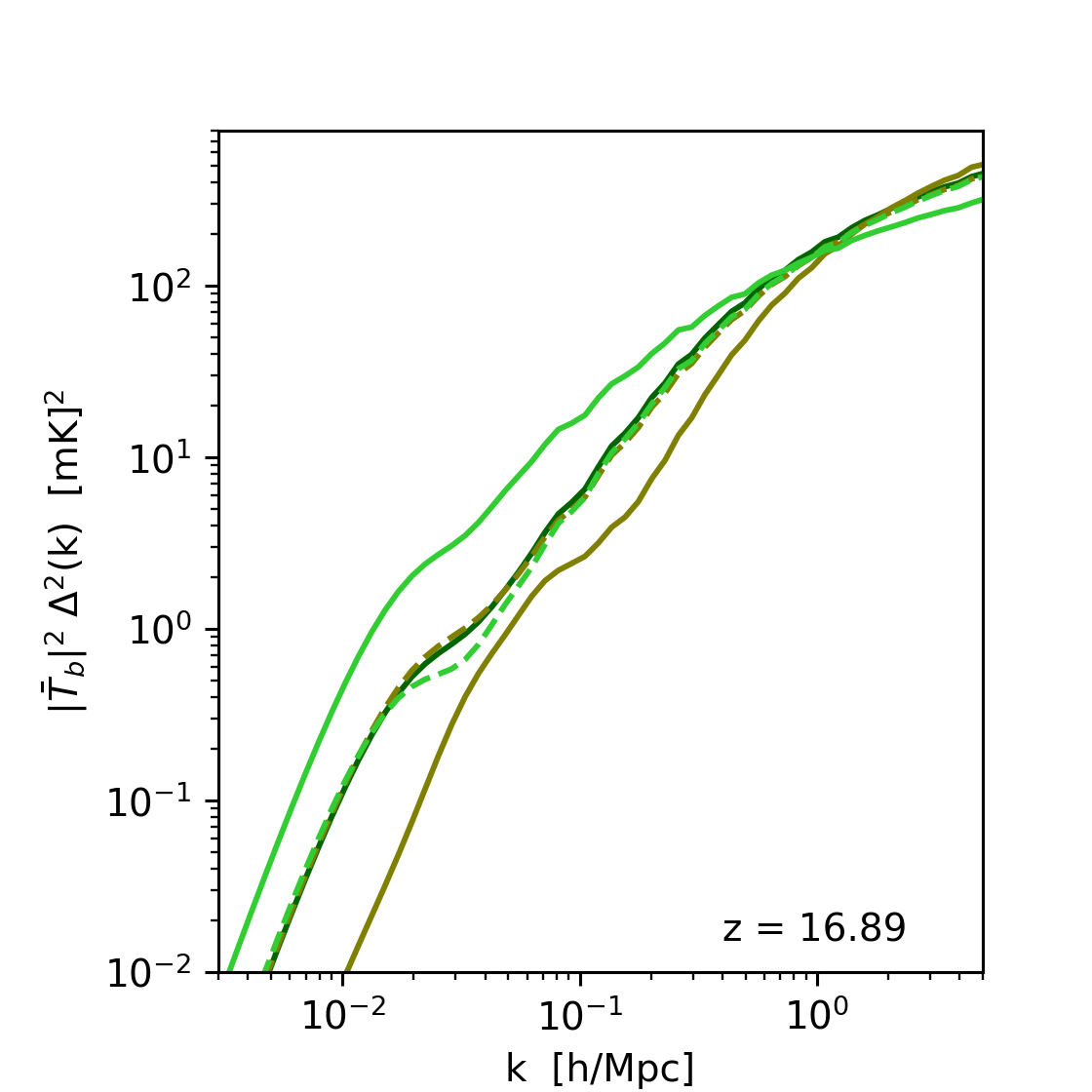}
\includegraphics[width=0.32\textwidth,trim=0.0cm 0.0cm 1.0cm 0.8cm,clip]{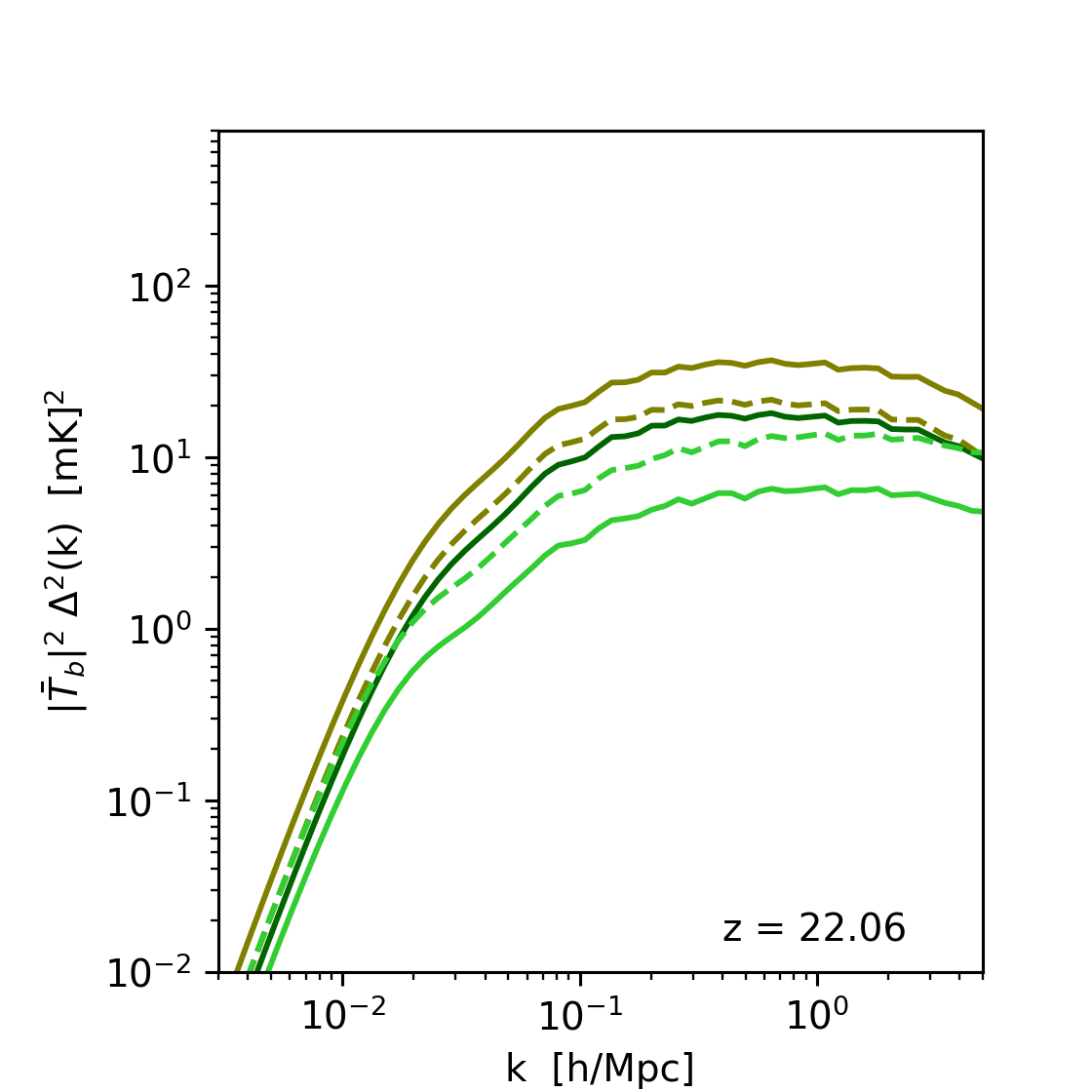}\\
\includegraphics[width=0.32\textwidth,trim=0.0cm 0.0cm 1.0cm 0.8cm,clip]{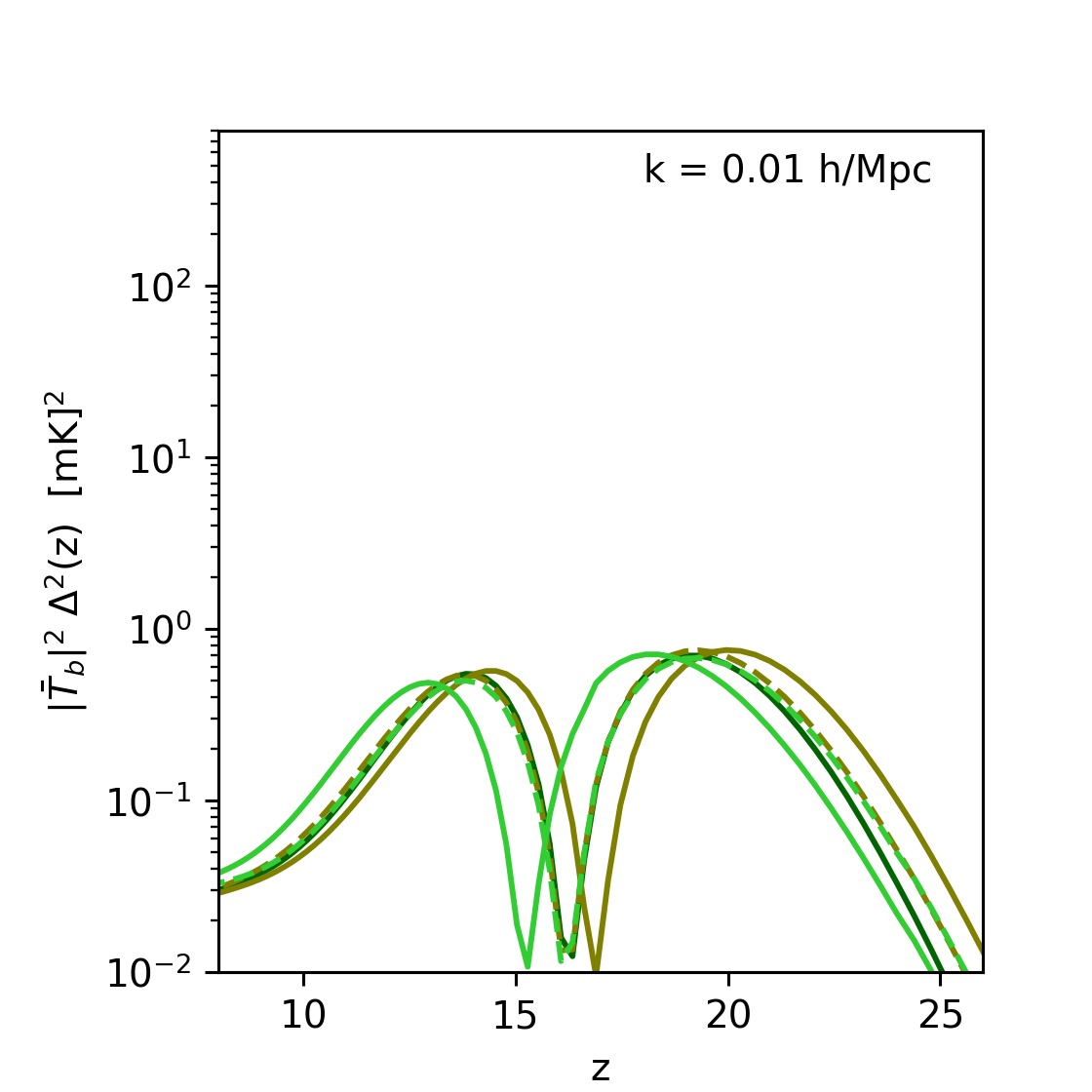}
\includegraphics[width=0.32\textwidth,trim=0.0cm 0.0cm 1.0cm 0.8cm,clip]{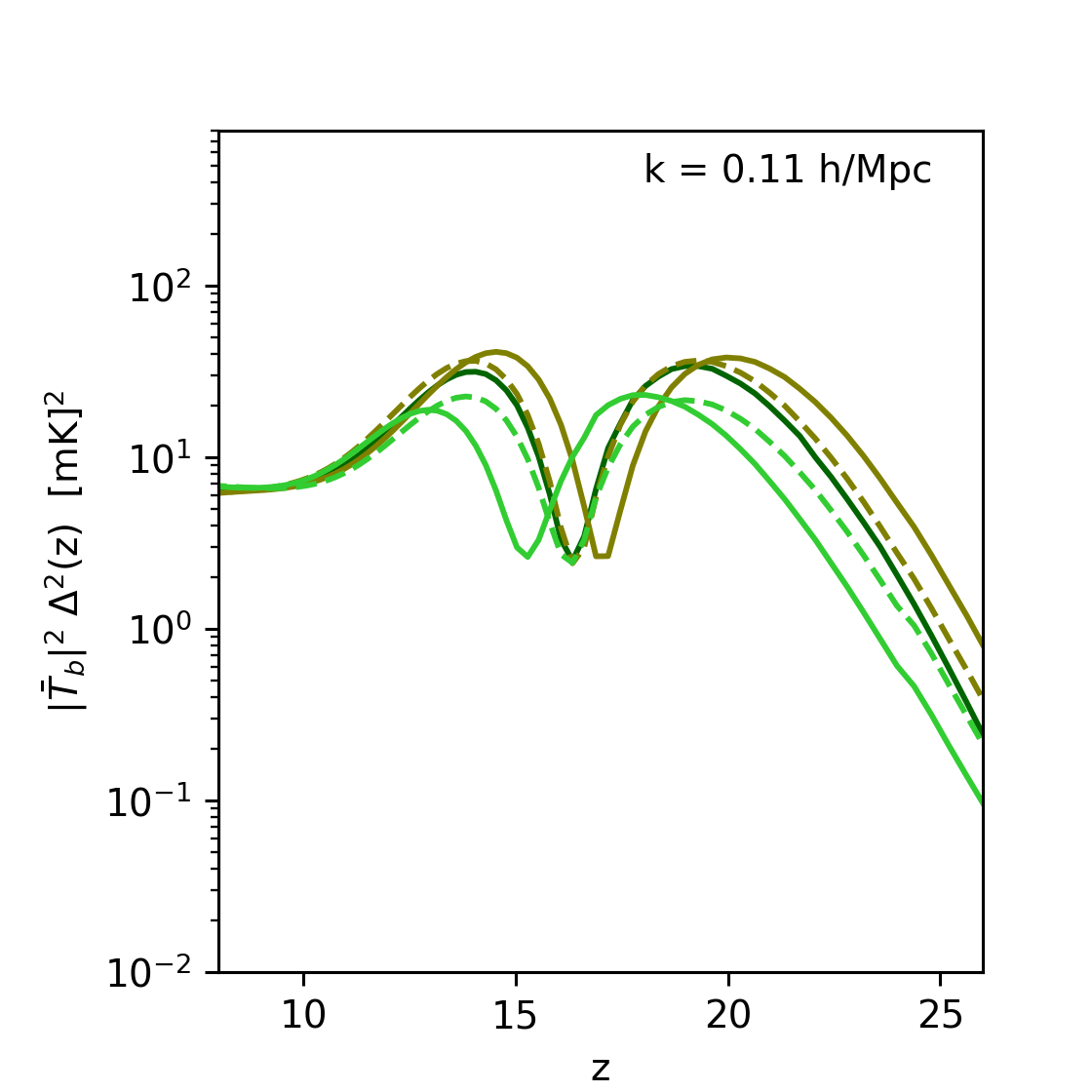}
\includegraphics[width=0.32\textwidth,trim=0.0cm 0.0cm 1.0cm 0.8cm,clip]{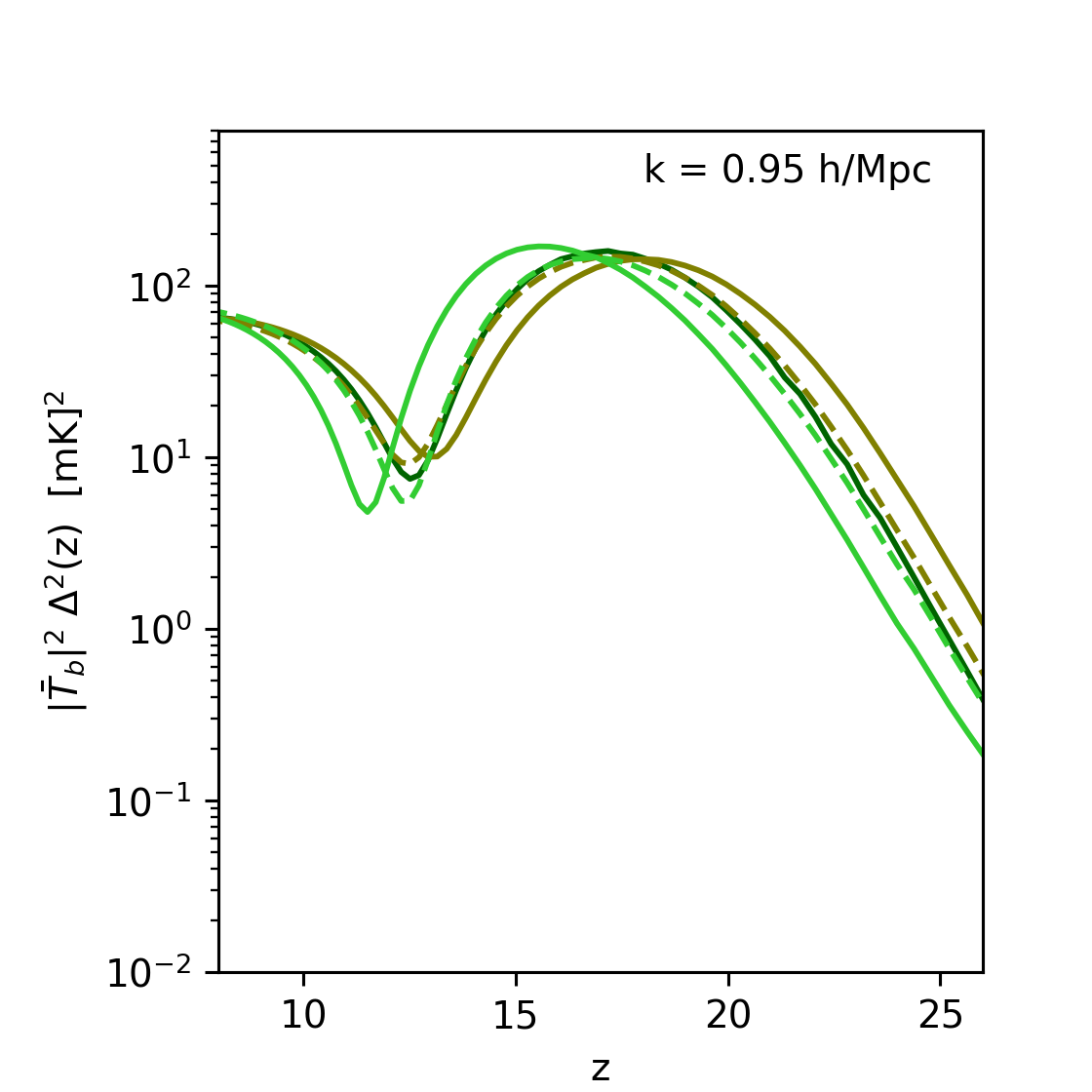}
\caption{Mass accretion modelling and how it affects the 21-cm signal assuming the source parameters of fiducial model B. \emph{Top-left:} Halo growth for the three accretion models EPS, EXP, and AM  (see Sec.~\ref{sec:MA}). \emph{Top-centre:} Star-formation rate density for the same models. \emph{Top-right:} Resulting global 21-cm signal, where the flux parameters of the EXP and AM models ($N_{\alpha}$, $f_X$) are either kept the same (solid lines) or where they are modified so that the minimum of the absorption signal lie at exactly the same redshift the the one from the EPS model (dashed lines). \emph{Centre and Bottom:} Resulting power spectra as a function of $k$-modes and redshift.}
\label{fig:resultsMAC}
\end{figure*}

\section{Effects from the mass accretion modelling}\label{app:MAC}
In Sec~\ref{sec:MA} we have investigated different modelling choices for the halo growth rate based on an extended Press-Schechter (EPS) prescription, an exponential accretion rate (EXP), and an abundance-matching technique (AM). We have compared these cases to simulations of B20 \citep{Behroozi:2020jhj} before selecting the EPS model as our standard method for the paper.

In this appendix we have a closer look at the EPS, EXP, and AM halo accretion rates, focusing on how they affect the 21-cm global signal and power spectrum. We thereby assume the fiducial model B for the source parameters. More details about model B can be found in Sec.~\ref{results3M}.

The top panels of Fig.~\ref{fig:resultsMAC} show the halo growth (left), the star-formation rate density (centre), and the global 21-cm signal (right) for the EPS, EXP, and AM models. Note that the absorption trough of the global signal is shifted towards higher redshifts when going from the AM to the EXP prescription.

In order to discriminate between effects originating from the global signal and the ones affecting the power spectrum, we allow the Lyman-$\alpha$ and X-ray flux parameters $N_{\alpha}$ and $f_X$ to be renormalised so that the global signal absorption troughs of the AM and EXP models are aligned with the ones from the EPS model. The resulting signal from such a renormalised flux analysis are given by the dashed lines in Fig.~\ref{fig:resultsMAC}.

The power spectrum as a function of $k$-modes and redshift obtained from the different accretion rate models are plotted in the middle and bottom panels of Fig.~\ref{fig:resultsMAC}. There are large differences of up to an order of magnitude between the models, clearly exceeding the differences observed in the global signal alone. The differences are smaller but still significant when looking at the models with matching global signals due to renormalised flux parameters (see dashed and dark-green solid lines). The power spectrum from these models can still differ by about a factor of 3 at most. We therefore conclude that the mass accretion rate has to be modelled at high accuracy in order to obtain reliable predictions of the 21-cm power spectrum.\\

\section{Effects from halo bias and mass function prescriptions}\label{app:MFB}
In Appendix~\ref{app:MAC} we have shown that modelling choices regarding the mass accretion rate may significant affect the 21-cm signal. Here we focus on another central modelling component: the halo mass function and corresponding bias prescription introduced in Eqs.~(\ref{massfct}-\ref{bias}).

\begin{figure*}[tbp]
\centering
\includegraphics[width=0.32\textwidth,trim=0.0cm 0.0cm 1.0cm 0.8cm,clip]{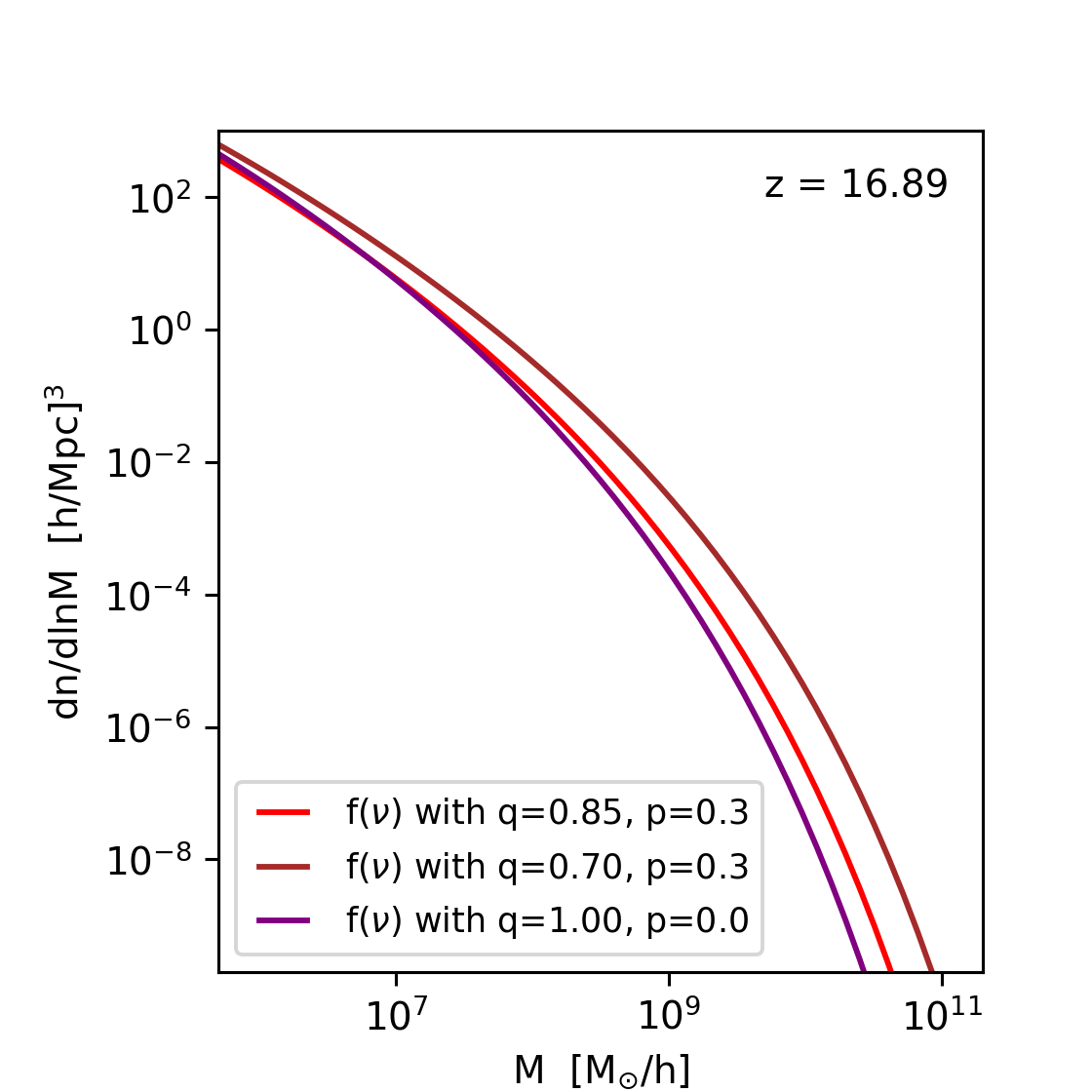}
\includegraphics[width=0.32\textwidth,trim=0.0cm 0.0cm 1.0cm 0.8cm,clip]{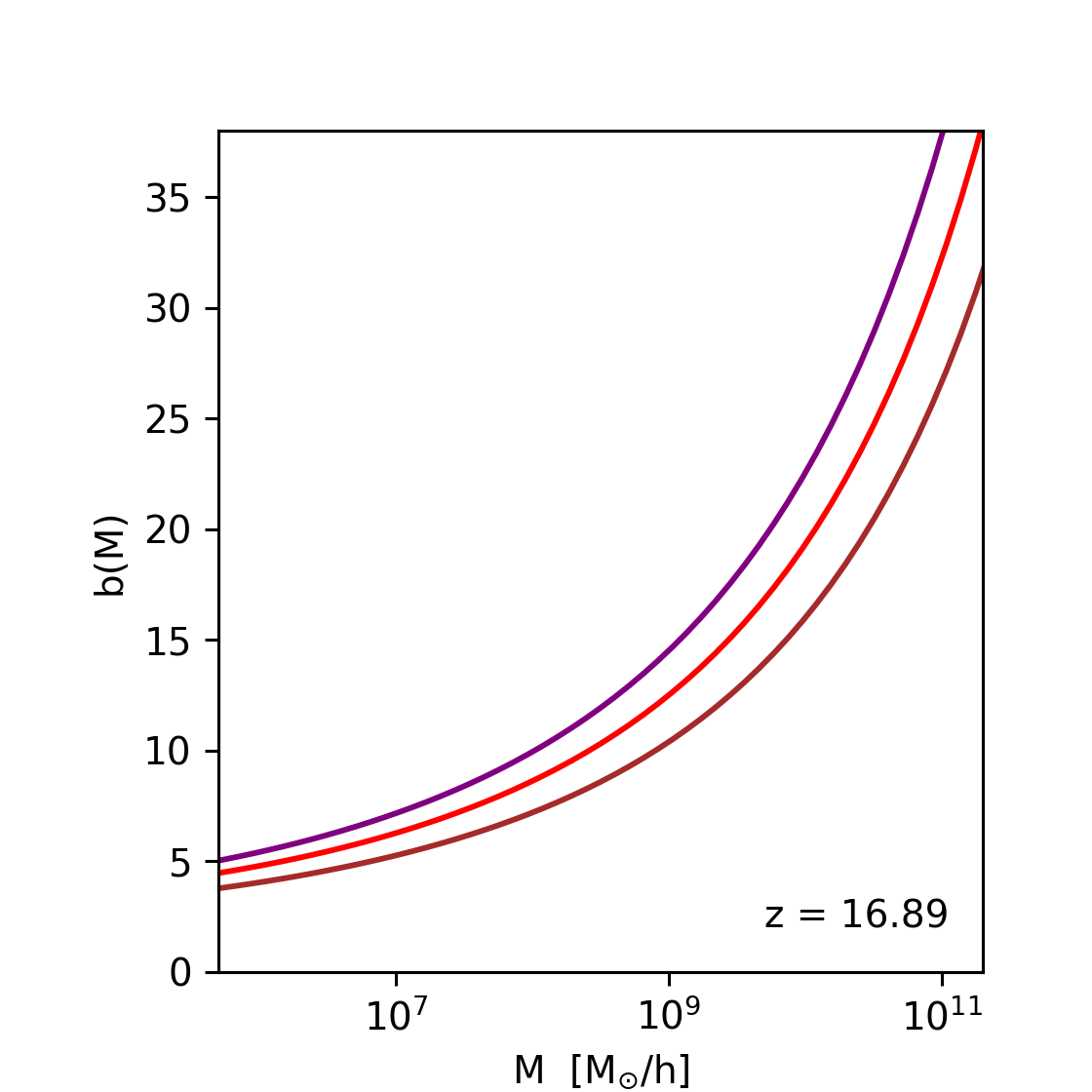}
\includegraphics[width=0.32\textwidth,trim=0.0cm 0.0cm 1.0cm 0.8cm,clip]{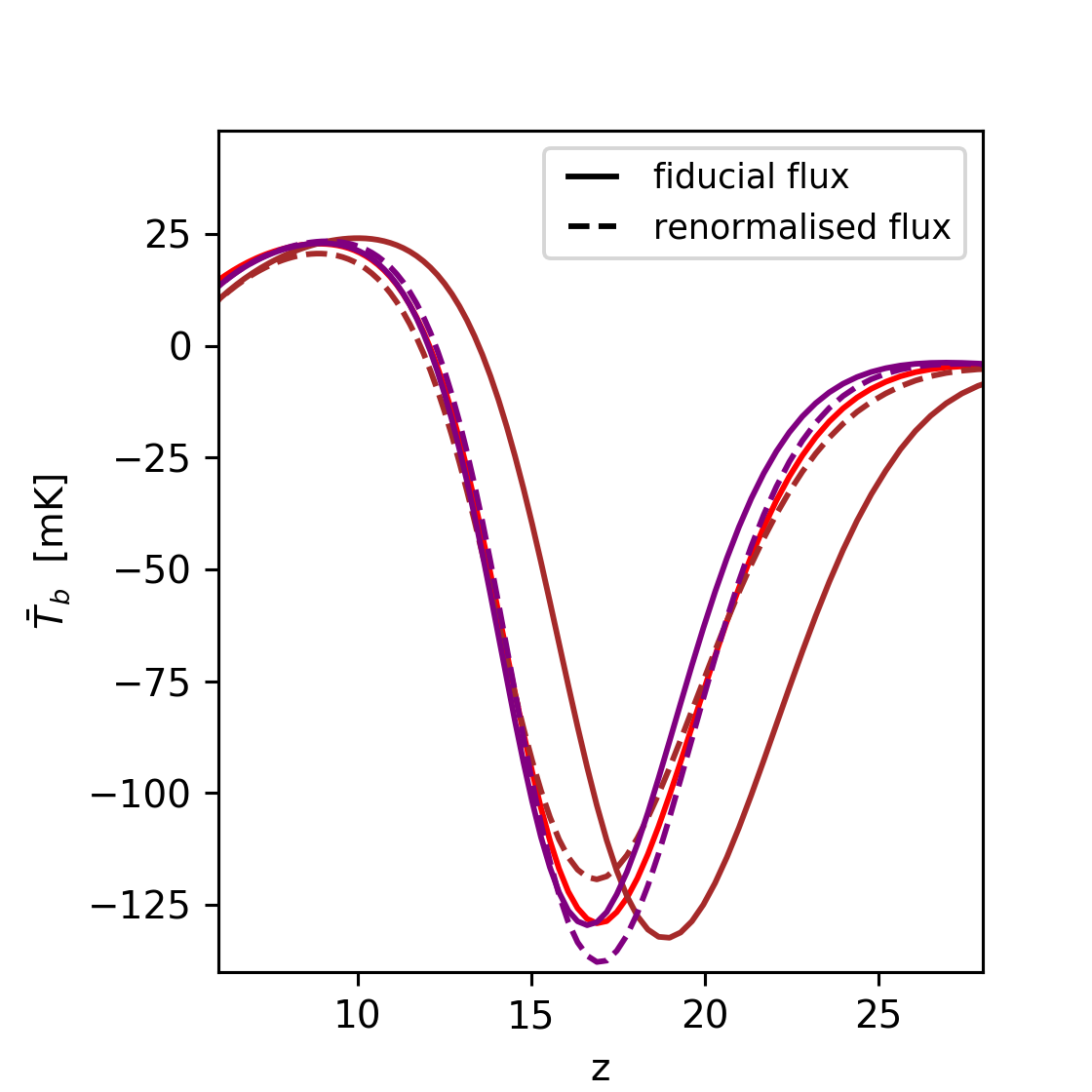}\\
\includegraphics[width=0.32\textwidth,trim=0.0cm 0.0cm 1.0cm 0.8cm,clip]{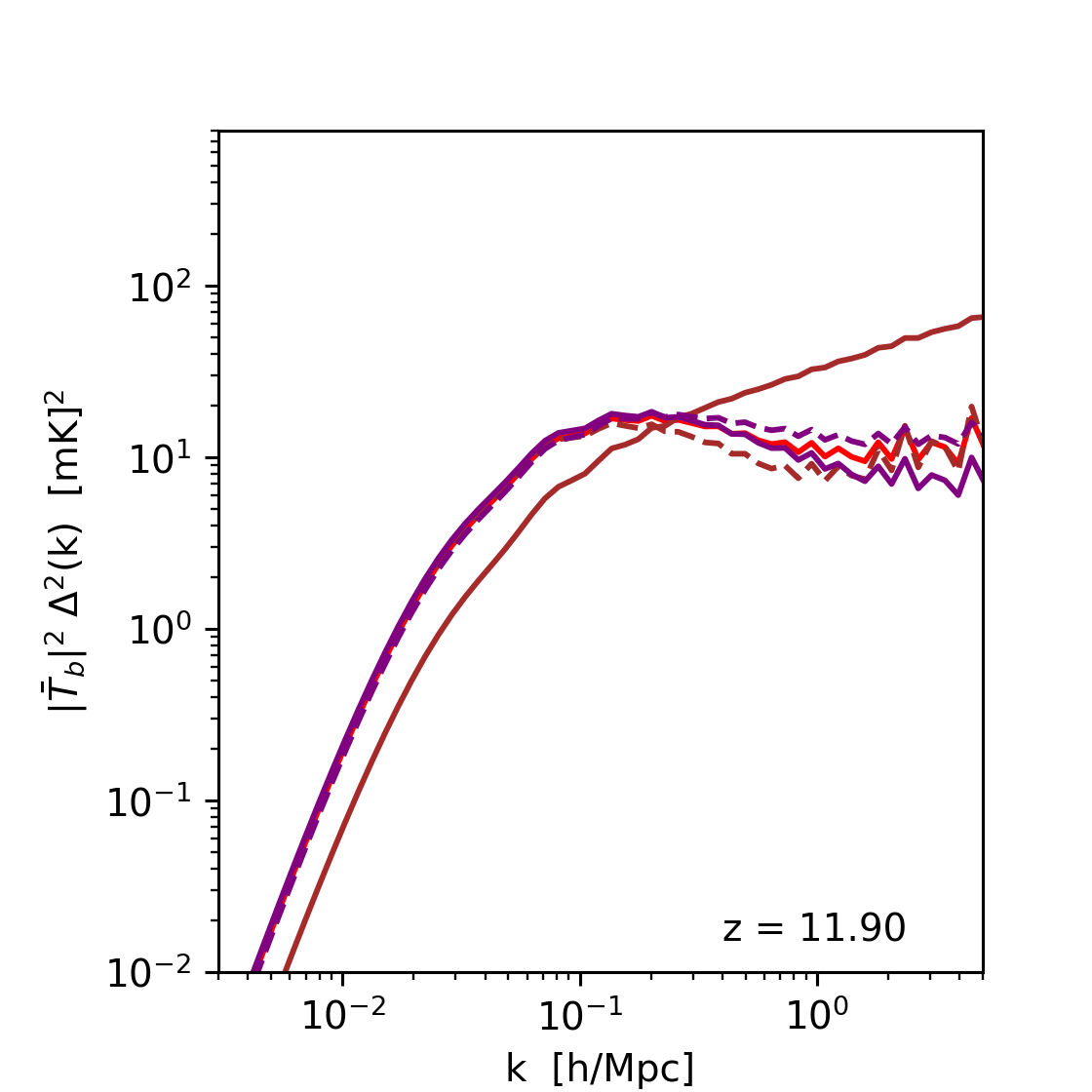}
\includegraphics[width=0.32\textwidth,trim=0.0cm 0.0cm 1.0cm 0.8cm,clip]{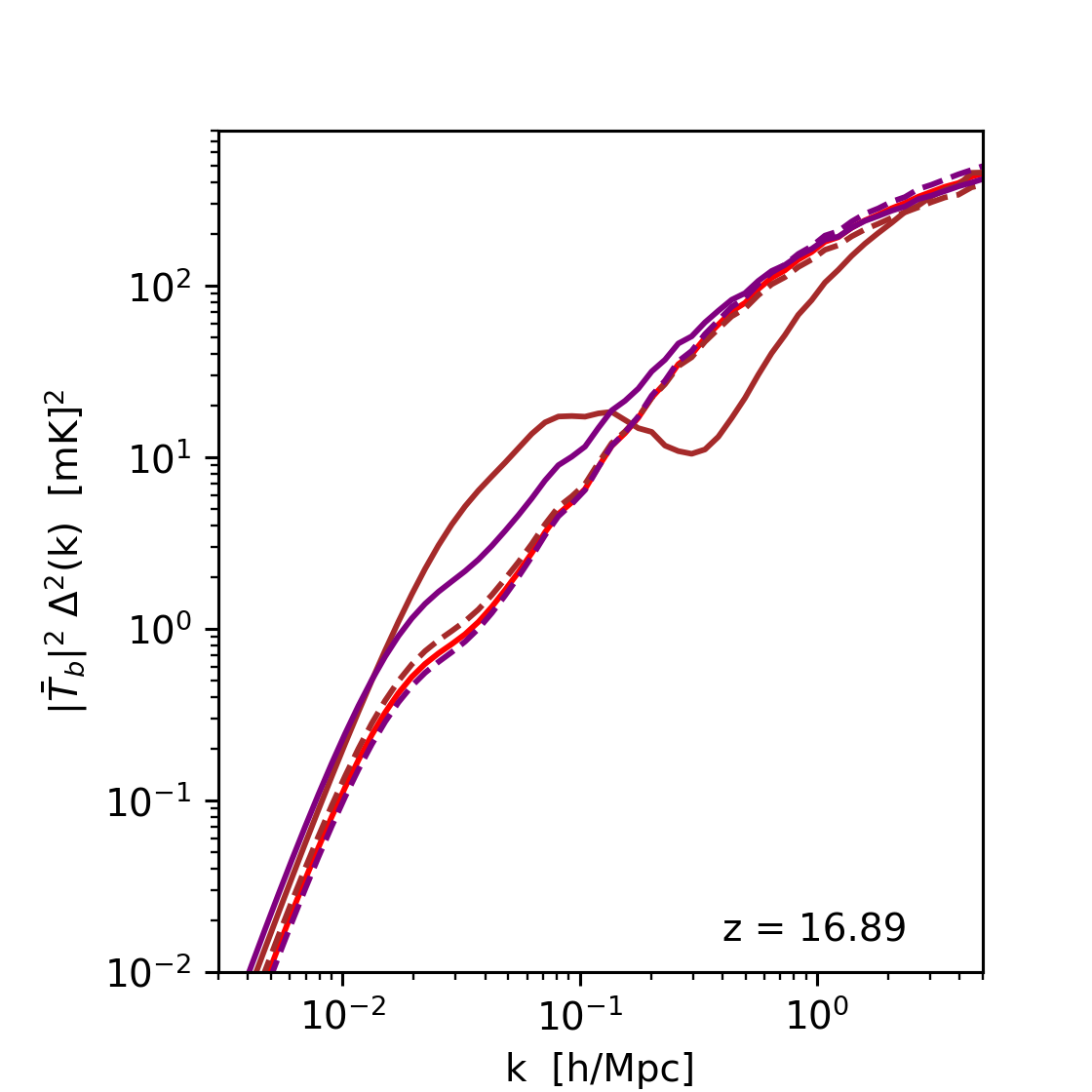}
\includegraphics[width=0.32\textwidth,trim=0.0cm 0.0cm 1.0cm 0.8cm,clip]{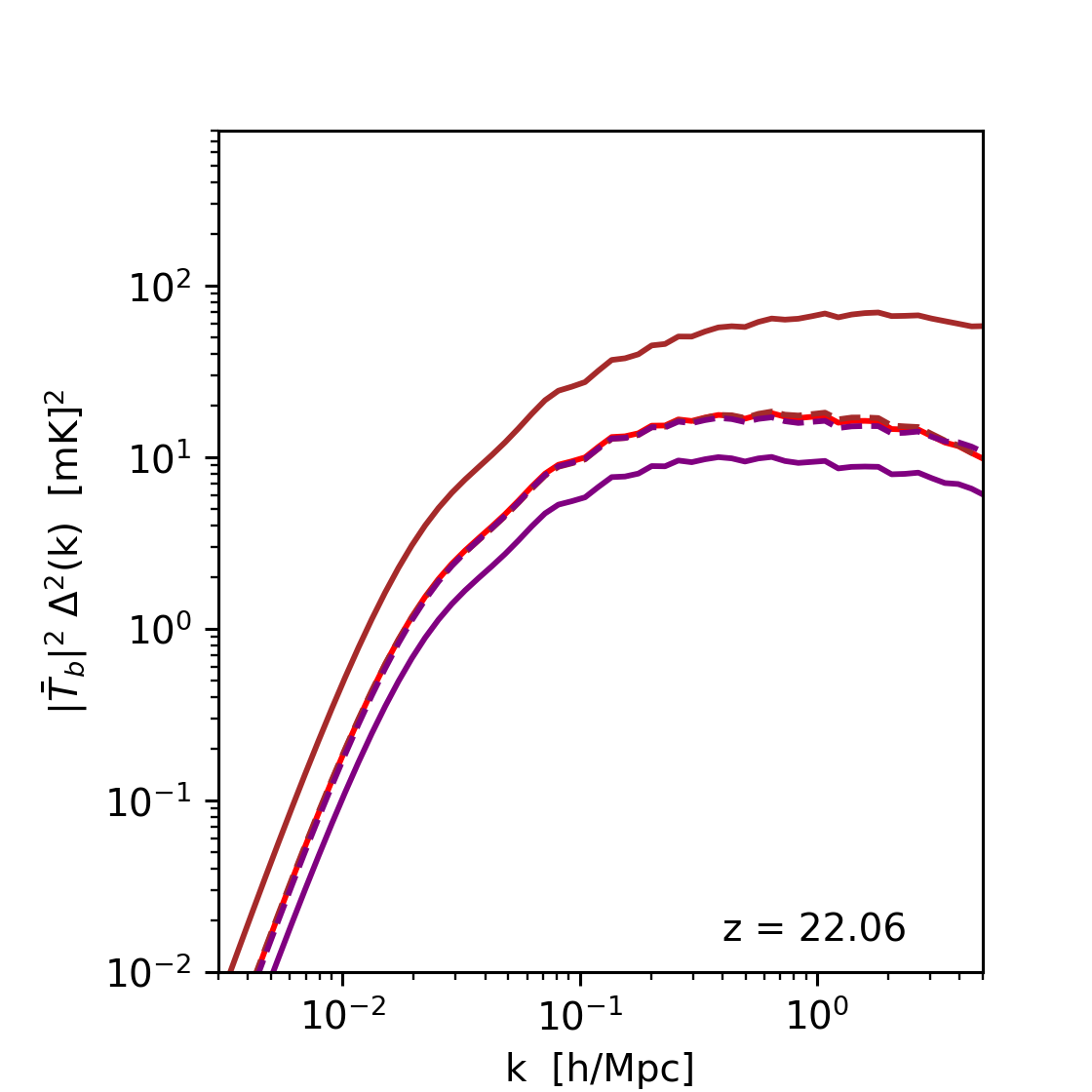}\\
\includegraphics[width=0.32\textwidth,trim=0.0cm 0.0cm 1.0cm 0.8cm,clip]{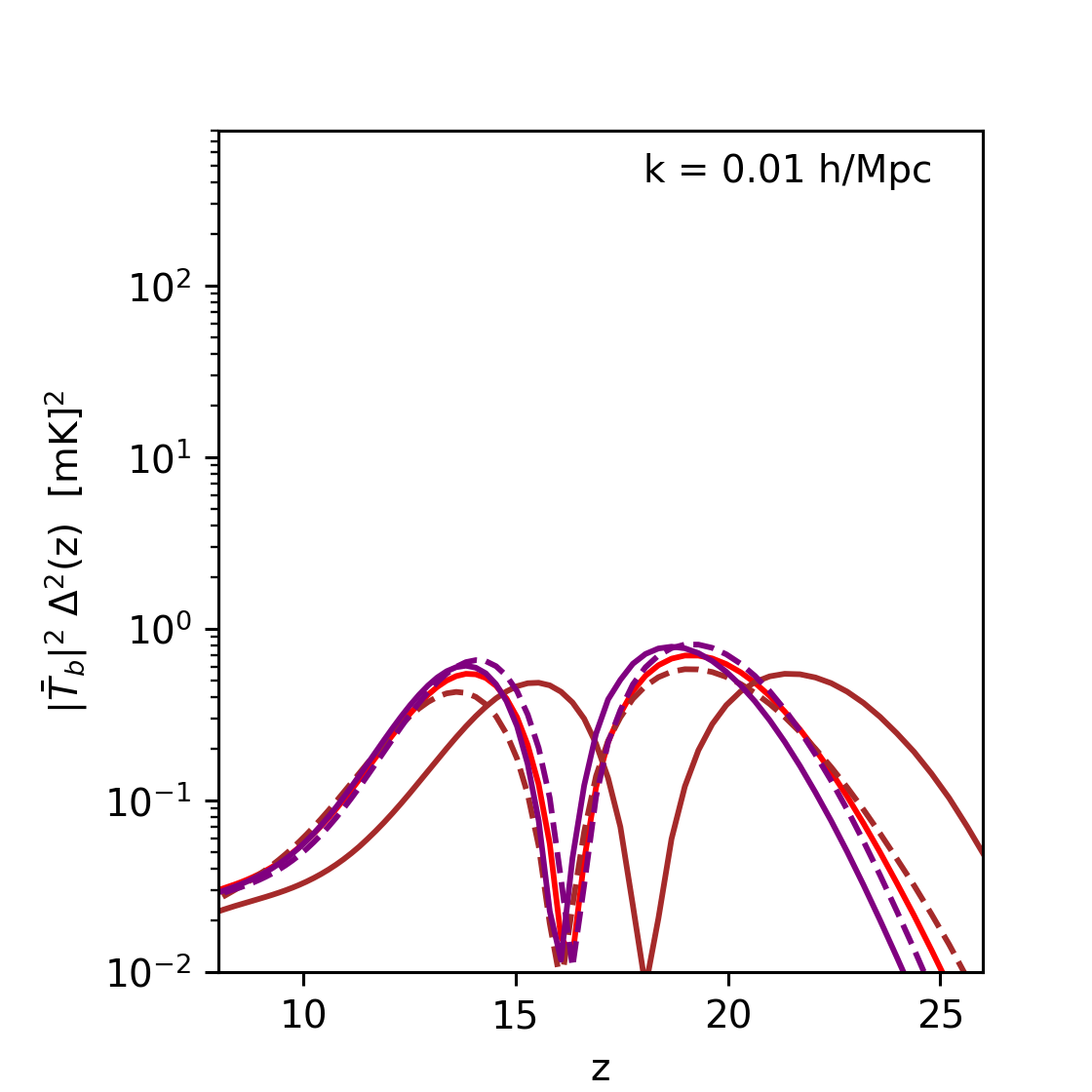}
\includegraphics[width=0.32\textwidth,trim=0.0cm 0.0cm 1.0cm 0.8cm,clip]{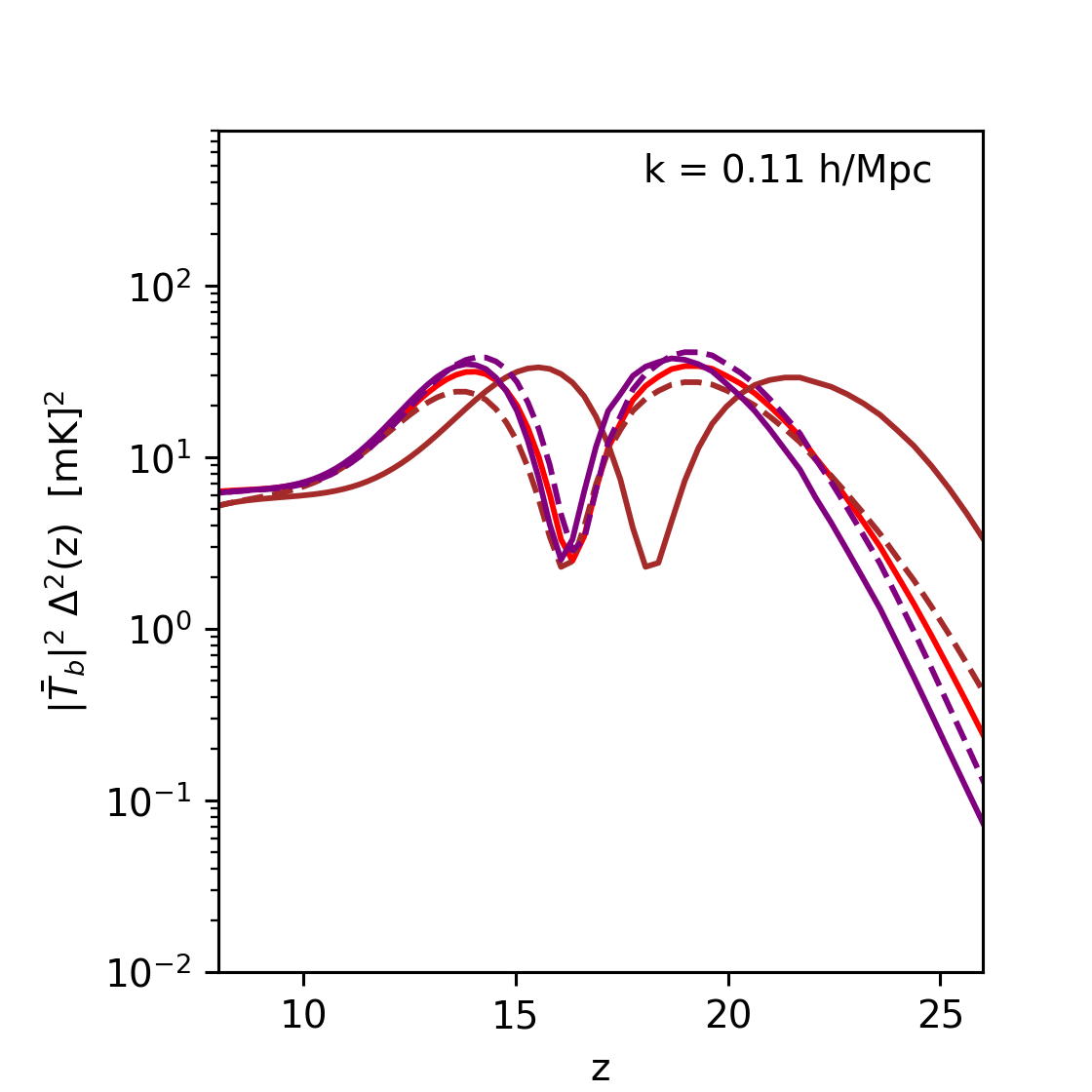}
\includegraphics[width=0.32\textwidth,trim=0.0cm 0.0cm 1.0cm 0.8cm,clip]{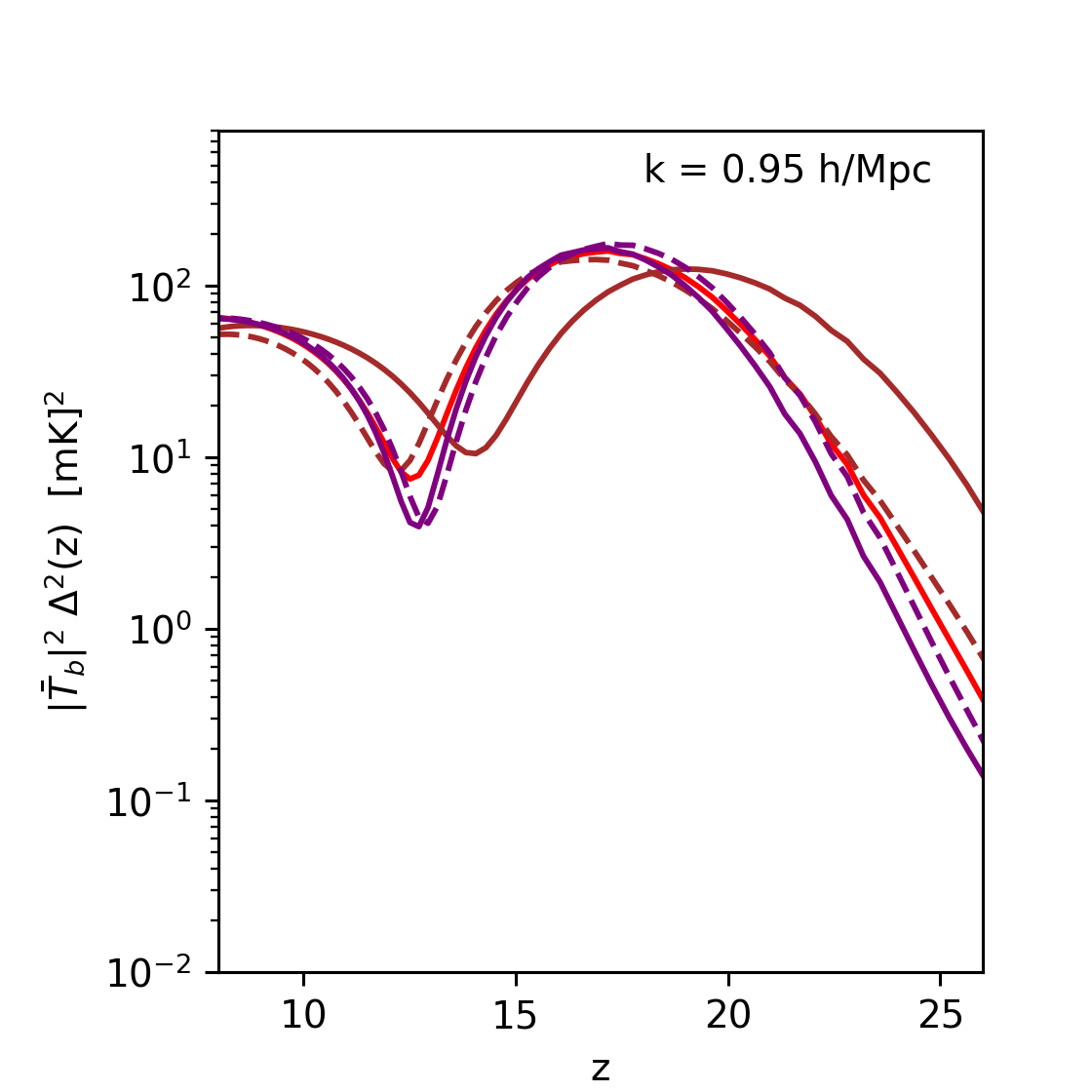}
\caption{Effects from the mass function and bias modelling on the 21-cm signal, assuming source parameters from the fiducial model B. \emph{Top-left and centre:} Halo bias and mass functions for different values of $q$ and $p$. The purple and brown lines correspond to the Press-Schechter ($q=1$, $p=0$) and Sheth-Tormen ($q=0.7$, $p=0.3$) mass functions, while the red line shows an case in-between ($q=0.85$, $p=0.3$) which is in better agreement with high redshift simulations. \emph{Top-right:} Global 21-cm signal resulting from these mass function and bias prescriptions, where the flux parameters ($N_{\alpha}$, $f_X$) are either kept the same (solid lines) or where they are changed so that the absorption troughs lie at the same redshift (dashed lines). \emph{Centre and Bottom:} Resulting power spectra as a function of $k$-modes and redshsifts.}
\label{fig:resultsMF}
\end{figure*}

In order to test the sensitivity of the 21-cm global signal and power spectrum, we vary the free model parameters $q$ and $p$ of the first crossing distribution $f(\nu)$ of Eq.~(\ref{fnu}). We thereby investigate three main cases, the original \citet{Press:1973iz} model  ($q=1$, $p=0$), the \citet{Sheth:2001dp} model ($q=0.707$, $p=0.3$), and an intermediate model ($q=0.85$, $p=0.3$) that is in best agreement with results from high-redshift simulations (see discussion in Sec.~\ref{sec:massfctbias}).

The corresponding halo bias and mass functions of the three models are shown in Fig.~\ref{fig:resultsMF} (top-left and top-centre panels). Note that the difference between the models become quite large in both cases, especially towards large halo masses. As a result, the global 21-cm signal is shifted substantially, the maximum of the absorption trough moving from $z\sim16$ to $z\sim18$ for the case of the ST instead of the PS prescription. In order to counteract this effect at the level of the global signal, we show two more models where the flux parameters $N_{\alpha}$ and $f_X$ are modified in order to bring the absorption trough back in line with the intermediate model (dashed lines). Note, however, that there are remaining differences in the shape of the absorption signal that cannot be absorbed by such a simple recalibration.

In the remaining panels of Fig.~\ref{fig:resultsMF}, we show the 21-cm power spectrum as a function of $k$-modes and redshift. There are significant differences, especially between the ST model (brown line) and the other two cases (solid red and purple lines). The recalibrated models, on the other hand, are much closer together, but differences of about a factor of two remain.

We conclude that next to the halo accretion rate, the halo bias and mass functions need to be known to good accuracy in order to avoid large modelling errors with the halo model approach. Note, that similar uncertainties also exist for semi-numerical methods or even simulation results that may only be trusted if they have realistic halo abundance and distributions.

\clearpage
\bibliography{ASbib}

\end{document}